\documentclass[useAMS,usenatbib]{mn2e}
\usepackage{charter}
\usepackage{graphicx}
\usepackage{grfext}
\PrependGraphicsExtensions*{.png,.PNG}
\usepackage{aas_macros}
\usepackage[usenames]{color}
\title[Variable Stars in Young Open Cluster NGC 2244]{Variable Stars in Young Open Cluster NGC 2244}
\author[G. Michalska]{G. Michalska$^{1}$\thanks{E-mail:
michalska@astro.uni.wroc.pl}\\
$^1$Instytut Astronomiczny, Uniwersytet Wroc{\l}awski, Kopernika 11, 51-622 Wroc{\l}aw, Poland\\
}
\begin{document}
\date{Accepted ... Received ...; in original form ...}
\pagerange{\pageref{firstpage}--\pageref{lastpage}} \pubyear{2018}
\maketitle
\label{firstpage}
\begin{abstract}
We present results of a $UBVI_{\rm C}$ variability survey in the young open cluster NGC\,2244. In total, we found 245 variable stars. Most of them, 211 stars, are variables with irregular variations. Furthermore, 23 periodic variables were found. 
We also detected four candidates for $\delta$ Scuti stars and 7 eclipsing binaries.

Based on the mid-infrared {\it Spitzer} and WISE photometry and near infrared $JHK_{\rm S}$ 2MASS photometry we classified 104 young stellar sources among our variables: 1 Class I object, 1 Class I/flat spectrum object, 4 flat spectrum objects, 91 Class II objects and 7 transition disk objects. This classification, together with $r^{\prime}i^{\prime}$H$\alpha$ IPHAS photometry and $JHK$ UKIDSS photometry, were used for identification of pre-main sequence stars among irregular and periodic variables. In this way, 97 CTTS candidates (96 irregular and one periodic variable), 68 WTTS candidates (54 irregular and 14 periodic variables) and 6 Herbig Ae/Be stars were found. 

For 223 variable stars we calculated membership probability based on proper motions from Gaia DR2 catalogue. Majority of them, 143 stars, are cluster members with probability greater than 70 percent. For only 36 variable stars the membership probability is smaller than 20 percent.

\end{abstract}


\section{Introduction}\label{sintro}

Stars in open clusters form in approximately the same chemical environment and have nearly the same age and distance. Multiwavelength photometry of young stellar clusters allows studying stars at the pre-main sequence (PMS) stage of evolution. The photometric variations in these stars are believed to originate from several mechanisms like rotation of a spotted star or obscuration by circumstellar matter \citep[see][and references therein]{herbst1994}. PMS stars are usually divided into two main groups: low mass ($<$\,2.5\,M$_{\sun}$) T Tauri stars (TTSs) and more massive (1.5\,--\,15\,M$_{\sun}$) Herbig Ae/Be (HAeBe) stars \citep{art2010}. Depending on the strength of the emission in the H$\alpha$ line, the TTSs are divided into weak line TTSs (WTTSs; with equivalent widths (EW), smaller than 10\,{\AA}) and classical TTSs (CTTSs; with EW\,$>$\,10\,{\AA}). Both WTTSs and CTTSs show brightness variations in X-ray, ultraviolet, optical and infrared domains.

Photometric variability of CTTSs is irregular. It may be attributed to unsteady accretion from the circumstellar disk as well as rotation of the surface hot and dark spots. The amplitudes range from few hundredths to several magnitudes in $V$ band \citep{herbst2000}. The light variations of WTTSs are periodic and related to the rotation of large dark spots on stellar surface.  Typical periods of such changes range between 0.5 and 18~days with amplitudes between 0.03 and 0.3~mag in $V$ band \citep{semkov2011}.

The HAeBe stars, defined by \citet{herbig1960}, have spectral types earlier than F5 and are characterized by broad emission lines and infrared excess due to a circumstellar disk. They are often located in obscured regions, in the vicinity of bright nebulae. Their light variability is irregular, connected with accretion processes and dust clumps in the surrounding disk. 
Light curves of many HAeBe stars with spectral types later than A0 and some TTS show 1\,--\,3~mag deep algol-like minima lasting several days to weeks. This class of variable is called UX~Orionis stars (UXors), after the prototype star UX Ori \citep{herbst1994}. 

The other interesting group of variable PMS stars are low-mass FU Orionis stars \citep{herbig1977} -- FUors. The characteristic feature of these objects is an increase in brightness by about 3\,--\,6~mag, lasting several months, followed by a slow decay over several years. The brightening is caused by a sudden change in the accretion rate from about 10$^{-7} M_{\sun}$\,yr$^{-1}$ (typical for TTSs) to 10$^{-4}M_{\sun}$\,yr$^{-1}$. Their spectral types change from typical for TTS to F or K-type supergiant. The third group of irregularly variable PMS stars are EXors, after the prototype, EX Lupi \citep{herbig1989}. These stars repeatedly undergo increase in brightness by about 1\,--\,4~mag, and, after a few weeks or months, fade back to their original brightness.

The observations (in particular, of stars in very young open clusters) show that some PMS stars pulsate or at least some candidates for PMS pulsating stars can be indicated. In recent years, the number of known PMS pulsating stars increased largely. The most common are PMS $\delta$~Sct-type \citep{zwintz2008,diaz2014} and $\gamma$~Dor-type stars \citep{zwintz2013}. A hybrid $\delta$~Sct/$\gamma$~Dor candidate was even found \citep{ripepi2011}. By now, no confirmed SPB or $\beta$ Cephei-type PMS star is known, however, although two suspected PMS SPB stars were discovered from the optical photometry obtained by MOST satellite in the region of NGC\,2244 \citep{gruber2012}. 

In this paper, we present the results of the search for variable stars in the very young open cluster NGC\,2244. The paper is organized as follows. First, we briefly introduce the cluster, our photometric observations, reduction process and transformation to the standard system (Sect.~\ref{subvi}). Variable stars found in the field we observed and their membership probability are presented in Sect.~\ref{svar} and Sect.~\ref{smem}, respectively. The identification of Young Stellar Objects based on {\it Spitzer} and 2MASS photometry is described in Sect.~\ref{syso}. The classification of PMS variables is discussed in Sect.~\ref{spms}. The results are summarized in Sect.~\ref{ssum}.

\section{The cluster and its $\bmath{UBVI_{\rm C}}$ photometry}\label{subvi}
\subsection{The cluster}

The very young open cluster NGC\,2244 is associated with the famous Rosette Nebula and located in the Perseus Arm of the Galaxy. Its age, based on isochrone fitting, was estimated for 2\,--\,4~Myr, the distance for 1.4\,--\,1.7~kpc, and the reddening, in terms of the colour excess $E(B-V)$, for 0.47~mag \citep{ogura,hens2000,parksung}. The cluster is embedded in H\,II region and is rich in O and B-type stars. Its photometric $UBVI_{\rm C}$ and H$\alpha$ study was performed by \citet{parksung}. They classified about 30 O and B-type cluster members. They also discovered 21 PMS stars with H$\alpha$ in emission (four of them are massive HAeBe stars) and six PMS candidates with detectable X-ray emission. A large population of PMS stars in NGC\,2244 was also reported by \cite{bonat} and \citet{berg2002}.

The cluster contains O-type $\beta$ Cephei type star \citep{bri2011}, double-lined eclipsing binary V578~Mon \citep{hens2000} and several other spectroscopic binaries. It also contains  one known Ap star, NGC\,2244-334 \citep{bagnulo2004}, having strong magnetic field. This star, however, was outside the field we observed.

\subsection[]{Observations and reductions}\label{sobs}

The photometric observations of NGC\,2244 were carried out with the Yale SMARTS 1-m telescope at Cerro Tololo Inter-American Observatory (CTIO) in Chile. The telescope was equipped with Y4KCam CCD camera covering about 20$^{\prime}\times$\,20$^{\prime}$ area in the sky. Between December 24, 2008 and January 8, 2009, we collected about 2000 frames in $V$ filter, 170 frames in $I_{\rm C}$ filter, 150 in $B$ filter, and 70 in $U$ filter. The exposure times ranged from 15 to 200 s, depending on the filter, seeing and sky transparency. All frames were calibrated with the use of Phil Massey scripts\footnote{http://www2.lowell.edu/users/massey/obins/y4kcamred.html.} designed to Y4KCam. Then, the images were reduced with the DAOPHOT II package (\citealt{stet1987}, \citealt{stet1992}). 

\subsection{\label{sstransf}Transformation to the standard system}

Our photometry was transformed to the standard system. Since many observed stars show irregular variability, we decided to calculate mean instrumental magnitudes and colours applying 2$\sigma$ clipping.
Then, these mean values were tied to the $UBVI_{\rm C}$ photometry of NGC\,2244 published by \citet{parksung}. The transformation was based on the photometry of about 400 stars common with these authors. The following transformation equations were obtained:
\begin{equation}
\label{eq_V}
V-v=(\mbox{0.009}\pm\mbox{0.003})\times(v-i)+(\mbox{13.0030}\pm\mbox{0.0017}),
\end{equation}
\begin{equation}
\label{eq_VI}
V-I_{\rm C}=(\mbox{0.993}\pm\mbox{0.002})\times(v-i)+(\mbox{0.9920}\pm\mbox{0.0009}),
\end{equation}
\begin{equation}
\label{eq_BV}
B-V=(\mbox{0.855}\pm\mbox{0.003})\times(b-v)+(\mbox{0.6754}\pm\mbox{0.0010}),
\end{equation}
\begin{equation}
\label{eq_UB}
U-B=(\mbox{1.107}\pm\mbox{0.008})\times(u-b)+(\mbox{0.104}\pm\mbox{0.003}),
\end{equation}
where $u$, $b$, $v$ and $i$ denote the mean instrumental magnitudes from our photometry, while $U$, $B$, $V$, and $I_{\rm C}$, the standard magnitudes. The residual standard deviations for the transformation equations were equal to 0.030, 0.016, 0.014 and 0.034~mag for Eq.~(\ref{eq_V}), (\ref{eq_VI}), (\ref{eq_BV}), and (\ref{eq_UB}), respectively.  

From the above equations, we computed $V$ magnitudes and the $(V-I_{\rm C})$ colour indices for 4058 stars, and the $(U-B)$ and $(B-V)$ colour indices for 989 and 1347 stars, respectively. Standard photometry of the variable stars is given in Table \ref{tubvi}\footnote{The full version of Table\,\ref{tubvi} is available in electronic form from the CDS.} in Appendix. The equatorial coordinates shown in this Table \ref{tubvi}, were determined by means of an astrometric transformation of mean stellar positions derived from images with the use of IRAF {\tt ccmap/cctran} tasks. As a reference catalogue, we used the UCAC4 catalogue \citep{zach2013}.

\begin{figure*}
\centering
\includegraphics[width=14.6cm]{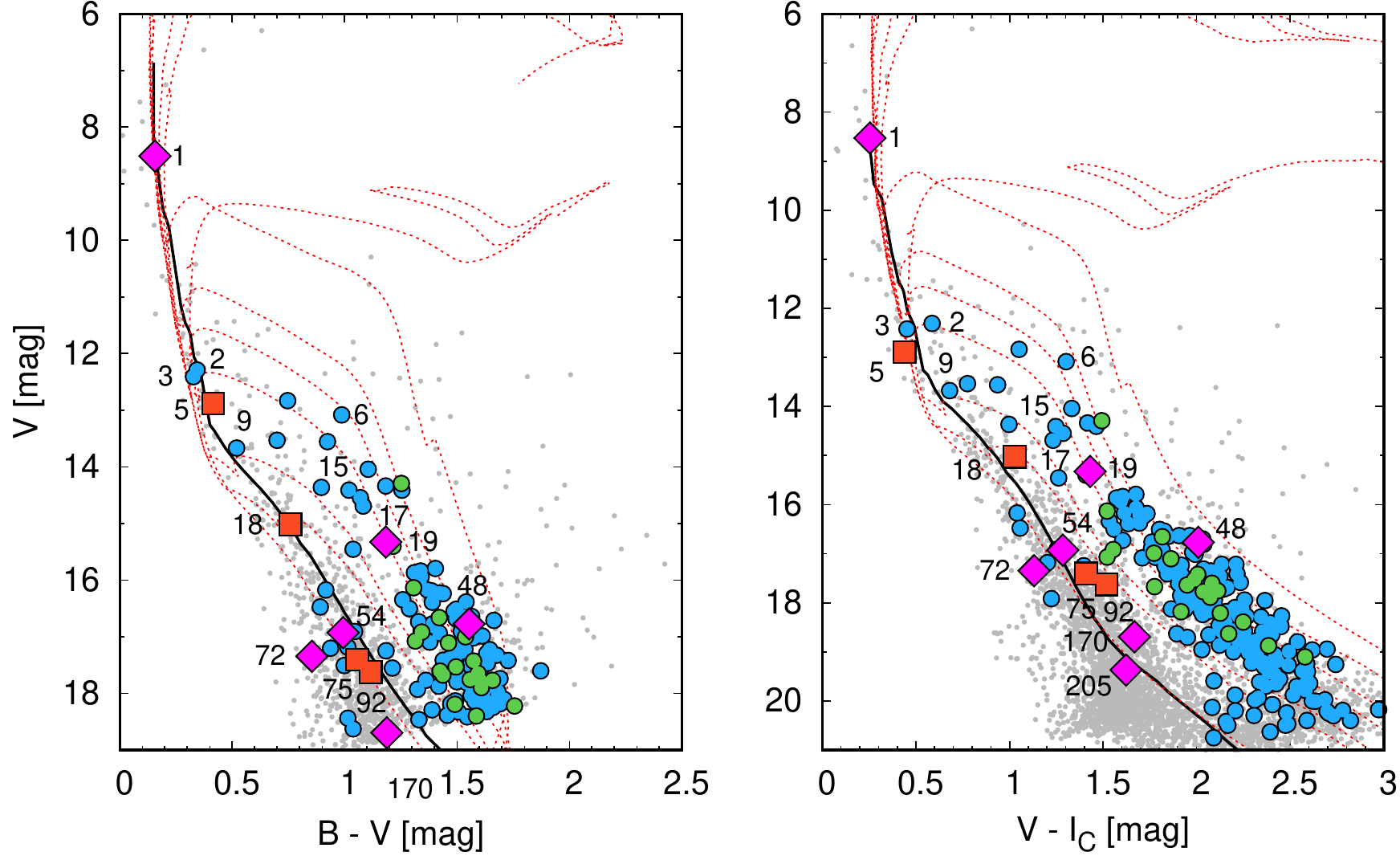}
\caption{Variable stars in $V$ vs. $(B-V)$ ({\it left}) and $V$ vs. $(V-I_{\rm C})$ ({\it right}) colour-magnitude diagrams for the observed field. The symbols indicate variable stars found in this paper: pulsating stars(red squares), eclipsing stars (pink diamonds), other periodic variables (green) and remaining variables (blue). Some variables discussed in text are labeled. The ZAMS relation (thick line) was taken from \citet{pec2013}. The isochrones (dotted lines) for 0.1, 0.5, 1, 2, 6, 10, and 100\,Myr were taken from \protect\cite{bress2012}. 
} 
\label{cmd}
\end{figure*}
\begin{figure*}
\centering
\includegraphics[width=14.6cm]{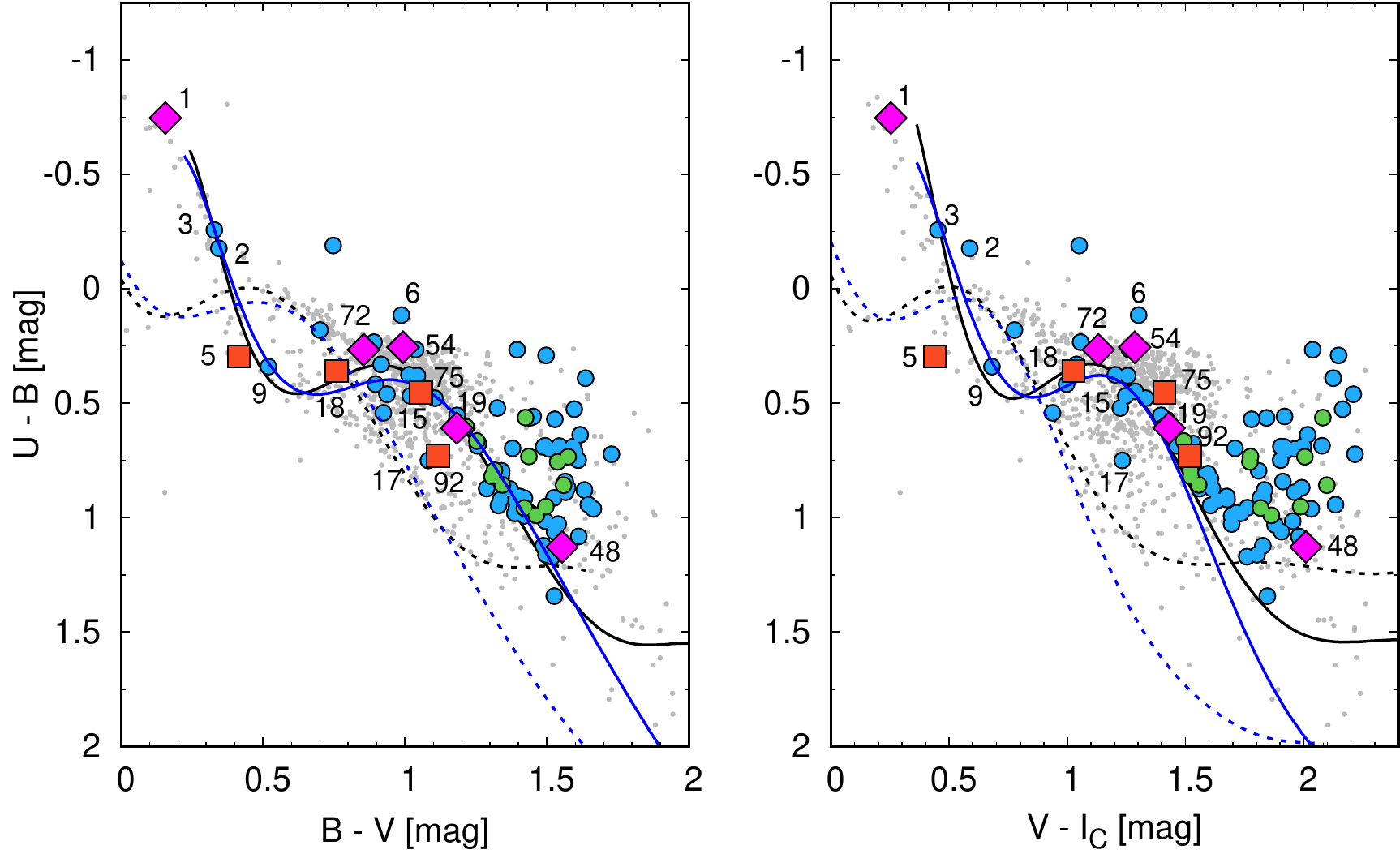}
\caption{Variable stars in $(U-B)$ vs.~$(B-V)$ ({\it left}) and $(U-B)$ vs. $(V-I_{\rm C})$ ({\it right}) colour-colour diagrams for the observed field. The symbols are the same as in Fig.~\ref{cmd}. The dashed lines represent intrinsic relations for dwarfs (black) and for giants (blue) taken from \citet{cald1993}. The solid lines are the relations shifted adopting $E(B-V ) = 0.47$ mag, $E(V-I_{\rm C})=0.6$ mag and $E(U-B)/E(B-V )=0.72$ \citep{mass1995}.} 
\label{ccd}
\end{figure*}

\subsection{\label{ssubvi}$\bmath{UBVI_{\rm C}}$ photometry}

The magnitudes and colours derived in Sect.~\ref{sstransf} are shown in the $V$ vs.~$(B-V)$ and $V$ vs.~$(V-I_{\rm C})$ colour--magnitude diagrams (Fig.~\ref{cmd}), and in the $(U-B)$ vs.~$(B-V)$ and $(U-B)$ vs.~$(V-I_{\rm C})$ colour--colour diagrams (Fig.~\ref{ccd}). 
The distance of 1.7 kpc, i.e.~$(m-M)_0=11.1$ mag, the mean colour excesses $E(B-V)=0.47$ mag, $E(V-I)=0.6$ mag and the total absorption in $V$, $A_V=1.46$ mag, were adopted from \cite{parksung}.
As can be seen in these figures, many stars are in their PMS phase of evolution and plenty of them are variable (coloured symbols). Their position spread between 0.5 and 10 Myr in the colour--magnitude diagrams may be caused by differential reddening towards NGC 2244, mentioned by \citet{berg2002}, or the stars were formed not at the same time.
The variable reddening towards the early-type stars in NGC 2244 was reported by \cite{mass1995} who obtained minimum and maximum values of $E(B-V)$ equal to 0.38 and 0.85, respectively. The estimation of masses and ages for PMS stars depends a lot on the adopted theoretical models of evolution \citep{bell2013}.

As can be seen in Fig.~\ref{cmd}, some stars lie close to or on the left side of ZAMS line. They are not cluster members. Among them, there are four eclipsing binaries, two $\delta$ Scuti candidates and several other variables. They are marked as ''nm'' in the last column in Table \ref{tubvi} in the Appendix.

\begin{figure}
\centering
\includegraphics[width=86mm]{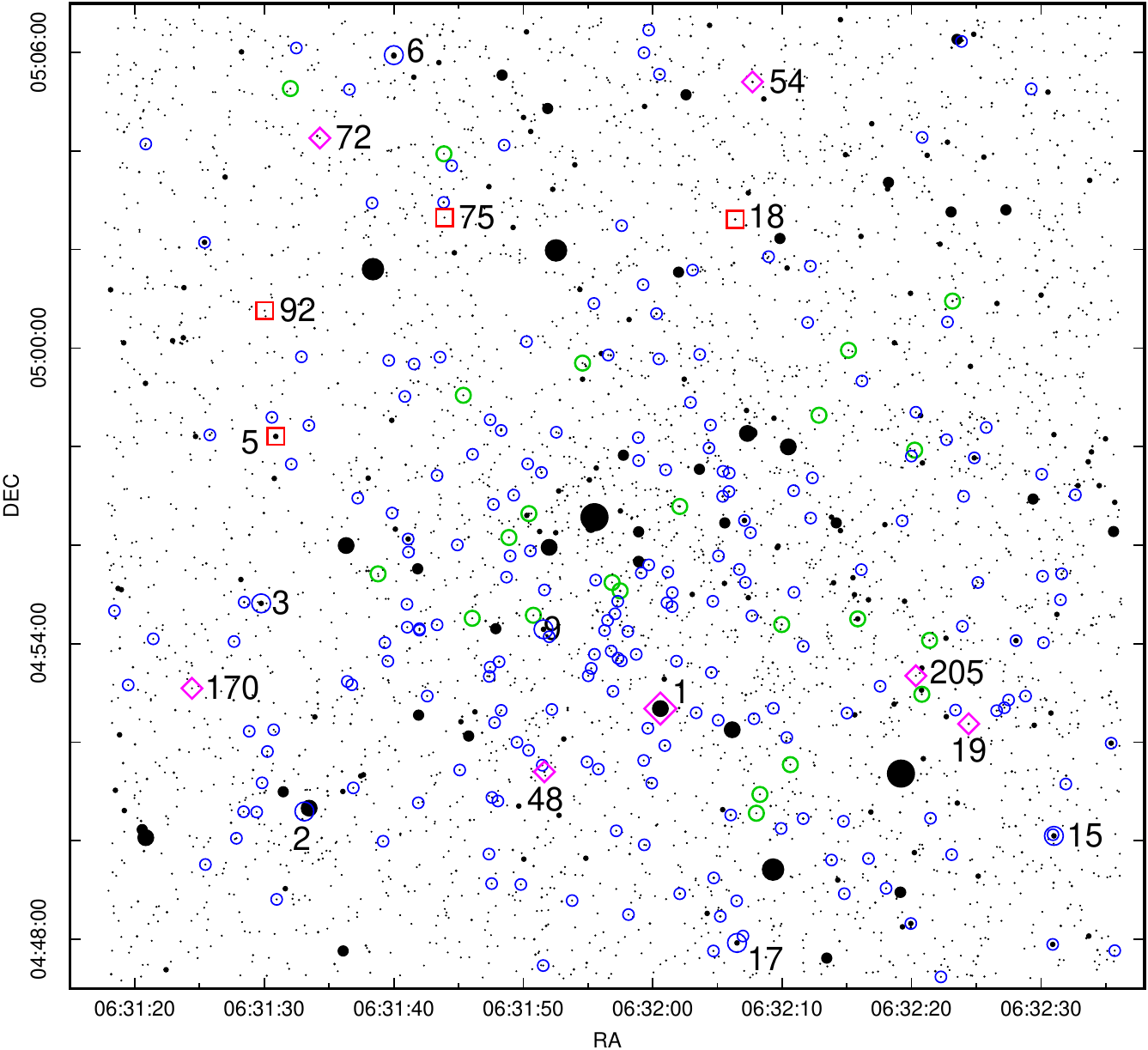}
\caption{Schematic map of the observed field. Variable stars are marked with colours: pulsating stars (red squares), eclipsing stars (pink diamonds), other periodic variables (green circles) and remaining variables (blue circles). Some variables discussed in text are labeled.} 
\label{xy}
\end{figure}

\section{Variable stars in the observed field}
\label{svar}

The variability search was based mainly on the Fourier periodograms, calculated up to 80 d$^{-1}$. In addition, the light curves, periodograms and phased light curves for all stars were inspected by eye.
Out of over 4000 stars in the observed field, 245 turned out to be variable. The brightest stars in our observations are strongly saturated. Since the Y4KCam is read out through four amplifiers, the saturation in one quadrant causes significant crosstalk, saturated in the remaining quadrants. This could have affected photometry of some stars. 

In this section we present the main properties of the variable stars which we divided into four groups: pulsating stars, binary stars, other periodic variables and the remaining variables. Their location in schematic map of observed field is shown in Fig.~\ref{xy}. Some examples of variable stars in the cluster field were already announced by \citet{gm}.

\subsection{\label{spuls}Pulsating variables}

\begin{table*}
 \centering
  \caption{\label{tampl}Parameters of the sine-curve fits to the $B$, $V$ and $I_{\rm C}$ differential magnitudes of the pulsating stars detected in the observed field. The numbers in parentheses denote the r.m.s.~errors of the preceding quantities with the leading zeroes omitted. The $\sigma_{\rm res}$ is the residual standard deviation, $N_{\rm obs}$ stands for the number of observations, $S/N$ is the signal-to-noise ratio.}
  \begin{tabular}{@{}crlccrcrrcl}
  \hline
star& \multicolumn{2}{c}{$f_i$} & Filter & $N_{\rm obs}$& \multicolumn{1}{c}{$A_i$} & \multicolumn{1}{c}{$T_{\rm max} - T_0$} & \multicolumn{1}{c}{$S/N$} &\multicolumn{1}{c}{$\sigma_{\rm res}$} & Spectal & Other\\
& \multicolumn{2}{c}{[d$^{-1}$]}& & & \multicolumn{1}{c}{[mmag]} & \multicolumn{1}{c}{[d]} & &\multicolumn{1}{c}{[mmag]}& type & name \\\hline\noalign{\vskip1pt} 
5  & 31.3106(12) & $f_1$  & $B$ &  130 &  5.7(5)  & 5195.6258(04) &  8.8 & 3.7 & A0/A2 & W69 \\
    &             & & $V$ & 1676 &  4.0(1)  & 5197.1584(02) & 20.8 & 3.7 & & \\
    &  2.0195(33)&  $f_x$ & $V$ &      &  1.5(2)  & 5197.1268(97) &  7.9  &  & &    \\
    & 16.9867(34) & $f_2$  & $V$ &      &  1.4(1)  & 5197.1210(09) &  7.1 & & & \\
    & 23.1527(54) & $f_3$  & $V$ &      &  0.9(1)  & 5197.1634(10) &  4.6 & & & \\
    & 20.6854(48) & $f_4$  & $V$ &      &  1.0(1)  & 5197.1412(10) &  5.0 & & & \\
     \noalign{\vskip1pt}\hline\noalign{\vskip1pt} 
18 &  8.4058(09) & $f_1$  & $V$ & 1664 & 18.7(5)  & 5197.1658(04) & 32.1 & 10.1 & & W1689 \\
    & 13.1906(15) & $f_2$  & $V$ &      & 11.0(5)  & 5197.1125(04) & 18.9 &  & &     \\
    &  9.1657(22) & $f_3$  & $V$ &      & 10.8(5)  & 5197.1433(07) & 18.4 &  & &     \\
    & 16.0898(24) & $f_4$  & $V$ &      & 14.0(8)  & 5197.1123(04) & 24.0 &  & &     \\
    &  9.6204(18) & $f_5$  & $V$ &      & 10.8(5)  & 5197.1683(06) & 18.5 &  &  &    \\
    & 13.6769(21) & $f_6$  & $V$ &      &  7.0(5)  & 5197.1124(04) & 12.0 &  & &     \\    
    & 16.3034(28) & $f_7$  & $V$ &      &  5.8(4)  & 5197.1163(06) &  9.9 &  & &     \\
    & 15.0569(52) & $f_8$  & $V$ &      &  5.9(8)  & 5197.1625(10) & 10.1 &  & &     \\
    & 16.8370(23) & $f_9$  & $V$ &      &  7.0(4)  & 5197.1385(05) & 12.0 &  &  &    \\
    &  7.0495(33) & $f_{10}$ & $V$ &      &  4.6(4)  & 5197.1327(20) &  8.0 &  & &     \\
    & 10.1020(44) & $f_{11}$ & $V$ &      &  5.1(4)  & 5197.1656(12) &  8.7 &  & &     \\
    &  2.1164(35) & $f_{x}$ & $V$ &      &  4.1(4)  & 5197.3522(72) &  7.1 &  &  &    \\
    & 10.5279(44) & $f_{12}$ & $V$ &      &  4.5(5)  & 5197.1171(15) &  7.8 &  & &     \\
    & 13.8981(41) & $f_{13}$ & $V$ &      &  4.1(4)  & 5197.1481(11) &  7.0 &  &  &    \\    
    & 16.9900(52) & $f_{14}$ & $V$ &      &  4.3(5)  & 5197.1508(09) &  7.3 &  &  &    \\    
    &  7.2741(44) & $f_{15}$ & $V$ &      &  3.9(5)  & 5197.2108(25) &  6.7 &  & &     \\  
\noalign{\vskip1pt}\hline\noalign{\vskip1pt} 
75 & 11.5336(13) &  & $B$ &  148 & 43.6(67) & 5196.262(3)   & 5.4 & 57.0  & & \\
    &             &  & $V$ & 2011 & 26.4(11) & 5198.6010(5)  & 17.5 & 31.9  &  &\\
    &             &  & $I_{\rm C}$ &  167 & 15.9(25) & 5196.959(2)   & 5.3 & 21.7  &  &\\  
\noalign{\vskip1pt}\hline\noalign{\vskip1pt} 
92 & 20.147(4)   &  & $V$ &  2033 &  8.7(11) & 5198.555(1) & 5.9 & 35.7  &  &\\
\noalign{\vskip1pt}\hline\noalign{\vskip1pt} 
\end{tabular}
\end{table*}  

\begin{figure}
\centering
\includegraphics[width=76mm]{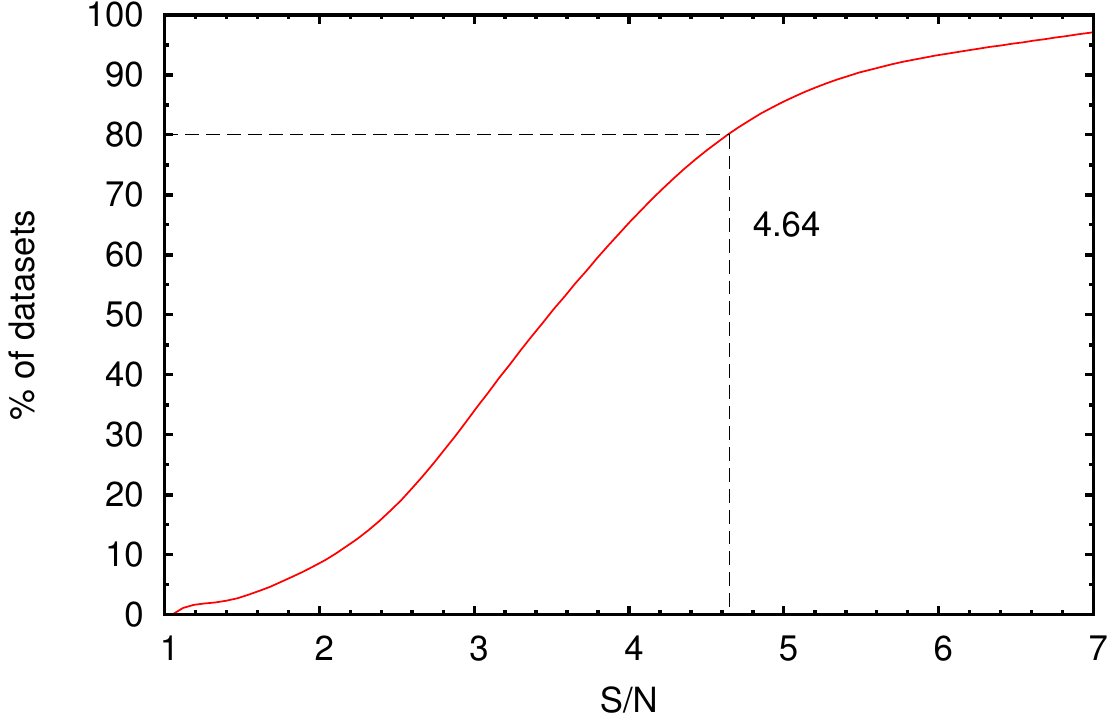}
\caption{The percentage of correct frequency detections for time series of observed pulsating stars vs.~signal to noise. Dashed line represents the confidence level equal to 80\% which corresponds to S/N=4.65.} 
\label{sn}
\end{figure}

\begin{figure}
\centering
\includegraphics[width=76mm]{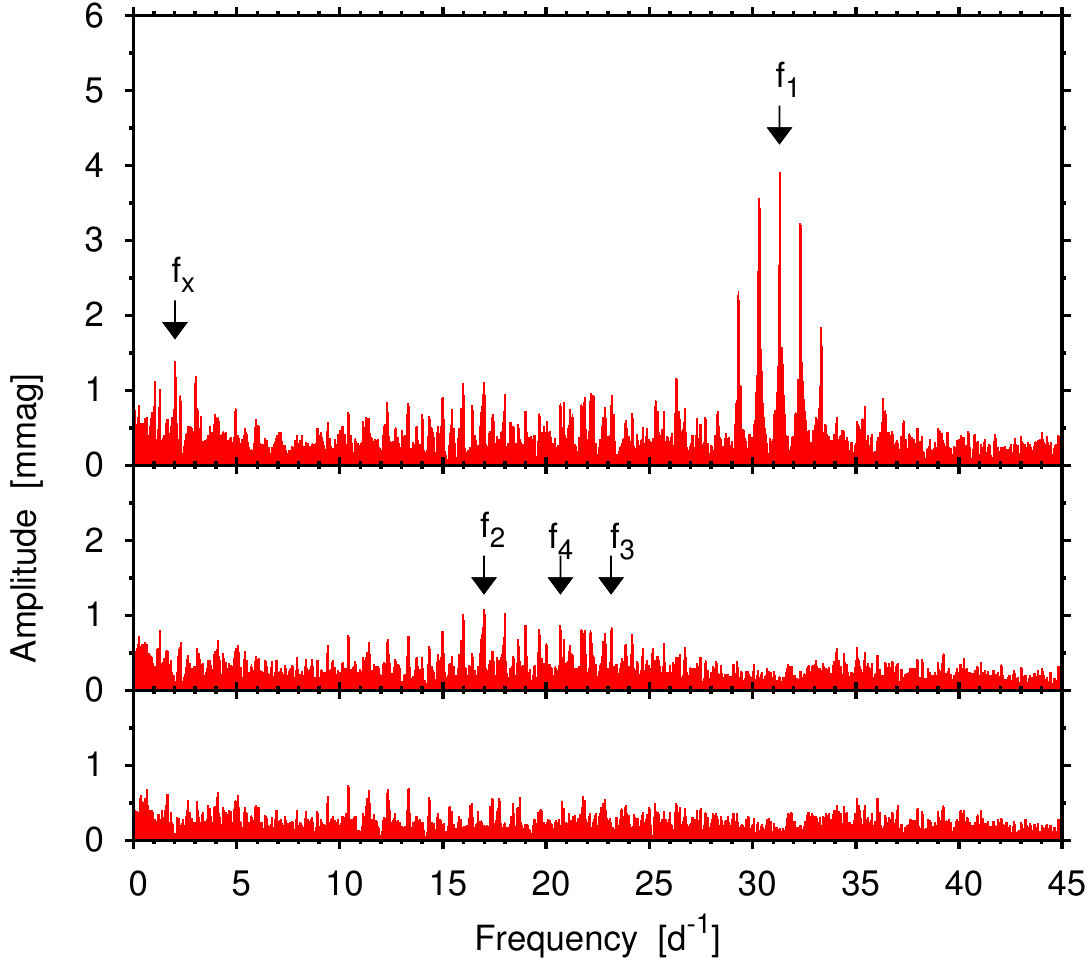}
\caption{Fourier frequency spectrum of the $V$-filter data of star 5: original data (top), after removing $f_1$ and $f_x$ (middle) and after removing $f_1$, $f_x$, $f_2$, $f_3$ and $f_4$ (bottom).} 
\label{trf99}
\end{figure}

There is one O-type star in NGC 2244, HD 46202 \citep*{mass1995}, which was reported to be $\beta$ Cephei type star \citep{bri2011}. Based on CoRoT data, the authors found eleven oscillations with frequencies between 0.5 and 5 d$^{-1}$ and amplitudes of about 0.1 mmag or less. We observed this star, but apparently the amplitudes were below our detection threshold.
We did not find any $\beta$ Cephei or SPB stars in our photometry, but it cannot be excluded that some of early type stars, saturated in our data, pulsate. Two SPB stars discovered by \citet{gruber2012} were outside our field of view.

Out of 245 variables detected in the observed field, four appear to be $\delta$~Scuti stars. The parameters of the sinusoidal terms (frequencies, $f_i$, amplitudes, $A_i$, and phases, $\phi _i$) for these stars, were derived by fitting the formula
\begin{equation}
{\label{eq}}
\langle m \rangle + \sum_{i=1}^{n} A_i \sin (\mbox{2} \pi f_i (t-T_0) + \phi _i)
\end{equation}
to the differential magnitudes. In Eq.~(\ref{eq}) $\langle m \rangle$ is the mean differential magnitude, $n$, the number of fitted terms, $t$, the time elapsed from the initial epoch $T_0=$ HJD\,2455000.  The parameters of the fit are listed in Table\,\ref{tampl}. Instead of $\phi _i$, we provide the time of maximum light, $T_{\rm max}$.

In order to evaluate signal-to-noise, S/N, corresponding to the   the probability of finding real signal equal to 80\%, we used the method described by \citet{baran2015}.
First, 1000 files with Gaussian noise for a given standard deviation were generated, and  the average noise in the periodogram was calculated. 
Then, a sinusoidal signal in the form of $A\sin(2\pi f t+\phi)$ was added. In this formula, $t$ represents the time of observation, $\phi$ is the phase taken as random value in the range [0,2$\pi$], and $A$ is amplitude of the signal covered the range of signal-to-noise (S/N) between 1 and 8, with the step of 0.05 mmag.
For each amplitude, we generated 1000 time-series datasets for which Fourier periodograms were calculated and the number of files with peak at the added frequency was counted. As can be seen in Fig.~\ref{sn}, the detection threshold with the 80\% confidence level corresponding to S/N=4.64. For this reason, in Table\,\ref{tampl} we show only the frequencies with S/N$\geq$4.6.

\begin{figure}
\centering
\includegraphics[width=76mm]{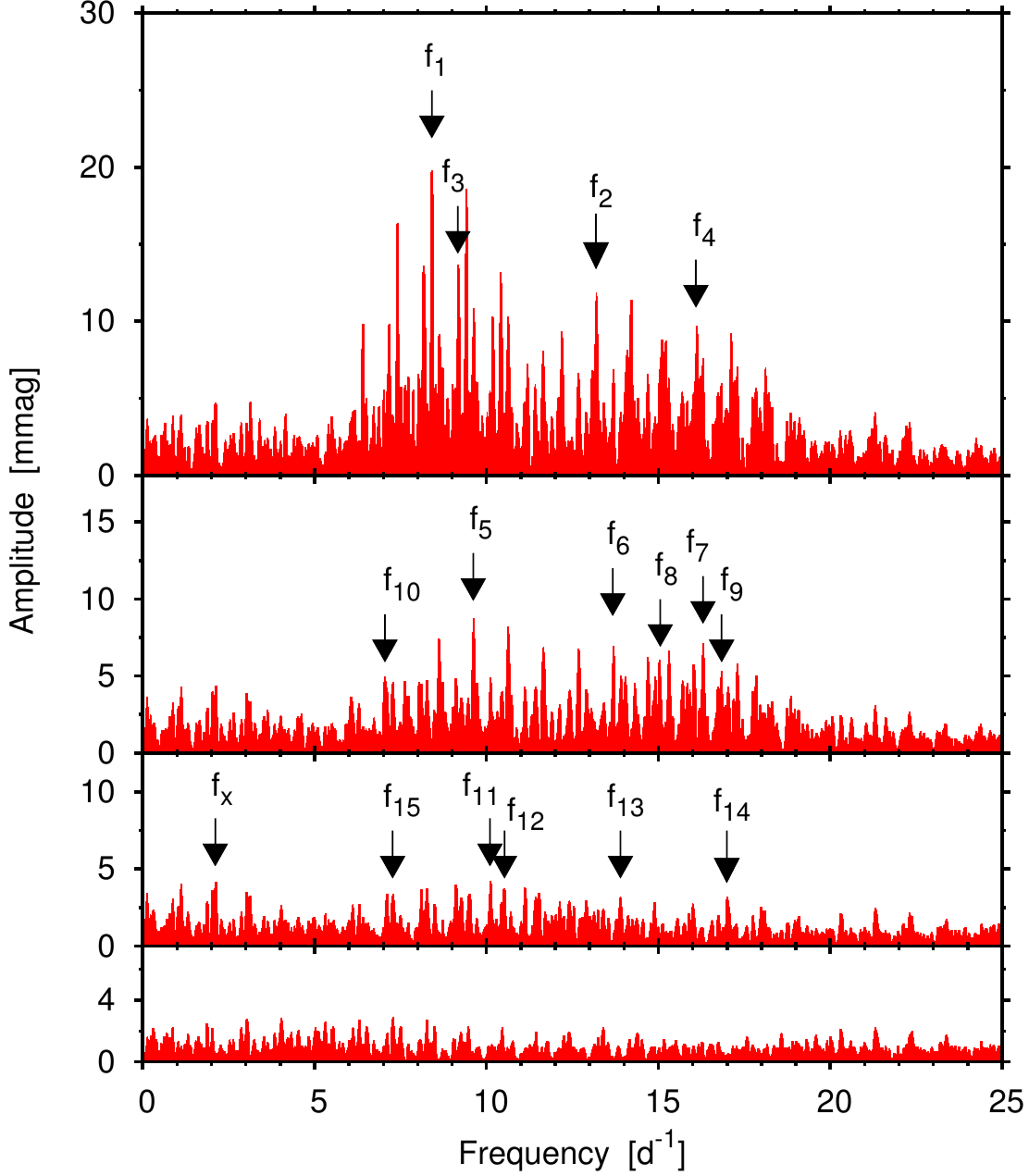}
\caption{Fourier frequency spectrum of the $V$-filter data of the $\delta$ Scuti candidate, star 18, at four steps of prewhitening. } 
\label{trf207}
\end{figure}

\begin{figure}
\centering
\includegraphics[width=76mm]{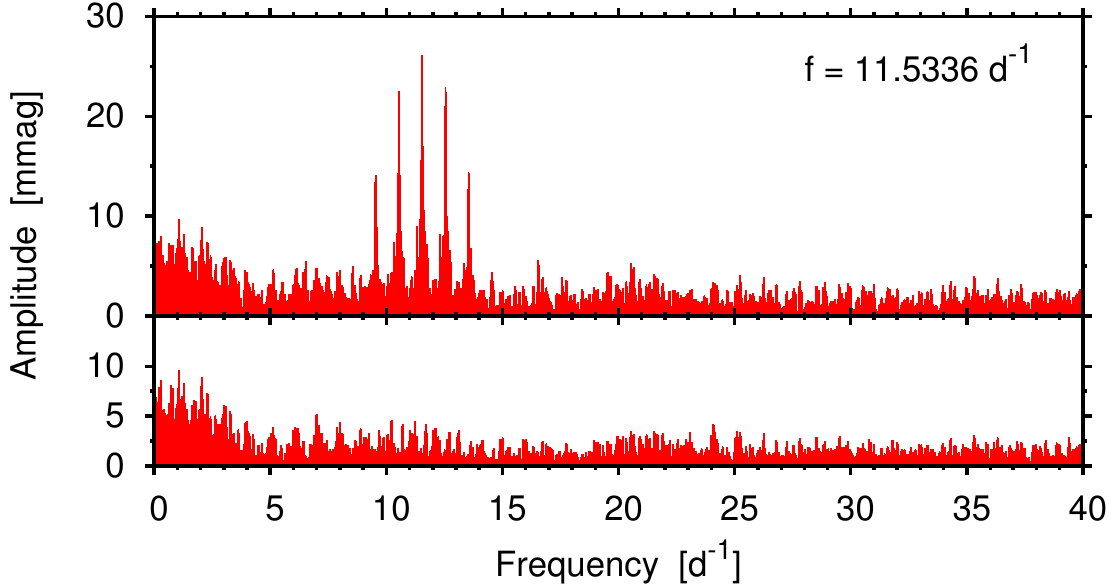}
\caption{Fourier frequency spectrum of the $V$-filter data of the $\delta$ Scuti candidate, star 75: original data (top), after removing $f=11.5336$ d$^{-1}$.} 
\label{trf768}
\end{figure}

One of $\delta$~Scuti candidates is star 5 (GSC 00154-01743). The Fourier frequency spectrum of its $V$-filter data (Fig.~\ref{trf99}) reveals frequency $f_1=$ 31.3106 d$^{-1}$. After prewhitening original data with this frequency, we found the frequency $f_x=$ 2.0195 d$^{-1}$ (likely instrumental) and three other significant terms with frequencies $f_2=$ 16.9867 d$^{-1}$,  $f_3=$ 23.1527 d$^{-1}$ and  $f_4=$ 20.6854 d$^{-1}$, typical for $\delta$~Scuti stars. The spectral type in WEBDA\footnote{The WEBDA database is available at http://www.univie.ac.at/webda/.} database, reported by \citet{kuz1986} is A0. In {\it Simbad} database the spectral type of star 5 is A2. The fitted spectral type, derived from synthetic photometry 
by \citet{pickles2010}, is B8V, however, it appears to be too early for $\delta$ Scuti star. The position in $(U-B)$ vs.~$(V-I_{\rm C})$ colour-colour diagram (Fig.~\ref{ccd}) indicates that this star is not a member of NGC 2244. The probability of its membership ranges between 0.16 \citep{marsch1982} and 0.88 \citep{chen2007}. Based on Gaia proper motions we obtained the probability equal to 0.27 (Sect.~\ref{smem}).

\begin{figure}
\centering
\includegraphics[width=76mm]{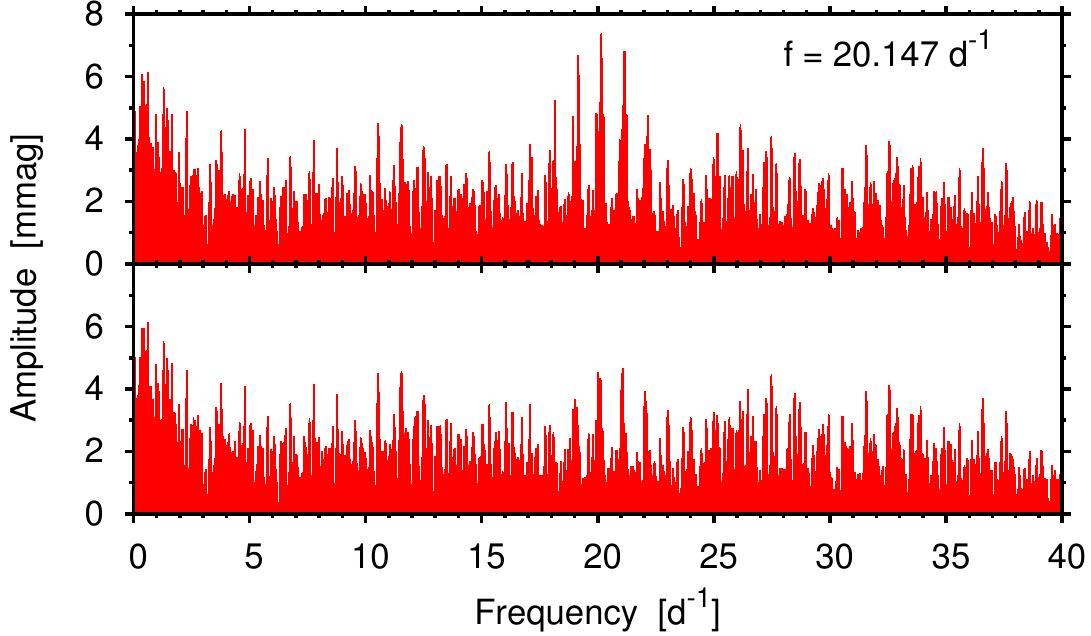}
\caption{Fourier frequency spectrum of the $V$-filter data of the $\delta$ Scuti candidate, star 92: original data (top), after removing $f=20.147$ d$^{-1}$.} 
\label{trf848}
\end{figure}

For three other $\delta$~Scuti candidates the spectral types are not available. Using prewhitening method with the subsequent strongest modes we found fifteen frequencies in Fourier periodogram of star 18 (Fig.~\ref{trf207}).
The periods, the multiperiodicity and the shape of light curve are typical for a $\delta$~Scuti star.
In each of two other stars, star 75 and 92, we found only one frequency of variability, characteristic for  $\delta$ Scuti-type variables: $f=$ 11.5336 d$^{-1}$ (Fig.~\ref{trf768}) and $f=$ 20.147 d$^{-1}$ (Fig.~\ref{trf848}), respectively. 

\begin{figure}
\centering
\includegraphics[width=76mm]{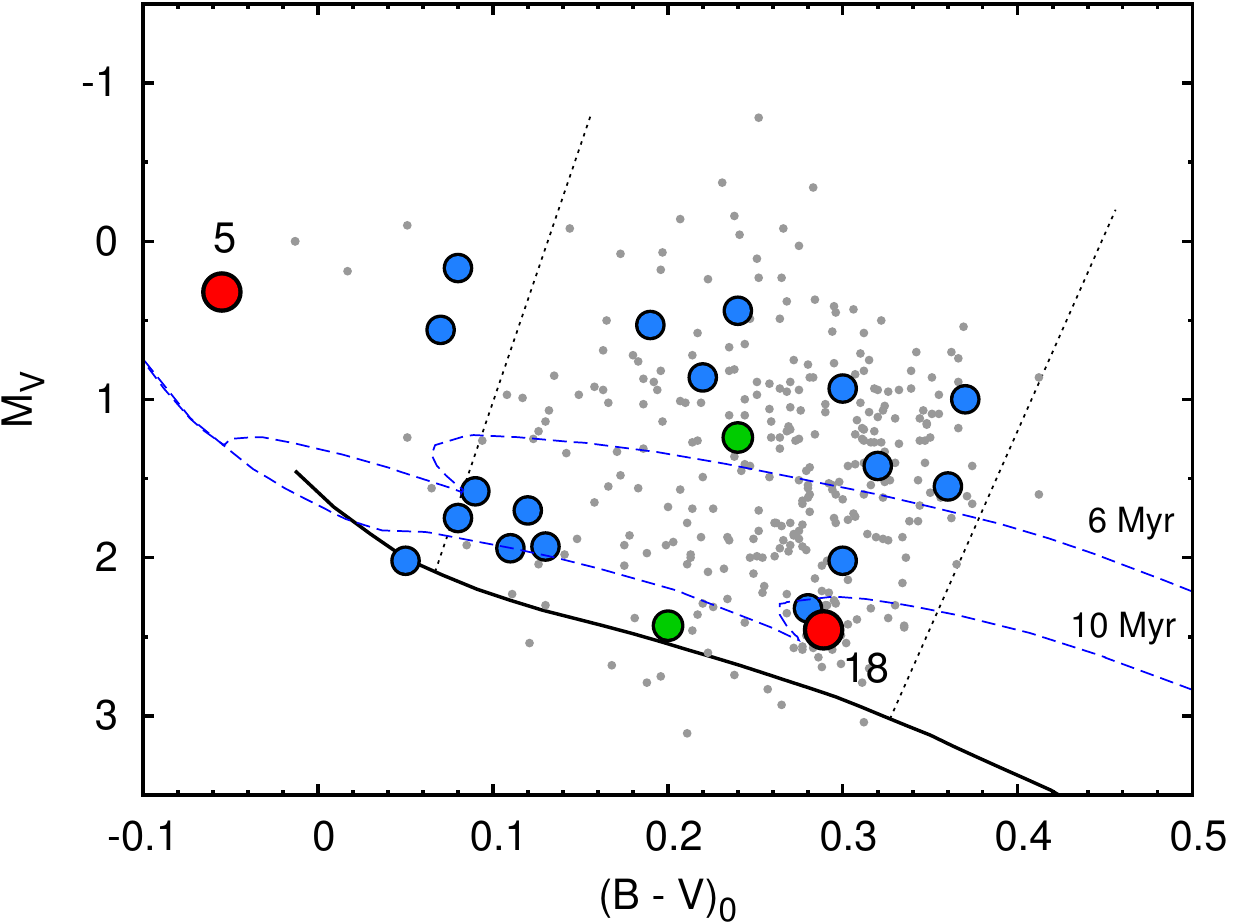}
\caption{Position of two $\delta$ Scuti candidates, star 5 and 18 (red dots) in HR diagram. Additionally, 18 pulsating PMS stars and candidates from clusters (blue dots) and  two pulsating PMS field stars (green dots) taken from \citet{zwintz2008} are plotted. The ZAMS (solid line), the borders of the classical $\delta$ Scuti instability strip (dotted lines) and classical $\delta$ Scuti stars (grey dots) were transformed from the data of \citet{rod2001}. The dashed line represents isochrones for 6 and 10 Myr taken from \protect\cite{bress2012}.} 
\label{pms}
\end{figure}

Assuming that all four $\delta$ Scuti candidates are cluster members, we calculated absolute magnitudes, $M_V$, and $(B-V)_0$ adopting $A_V=1.46$ mag, $E(B-V)=0.47$ mag and the distance modulus equal to 11.1 mag found by \citet{parksung}. 
These values for two $\delta$ Scuti candidates, star 5 and 18, are plotted in colour-magnitude diagram (Fig.~\ref{pms}) together with 18 pulsating PMS stars and candidates from clusters (blue dots) and two pulsating PMS field stars with best parallaxes (green dots), taken from Table 3 and 4 of \citet{zwintz2008}, respectively. The author notified that PMS pulsators occupy the same instability region in H-R diagram  as classical $\delta$ Scuti stars, marked with grey dots in Fig.~\ref{pms}. We derived their $(B-V)_0$ from $(b-y)_0$ colours published by \citet{rod2001}\footnote{http://www.iaa.es/~eloy/dsc00.html} using transformation given by \citet{cald1993}. The instability strip (dotted line) and ZAMS (solid line) were also transformed from the data of \citet{rod2001}.
As can be seen in this diagram, star 18 is situated inside the instability strip. Its position is quite close to ZAMS and it appears to be older than cluster members. The membership probability determined in Sect.~\ref{smem} is equal to 62.2\%. 
Furthermore, its position in $V$ vs.~$(B-V)$ and $V$ vs.~$(V-I_{\rm C})$ colour-magnitude diagrams (see Fig.~\ref{cmd}) indicates this star is rather old.

Since the $(B-V)_0$ colours of star 75 and star 92 are greater than 0.5 and $M_V$ is close to 5 mag, we did not show them in Fig.~\ref{pms}. If these stars are of $\delta$ Scuti-type, they are field stars, located beyond NGC 2244. Their membership probability determined in Sect.~\ref{smem} is equal to 1.1\% and 0\% for star 75 and 92, respectively. The star 5 have $(B-V)_0=-0.05$ and it is outside the instability strip borders. This indicates it is likely a foreground star. 

\subsection{\label{secl}Binary stars}
\begin{figure}
\centering
\includegraphics[width=76mm]{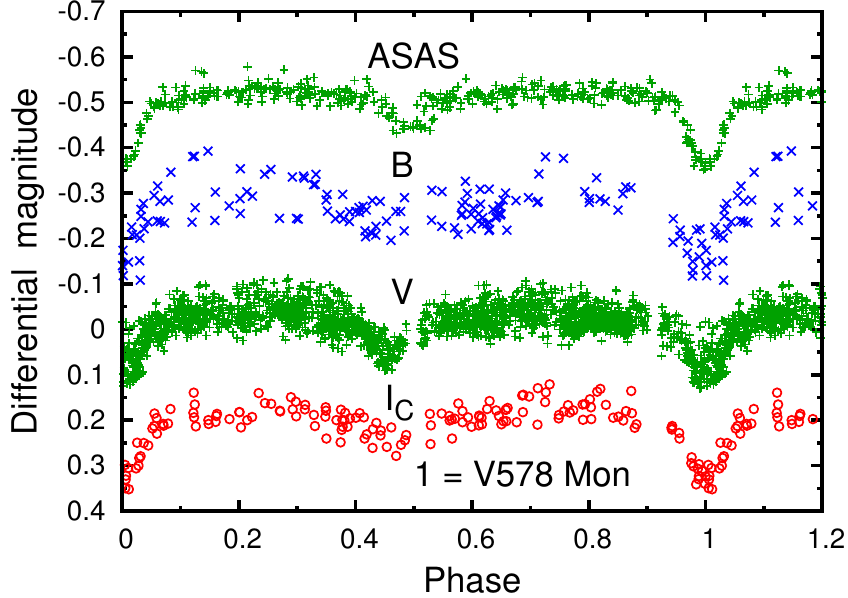}
\caption{Phase diagrams of the $V$-filter ASAS observations and our $BVI_{\rm C}$ observations of eclipsing system V578 Mon. Offsets were applied to separate light curves in different bands.} 
\label{lcecl16}
\end{figure}

\begin{figure*}
\centering
\includegraphics[width=42pc]{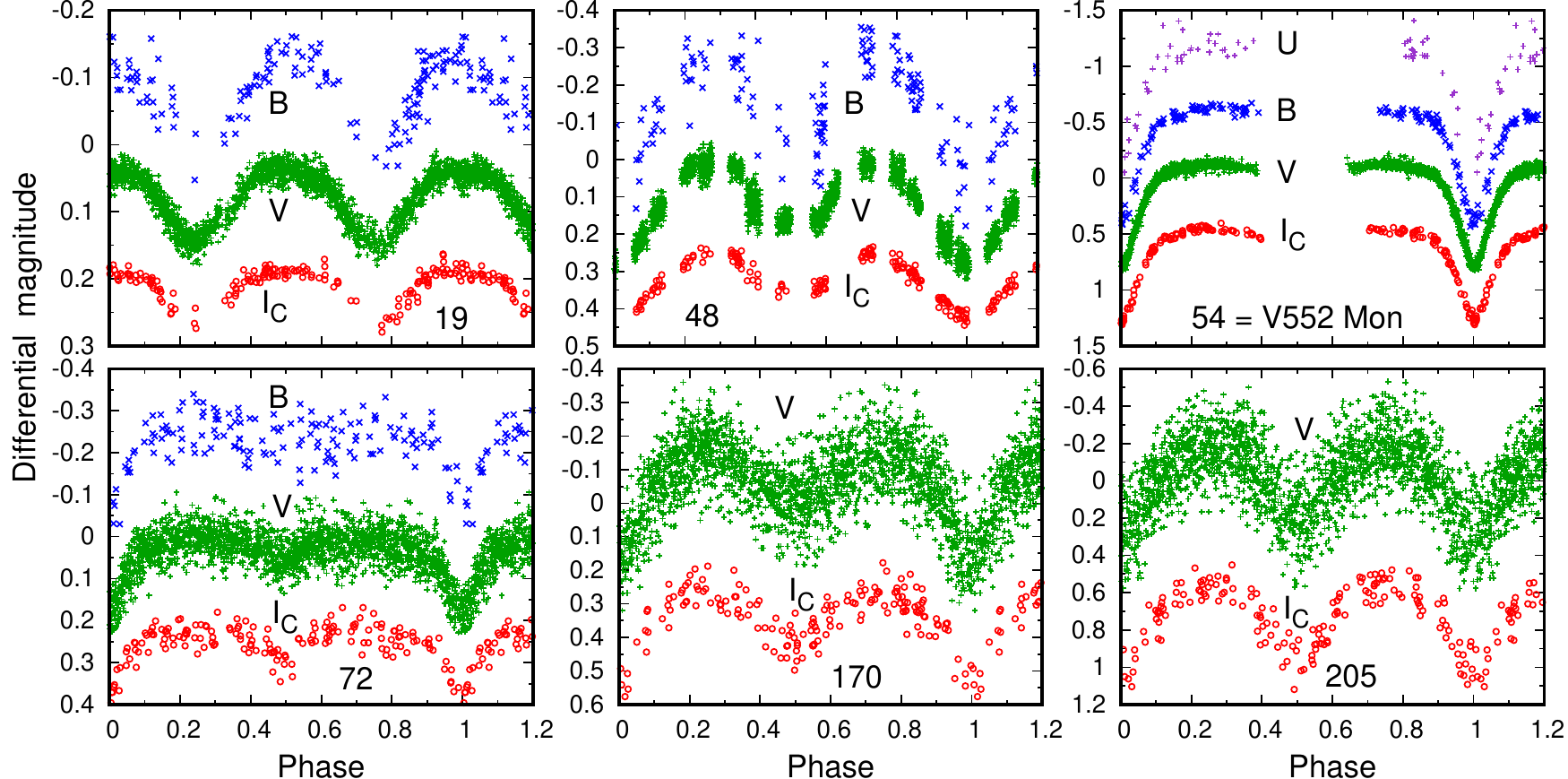}
\caption{Phase diagrams of six eclipsing systems (star 19, 48, 54, 72, 170 and 205). Offsets were applied to separate light curves in different bands.} 
\label{lcecl}
\end{figure*}

\begin{table}
\centering
\caption{\label{tmin}Parameters of eclipsing binary stars found in the observed field. The depths of eclipses are given for the $V$-filter observations.
The numbers in parentheses denote the r.m.s.~errors of the preceding quantities with the leading zeroes omitted.}
  \begin{tabular}{@{}rlrcc}
  \hline
Star  & \multicolumn{1}{c}{Period} & \multicolumn{1}{c}{$T_{\rm min} \mbox{I}$ [d]} & \multicolumn{2}{c}{Eclipse depth} \\
 & \multicolumn{1}{c}{[d]} & \multicolumn{1}{c}{HJD-2450000} & I [mag] & II  [mag] \\
\noalign{\vskip1pt}\hline\noalign{\vskip1pt}
1~~  & 2.411(1)  & 5203.7289(33) & 0.11 & 0.08 \\
1$^\star$ & 2.408544(1)& 3142.0074(04) & 0.16 & 0.07\\
19~~ & 1.7789(17) & 5199.4827(49) & 0.12 & 0.12\\
48~~ & 7.3415(86) & 5201.146(24) & 0.27 & 0.17\\
54~~ & 0.50241(1) & 5197.1875(06) & 0.92 &  --   \\
72~~ & 0.72276(10)& 5193.07235(6) & 0.16 & 0.04 \\
170~~ & 0.58681(11)& 5197.9792(22) & 0.30 & 0.15 \\
205~~ & 0.26582(2) & 5194.2811(03) & 0.44 & 0.38 \\
\noalign{\vskip1pt}\hline
\multicolumn{5}{l}{$^\star$ for $V$-filter ASAS data}\\
\end{tabular}
\end{table}

We identified seven stars (WEBDA numbers: 59, 80, 119, 128, 133, 194, and 201), previously known as single lined spectroscopic binaries \citep{huang2006}. All they are not variable in our data.
In the observing field there are also two double-lined spectroscopic stars, HD 46149 \citep{mahy2009} and HD 46150 \citep{chini2012}. The primary components of both systems are O-type stars. Unfortunately, they are saturated in our photometry.

NGC 2244 contains a close visual pair, HD 46180 ($V_A=9.2$, $V_B=10.3$, $\rho=5.1$ arcsec and $\theta=82\degr$; \citealt{mason2001}). Based on DDO (David Dunlap Observatory) spectroscopy obtained for MOST variables, \citet{prib2009} discovered that HD 46180 is a quadruple system composed of two binary stars. One pair is an eclipsing system with the orbital period of about 3.09 d and amplitude 0.016 mag \citep{prib2010}.
Unfortunately, HD 46180 and the other star are not resolved in our data, and they are saturated. Consequently we are not able to detect any eclipses.

There is also a well-known eclipsing system V578 Mon in the observed field (star 1 in our list of variable stars).
According to \citet*{hens2000}, this binary has an orbital period equal to 2.40848 d, and consists of two massive ($M_1=14.54\pm0.08$ M$_{\sun}$ and $M_2=10.29\pm 0.06$ M$_{\sun}$) early B-type stars (B1V+B2V). The age of the system and the distance derived by these authors are equal to $2.3\pm 0.2$ Myr and $1.39\pm 0.1$ kpc, respectively.
The eccentricity of about 0.077 and apsidal motion equal to 0.071 deg cycle$^{-1}$ (apsidal period equal to 33.48 yr) were determined by \citet{garcia2011}. The absolute dimensions of V578 Mon were recently determined by \citet{garcia2014}. They found the radii $R_1=5.41\pm 0.04 R_{\sun}$ and $R_2=4.29\pm 0.05 R_{\sun}$, and effective temperatures $T_1=30000\pm 500$ K and $T_2=25750\pm 435$ K.

The $BVI_{\rm C}$ phased diagram of V578 Mon is shown in Fig.~\ref{lcecl16}.
Since this bright system ($V=8.5$ mag) is partially saturated in our photometry, we were not able to determine the period accurately. For this purpose, we used ASAS data \citep{pojm2003} and we got $P=2.408544$ d which is consistent with the period of \citet*{hens2000}.

V578 Mon was also observed by MOST. Based on this photometry, \citet{prib2010} found the orbital period $P=2.40860$ d. Adopting mass ratio $q=0.71$ from spectroscopy, they obtained inclination $i=73.1\degr$, the relative radii $r_1=0.254$ and $r_2=0.178$, the eccentricity $e=0.066$ and surprisingly low for early B-type stars the effective temperatures, $T_1=16980$ K and $T_2=13310$ K.

The other eclipsing system found in our data is V552 Mon (star 54 in our list of variable stars). In General Catalog of Variable Stars (GCVS) its type of variability has been assigned as IN:, i.e.~probable irregular eruptive star \citep{GCVS}. There is no information about variability and spectral type of five other eclipsing binaries (19, 48, 72, 170, 205). The periods, the times of minimum, and the eclipse depths for each eclipsing binary are given in Table \ref{tmin}. The phase diagrams are shown in Fig.~\ref{lcecl}.

The positions of systems 19 and 48 in the $V-(V-I_{\rm C})$ colour-magnitude diagram (Fig.~\ref{cmd}) indicate that they can be cluster members. Four other binaries (54, 72, 170 and 205) are close to ZAMS in this diagram, so they are probably non-members. It cannot be excluded, however, that the changes in their light curves some of them are caused by cool spots. This type of variability is described in Sect.~\ref{ssin}.

We found many variables with evident eclipse in their light curve, showing also another type of variability. For this reason we describe them in Sect.~\ref{sother}, together with other irregular variables.

\subsection{\label{ssin}Other periodic variables}
In addition to pulsating stars (Sect.~\ref{spuls}) and eclipsing systems (Sect.~\ref{secl}) we found a group of 23 other periodically variable stars. Their periods, times of the maximum and amplitudes are listed in Table \ref{tsin}. The phase diagrams of these stars are available online in electronic version. 
In colour-magnitude (Fig.~\ref{cmd}) and colour-colour (Fig.~\ref{ccd}) diagrams these stars are marked as green dots. Almost all these stars appear to be cluster members. The shape of light curves and the position in colour-magnitude diagram indicate that most of these periodic stars may be WTTS stars. 

The WTTSs do not have active accretion disks and the changes in their light curves are mainly caused by cool spots (\citealt{bouvier1993}, \citealt{herbst1994}). As we mentioned in the Introduction, the typical periods of such variability are 0.5--18 d \citep{semkov2011}. Our observations do not allow to derive the longest periods from this range, so that in Table \ref{tsin} we list stars only with periods shorter than half of the observing time. We did not find any star with period shorter than one day, although, in other clusters, e.g.~NGC 2282, there are up to 50\% of such stars \citep{dutta2018}.

\begin{table}
\centering
\caption{\label{tsin}Parameters of 23 periodic variable stars found in the observed field determined from the $V$-filter observations ($T_0=2450000$). The numbers in parentheses denote the r.m.s.~errors of the preceding quantities with the leading zeroes omitted.}
  \begin{tabular}{@{}cllr}
  \hline
Star  & \multicolumn{1}{c}{Period} & \multicolumn{1}{c}{$T_{\rm max}-T_0$} & \multicolumn{1}{c}{Amplitude} \\
 & \multicolumn{1}{c}{[d]} & \multicolumn{1}{c}{[d]} &  [mmag] \\
\noalign{\vskip1pt}\hline\noalign{\vskip1pt}
11 & 1.39484(45) & 5198.3285(18) & 25.0(02)\\
20 & 2.3533(10) & 5198.2599(24) & 63.5(04)\\
27 & 2.9650(30) & 5197.988(64) & 37.8(61)\\
43 & 2.9207(21) & 5197.1121(78) & 170.3(24)\\
53 & 5.715(16)  &  5196.617(11) & 69.7(07)\\
57 & 1.73898(41) & 5197.36557(91) & 361.9(10)\\
59 & 1.7338(11) & 5199.1084(41) & 58.4(09)\\
62 & 3.8134(68) & 5196.9136(65) & 92.5(10)\\
79 & 2.2971(17) & 5198.6714(39) & 100.5(12)\\
85 & 4.1615(99) & 5200.3439(92) & 82.2(12)\\
90 & 2.72813(65) & 5199.5292(16) & 388.3(14)\\
93 & 1.58577(91) & 5198.5312(26) & 132.9(13)\\
95 & 1.5282(13) & 5198.6391(34) & 89.3(12)\\
100 & 1.8069(14) & 5198.3071(37) & 111.1(14)\\
103 & 1.6836(16) & 5198.3854(59) & 58.96(13)\\
105 & 5.212(18)  & 5196.195(13)  &  86.4(15)\\
113 & 1.10223(34) & 5198.5434(21) & 174.3(19)\\
135 & 1.07587(73) & 5199.3193(33) & 95.9(18)\\
137 & 4.4913(51) & 5199.7514(99) & 231.0(34)\\
151 & 2.3023(24) & 5196.3062(51) & 187.4(24)\\
166 & 5.142(17) & 5198.089(13) & 173.1(27)\\
182 & 4.9095(86) & 5196.773(14) & 257.3(46)\\
192 & 6.880(23) & 5196.916(20) & 176.4(34)\\\hline
\end{tabular}
\end{table} 

\begin{figure}
\centering
\includegraphics[width=70mm]{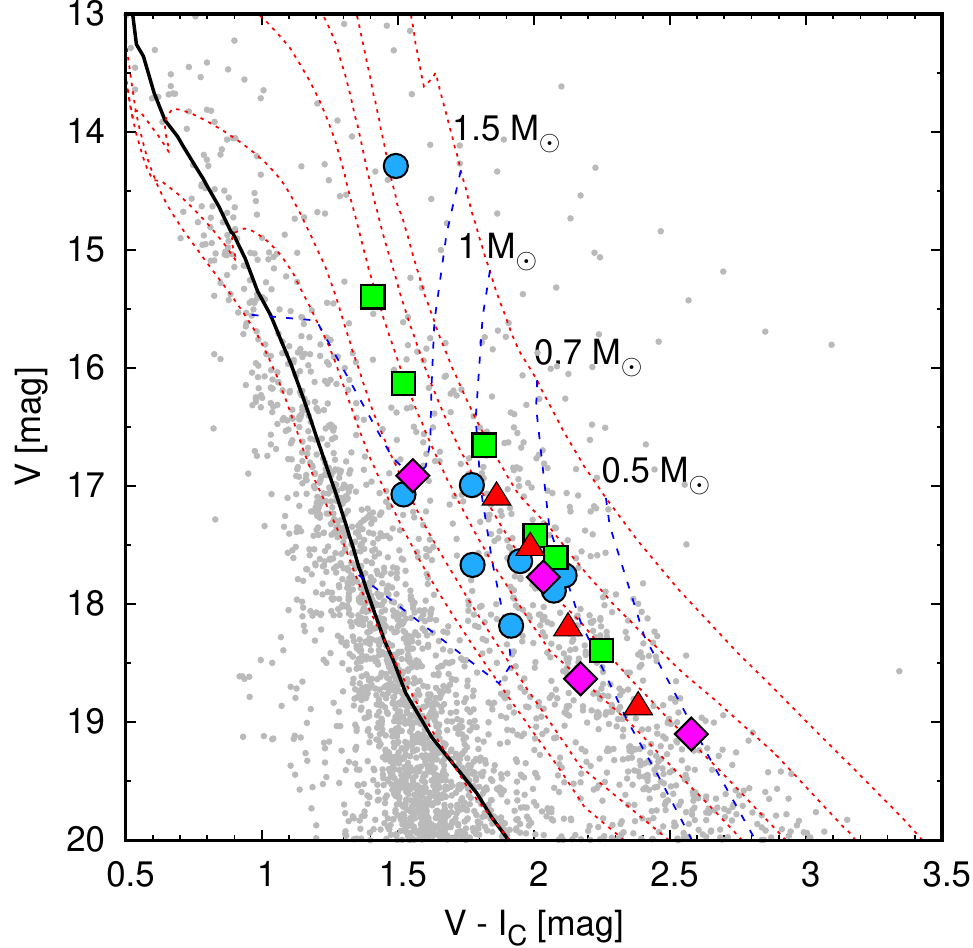}
\caption{Position of the periodic variables in the colour-magnitude diagram. The ZAMS relation for Z=0.02 (thick line) was taken from \citet{pec2013}. The symbols represent four ranges of periods: 1--2 days (blue dots), 2--3 days (green squares), 3--5 days (red triangles) and more than 5 days (magenta diamonds). The isochrones (dotted lines) for 0.1, 0.5, 1, 2, 6, 10, and 100\,Myr and mass tracks (dashed lines) were taken from \protect\cite{bress2012}. We adopted here mean values of $E(V-I_{\rm C})=0.6$ mag, $A_V=1.46$ mag and $(m-M_{\rm V})_0=11.1$ mag.} 
\label{periodic}
\end{figure}
The faster rotation of the stars with decreasing masses is clearly seen in old open clusters \citep{hen2012}. Some authors, however, show correlation between periods and masses  of stars in young open cluster. \citet{lata2012}, who analyzed PMS variable stars of NGC 7380, NGC 1893 and Be59, show that stars with masses greater than 2\,M$_{\sun}$ rotate faster. The opposite dependence was found by \citet{lamm2005} who studied the period distribution of PMS variables in Orion Nebula Cluster and in NGC 2264. They show that low mass stars rotate faster, on average, then high mass stars. In order to verify  if there is any relation in our sample of periodic variables, we checked their position in the colour-magnitude diagram shown in Fig.~\ref{periodic}. The coloured symbols in this diagram represent four ranges of periods: 1--2 days (blue dots), 2--3 days (green squares), 3--5 days (red triangles) and more than 5 days (magenta diamonds). The isochrones and mass tracks were taken form \citet{bress2012}. As can be seen in this figure, the stars with different periods are distributed in a wide range of masses.
However, we note that in our sample there are no stars with periods of less than 1 day and more then 7 days.
\begin{figure*}
\includegraphics[width=13.6cm]{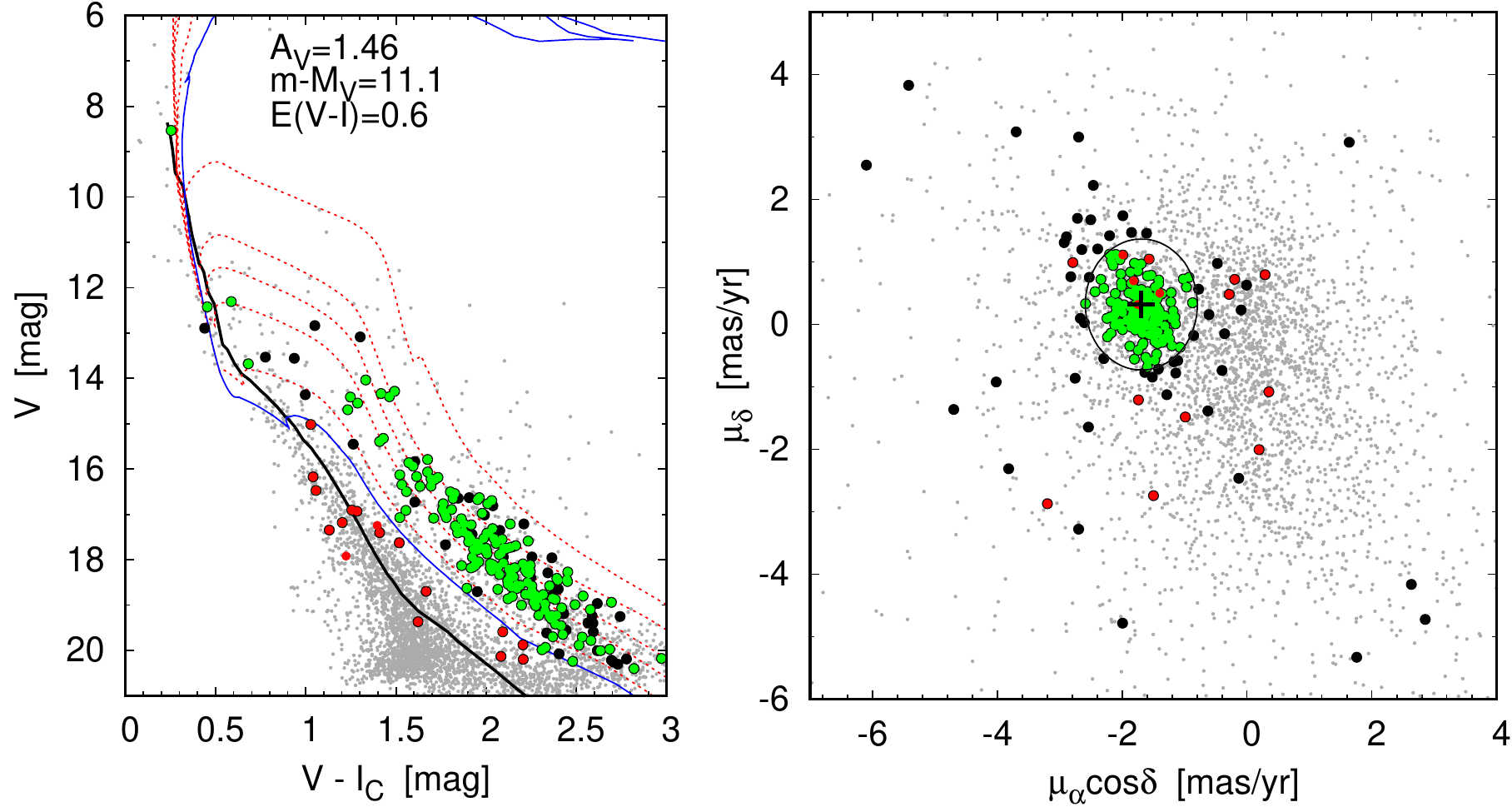}
\caption{{\it Left}: Colour-magnitude diagram for about 3600 stars in the observed field. {\it Right}: Proper motions distribution for about 3600 stars in the observed field. Green circles represent variable stars falling into the oval area around mean proper motion of NGC 2244 (black cross), red circles -- variable stars with $V$ and $(V-I_{\rm C})$ below isochrone 10 Myr (blue line), black circles -- other variable stars.} 
\label{pm}
\end{figure*}

\subsection{Irregular variables}
\label{sother}

The remaining 211 variable stars are classified as irregular. Their light curves are available online in electronic version.
As can be seen in these figures, the range of light variation of some stars is greater than 0.5 mag (e.g.~stars 149, 186, 216, and 232). Some of them show an increase in the light similar to those observed in FUors (e.g.~variable 34 and 131) or EXors (e.g. variable 83, 211) stars, but the time scale of variations is much shorter. 

Almost all irregular variables lie between 0.1 and 6~Myr isochrones in the $V$ vs.~$(V-I_{\rm C})$ colour-magnitude diagram (Fig.~\ref{cmd}), in the region of pre-main sequence stars. The variability of these stars can be therefore caused by changes in the accretion disks and hot spots which is typical for CTTSs (\citealt{herbst1994}, \citealt{kuro2013}). 
The eclipses in the light curves which can be seen in some of them (82 and 149) indicate that these stars are obscured by the dust clumps although they can be members of binary or multiple systems.

Five irregular variable (star 2, 3, 9, 50, and 67) have strong H$\alpha$ emission \citep{parksung}. Three of them, star 2 (spectral type B4 Ve), star 3 (B7 Ve) and star 9, are marked by these authors as HAeBe candidates. Strong emission in H$\alpha$ line of stars 2, 3 and 6 was notified by \citet{li2002} who classified star 3 as Herbig Be and star 6 (spectral type F3 Ve) as HAeBe candidate.

In addition to CTTSs and HAeBe, among irregular variables, we can also find WTTS stars. Some more details about classification of PMS variable stars can be found in Sect.~\ref{sspms}.

\section{Membership}
\label{smem}

An important step in studying a cluster is the determination of the membership in its field. Some non-members can be identified using the colour-magnitude diagram. The $V-(V-I_{\rm C})$ colour-magnitude diagram for the observed stars is shown in the left panel of Fig.~\ref{pm}. As we mentioned in Sect.~\ref{subvi}, the age of NGC 2244 was estimated at 2\,--\,4~Myr so that we considered stars below the 10 Myr isochrone (blue line in Fig.~\ref{pm}) as non-members.

The other way of membership determination is the use of proper motions. A statistical method  was proposed by \citet{vas1958} who used the distribution functions of proper motions for cluster and field stars to derive membership probability for individual stars. 
The method was modified over the years (\citealt{sanders1971}, \citealt{zhao1990}). In this paper, we applied the method used by \citet{sariya2012} who defined the membership probability of the $i^{th}$ star as
{\setlength\arraycolsep{2pt}  
\begin{eqnarray}
P_{\mu}(i)& =& \frac{n_c~.~\phi_c^{\nu}(i)}{n_c~.~\phi_c^{\nu}(i)+n_f~.~\phi_f^{\nu}(i)},\\\nonumber
\end{eqnarray}}

\noindent where $n_{c}$ and $n_{f}$ are the normalized numbers of stars for cluster and field 
($n_c + n_f = 1$) and $\phi_c^{\nu}$ and $\phi_f^{\nu}$ are the frequency distribution functions for the cluster and field stars, defined as: 
{\setlength\arraycolsep{2pt}  
\begin{eqnarray}
    \phi_c^{\nu} &=&\frac{1}{2\pi\sqrt{{(\sigma_{xc}^2 + \epsilon_{xi}^2 )} {(\sigma_{yc}^2 + \epsilon_{yi}^2 )}}}\nonumber\\
   &&\times \exp\left\{{-\frac{1}{2}\left[\frac{(\mu_{xi} - \mu_{xc})^2}{\sigma_{xc}^2 + \epsilon_{xi}^2 } + \frac{(\mu_{yi} - \mu_{yc})^2}{\sigma_{yc}^2 + \epsilon_{yi}^2}\right] }\right\}
\end{eqnarray}}
and
{\setlength\arraycolsep{2pt}  
\begin{eqnarray}
\Phi_f^{\nu}&=&\frac{1}{2\pi \sqrt{(1-\gamma^2)}\sqrt{ (\sigma^2_{xf}+\epsilon^2_{xi})
(\sigma^2_{yf}+\epsilon^2_{yi})}}\nonumber\\
  && \times\exp \left\{-\frac{1}{2(1-\gamma^2)} \left[ \frac{(\mu_{xi}-\mu_{xf})^2}{\sigma^2_{xf}+\epsilon^2_{xi}} \right. \right.\nonumber \\
 &&\left. \left. -\frac{2\gamma(\mu_{xi}-\mu_{xf})(\mu_{yi}-\mu_{yf})} {\sqrt{(\sigma^2_{xf}+\epsilon^2_{xi})(\sigma^2_{yf}+\epsilon^2_{yi})}} + \frac{(\mu_{yi}-\mu_{yf})^2}{\sigma^2_{yf}+\epsilon^2_{yi}} \right]\right\}, 
\end{eqnarray}}

\noindent where $\mu_{xi}$ and $\mu_{yi}$ are the proper motions of the
$i^{\rm th}$ star ($\mu_{xi}=\mu_{\alpha}\cos\delta$ and $\mu_{yi}=\mu_{\delta}$ of the
$i^{\rm th}$ star) with errors $\epsilon_{xi}$ and $\epsilon_{yi}$, respectively. The parameters $\mu_{xc}$ and $\mu_{yc}$ define  the cluster proper motion centre with dispersions $\sigma_{xc}$ and $\sigma_{yc}$, while $\mu_{xf}$ and $\mu_{yf}$ are the field proper motion centre with dispersions $\sigma_{xf}$ and $\sigma_{yf}$, respectively. The correlation coefficient $\gamma$ was calculated by\\
{\setlength\arraycolsep{2pt}  
\begin{eqnarray}
\gamma &=& \frac{(\mu_{xi} - \mu_{xf})(\mu_{yi} - \mu_{yf})}{\sigma_{xf}\sigma_{yf}},
\end{eqnarray}}

In order to find the membership probability for each star we observed, we used the proper motions from the Gaia DR2 catalogue (\citealt{gaia2016}, \citealt{gaia2018}). For about 4000 stars in the observed field, we searched for Gaia counterparts in the search radius of 1\arcsec.
Using the sample of about 50 stars with membership probability greater than 70\% from the catalogue of \citet{parksung}, the mean proper motion of the cluster were calculated using 3$\sigma$ clipping. We obtained $\mu _{\alpha}\cos\delta=-1.69$ mas yr$^{-1}$ and $\mu _{\delta}= 0.32$ mas yr$^{-1}$ with standard deviation $\sigma_{\mu _{\alpha}}=0.30 $ and $\sigma_{\mu _{\delta}}=0.35$, respectively. The proper motions from Gaia DR2 catalogue are shown in the right panel of Fig.~\ref{pm}. As we can see, most of our variable stars have proper motions concentrated around the calculated center (black cross). We assumed that cluster stars are within an ellipse with semi-axis equal to $3\sigma_{\mu _{\alpha}}$ and $3\sigma_{\mu _{\delta}}$. We find 426 stars within this area, for which $V$ magnitudes and $(V-I_{\rm C})$ colours placed them above 10 Myr isochrone in $V-(V-I_{\rm C})$ colour-magnitude diagram. For this sample we calculated mean proper motion of the cluster $\mu_{xc}=-1.67$ and $\mu_{yc}=0.23$ mas\,yr$^{-1}$ with dispersions $\sigma_{xc}=0.34$ and $\sigma_{yc}=0.41$ mas\,yr$^{-1}$. For the remaining 3546 stars we calculated the mean proper motion of field stars $\mu_{xf}=-0.14$ and $\mu_{yf}=-1.22$ mas\,yr$^{-1}$ with dispersions $\sigma_{xf}=3.87$ and $\sigma_{yf}=4.00$ mas\,yr$^{-1}$. These values allow to calculate frequency distribution functions $\phi_c^{\nu}$ and $\phi_f^{\nu}$, and then the probability $P_{\mu}(i)$ for each star. The probabilities for variable stars are shown in Table \ref{tubvi}.
	
\section{Identification of Young Stellar Objects}
\label{syso}

Before we classify the PMS variable stars, described in Sect.~\ref{ssin} and \ref{sother}, first we deal with the classification of young stellar objects (YSOs). The YSOs comprise both protostars and PMS stars. The YSOs are frequently embedded in their parental molecular clouds or have circumstellar disks. The presence of the embedding matter makes classification based on the infrared data quite obvious. In this paper, we use 2MASS, Spitzer (IRAC and MIPS), and WISE infrared photometry to identify which stars we detected in NGC\,2244 are YSOs.

\cite{lada1987} divided YSOs into three classes, I, II, and III. Class I sources are objects deeply embedded in the surrounding cloud with emission dominated by the cold dust of their envelopes (protostars). They practically cannot be detected in the visual domain. Class II sources having accretion disks are identified with TTSs and are visible in the optical range although they have infrared excesses. Finally, Class III are post-T Tauri objects with no or little infrared excess. Their photometric properties are similar to the normal main-sequence stars.

Three other classes of YSOs were proposed in the literature. Using observations in the submillimeter domain, \citet{andre1993} found extremely young object and proposed a fourth class of YSOs, Class 0. The Class 0 source cannot be seen even in near- or mid-infrared. The fifth class, called ``flat spectrum" (hereafter FS), was proposed by \cite{greene1994} based on the slope of SED calculated between 2.2 and 10\,$\mu$m, defined as $\alpha = d\log(\lambda F_{\lambda})/d\log\lambda$, which is between $-$0.3 and 0.3 for these stars.  Class I YSOs have $\alpha >$ 0.3, Class II, $-$1.6 $<\alpha < -$0.3, and Class III, $\alpha <-$1.6. Based on the slope of SED calculated between 3.6 and 8\,$\mu$m,  \citet{lada2006} found that objects with thick disks have spectral index $\alpha > -$1.80, the objects with ``anemic disks'' (transition disk objects, objects with thin disks) have $-$2.56$<\alpha < -$1.8 and objects without disks $\alpha <-$2.56.

The YSOs of Class I and II can be easily distinguished using mid-infrared photometry, e.g.~from Spitzer satellite \citep{mege2004,allen2004,allen2007}. Based on Spitzer IRAC (Infrared Array Camera) and MIPS (Multiband Imaging Photometer for Spitzer) observations, \cite{balog2007} found 25 Class I objects and 337 Class II objects in NGC\,2244.

For the identification of YSOs among our objects, we used the method described in Appendix A of \cite{guter2009}, developed by \cite{mege2012}. They utilized {\it Spitzer\/} IRAC and MIPS photometry, and 2MASS photometry for selection Class I, Class II and transition disk objects. 

\begin{figure*}
\centering
\includegraphics[width=174mm]{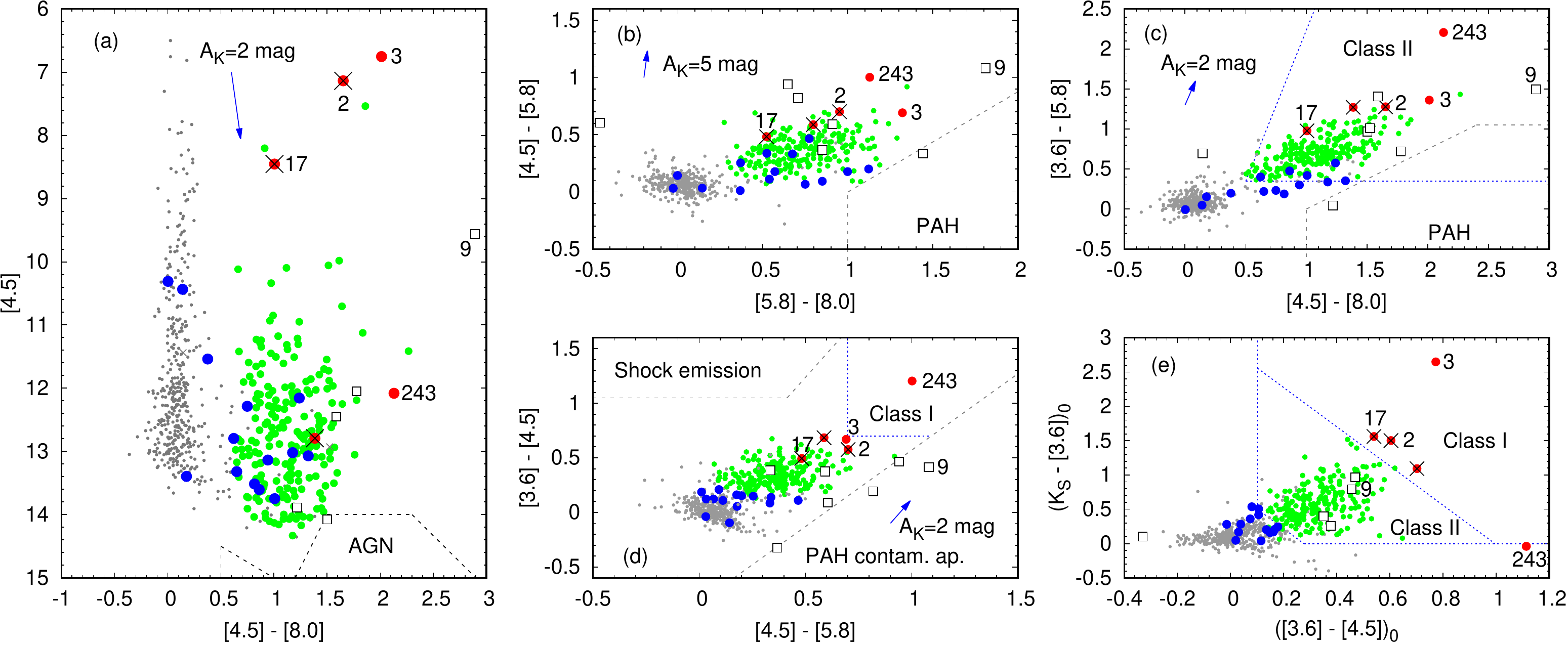}
\caption{{\it Spitzer\/}/IRAC colour-magnitude diagram and colour-colour diagrams used for   the selection of Class I (red dots) and Class II (green dots) objects and for the isolation of the extragalactic sources (white squares, see Sect.~\ref{ssyso1} for more details).
Dark dots represent Class III objects and/or field stars. Three crossed Class I objects does not fulfil the colour restriction for being a protostar ([4.5]--[24]$>$4.761). Transition disk objects are marked with blue dots. The arrows show the reddening vectors \citep{flah2007}. Some variables discussed in text are labeled.} 
\label{ccd-58-80-45-58-3}
\end{figure*}

\subsection{\label{ssyso1}\textbfit{Spitzer} and 2MASS photometry}

The {\it Spitzer\/} data were taken from catalogue published by \citet{balog2007}.
We found 674 counterparts in this catalogue putting an offset of 2\arcsec \,between our and {\it Spitzer\/} positions. Almost all of them have photometric uncertainties $\sigma < 0.1$ mag. All of them have photometry in all four IRAC bands ([3.6], [4.5], [5.8] and [8.0] $\mu$m) but only 167 stars have MIPS photometry ([24] $\mu$m). 

At the beginning, according to the criteria of \cite{guter2009} and \cite{mege2009}, we identify extragalactic sources that may be misclassified as YSOs. They are marked with white squares in Fig.~\ref{ccd-58-80-45-58-3}$-$\ref{cmd-K-24-K}. The other symbols shown in Fig.~\ref{ccd-58-80-45-58-3}$-$\ref{cmd-K-24-K} denote our final classification of the YSOs, described in Sect.~\ref{syso}.
We found only one object that satisfy colour and brightness criteria (marked with dashed line in Fig.~\ref{ccd-58-80-45-58-3}a) for active galactic nuclei (AGNs). Two objects fall in the region of galaxies with bright PAH (polycyclic aromatic hydrocarbon) emission: one in [4.5]$-$[5.8] {\it vs.} [5.8]$-$[8.0] colour-colour diagram (Fig.~\ref{ccd-58-80-45-58-3}b) and one in [3.6]$-$[5.8] {\it vs.} [4.5]$-$[8.0] colour-colour diagram (Fig.~\ref{ccd-58-80-45-58-3}c).
We did not find any shock emission knots. Five objects, according to criteria of \citet{guter2009}, may be contaminated by saturated PAH emission objects (Fig.~\ref{ccd-58-80-45-58-3}c). All these objects, are not included in any catalogue of AGNs or galaxies and they are very close to restriction lines in colour-magnitude and colour-colour diagrams, so they are probably not extragalactic sources. 

The $JHK_{\rm S}$ photometry was taken from 2MASS All-Sky Catalog of Point Sources of \citet{curti2003}. For 3098 our stars we found counterparts among 2MASS sources in the search radius of 2\arcsec, 1517 of them have photometric uncertainty $\sigma < 0.1$ mag in all three bands, and 653 have counterparts in the catalogue of \citet{balog2007} (605 of them have $\sigma < 0.1$ mag). 

The {\it Spitzer\/} and 2MASS magnitudes for variable stars are shown in Table \ref{tir}\footnote{The full version of Table\,\ref{tir} is available in electronic form from the CDS.} in Appendix.

\subsubsection{\label{ssyso2}Selection of Class I and Class II objects}

After checking for extragalactic objects, we identified Class II objects using restrictions in [4.5]--[8.0] and [3.6]$-$[5.8] colours (shown in Fig.~\ref{ccd-58-80-45-58-3}c), and with [3.6]$-$[4.5]$-\sigma$([3.6]$-$[4.5])$>$0.5. From this sample, objects with [4.5]$-$[5.8] and [3.6]$-$[4.5] colours greater than 0.7 were classified as Class I. We found one such source,variable 243\footnote{The numbers indicate the variables we found, listed in Table \ref{tubvi}}, that fulfil these colour restrictions (Fig.~\ref{ccd-58-80-45-58-3}d). 

Some additional YSOs can be found from the constraints of the dereddened ([3.6]$-$[4.5])$_0$ and ($K_{\rm S}-$[3.6])$_0$ colours derived according to procedure described in Appendix A.2.~of \citet{guter2009}. Originally, this method was applied to the stars without [5.8] and/or [8.0] $\mu$m photometry which have good quality ($\sigma < 0.1$ mag) $H$, $K_{\rm S}$ and/or $J$ photometry. 
Although all our stars with {\it Spitzer\/} data have photometry in all four IRAC bands, we apply this method to check if we can find some additional Class I and Class II objects.

This way we found 4 additional Class I candidates (Fig.~\ref{ccd-58-80-45-58-3}e) and 23 additional Class II candidates. Out of these 23 objects, 12 stars which have $[4.5]-[24]>2.5$ mag and $[5.8]-[24]>2.5$ mag (or $[4.5]-[8.0]>0.5$ mag for objects without [24] $\mu$m band photometry), we classified as Class II objects. 

\begin{figure}
\centering
\includegraphics[width=68mm]{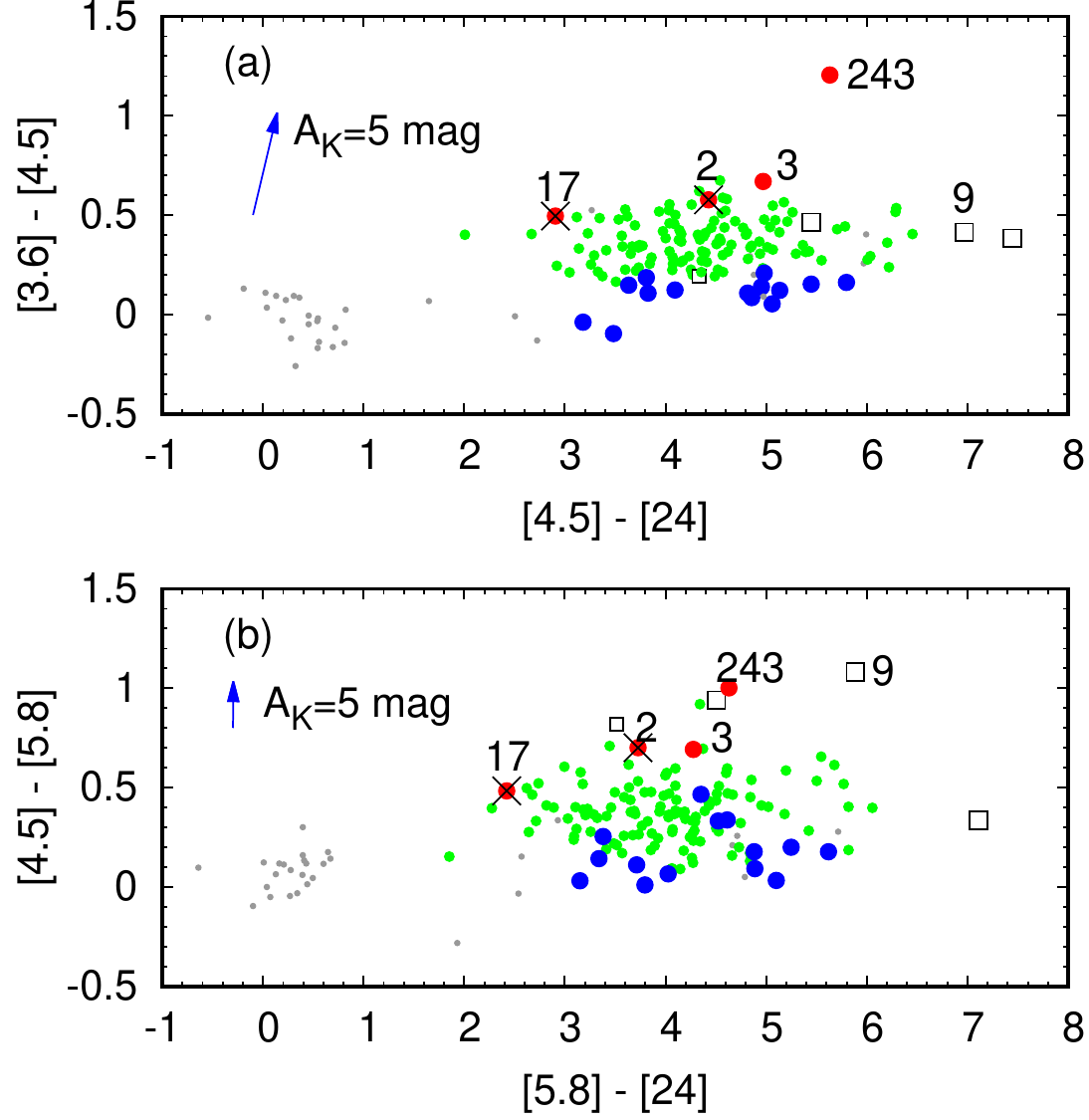}
\caption{{\it Spitzer\/}/IRAC and MIPS colour-colour diagrams used for the selection of transition disk objects (blue circles). The coloured symbols are the same as in Fig.~\ref{ccd-58-80-45-58-3}. The arrow shows the extinction vector for $A_K=5$ mag \citep{flah2007}.} 
\label{ccd-240}
\end{figure}

All objects selected as protostars must have $[5.8]-[24]$ colour greater than 4 mag. Two objects (variables 2 and 17) found previously as Class I do not meet this condition and for one object there are no [24]-band photometry. They are marked with crosses in Fig.~\ref{ccd-58-80-45-58-3}$-$\ref{cmd-K-24-K}. Finally, using {\it Spitzer\/} and 2MASS photometry, we found two Class I objects (variable 3 and 243) and 227 Class II objects.

\subsubsection{\label{ssyso3}Selection of transition disk objects}

Another class of YSOs are transition disk objects which do not show near- and mid-infrared excess. These sources can be found with the use of [24] $\mu$m MIPS photometry. We used the approach of \cite{mege2012} who considered objects with [8]--[24] colours greater than 2.5 mag and the dereddened [3.6]--[4.5] colours smaller than 0.2 mag as transition disk object. This way, among 167 objects with MIPS photometry, we found 14 transition disk objects (7 of them are variable). They are marked in blue in Fig.~\ref{ccd-58-80-45-58-3}--\ref{cmd-K-24-K}. 

\subsubsection{\label{ssyso4}Selection of Class III and flat spectrum objects}

Most of the remaining 426 objects which were not classified as Class I, Class II or transition disks are Class III objects (diskless YSOs) and/or field stars (pure photosphere), 84 of them are variable. 
It is difficult to distinguish between these two groups. One way was presented by \citet{rebull2007} who used $K_{\rm S}$ vs. $K_{\rm S} - [24]$ colour magnitude diagram.
The authors established the ranges of $K_{\rm S} - [24]$ colour for Class I (8.31$-$14 mag), flat spectrum (6.75$-$8.31 mag), Class II (3.37$-$6.75 mag) and Class III ($<$3.37 mag) sources. In order to separate Class III objects from foreground/background stars, they set a boundary colour $K_{\rm S} - [24]=2$ mag. 
\begin{figure}
\centering
\includegraphics[width=68mm]{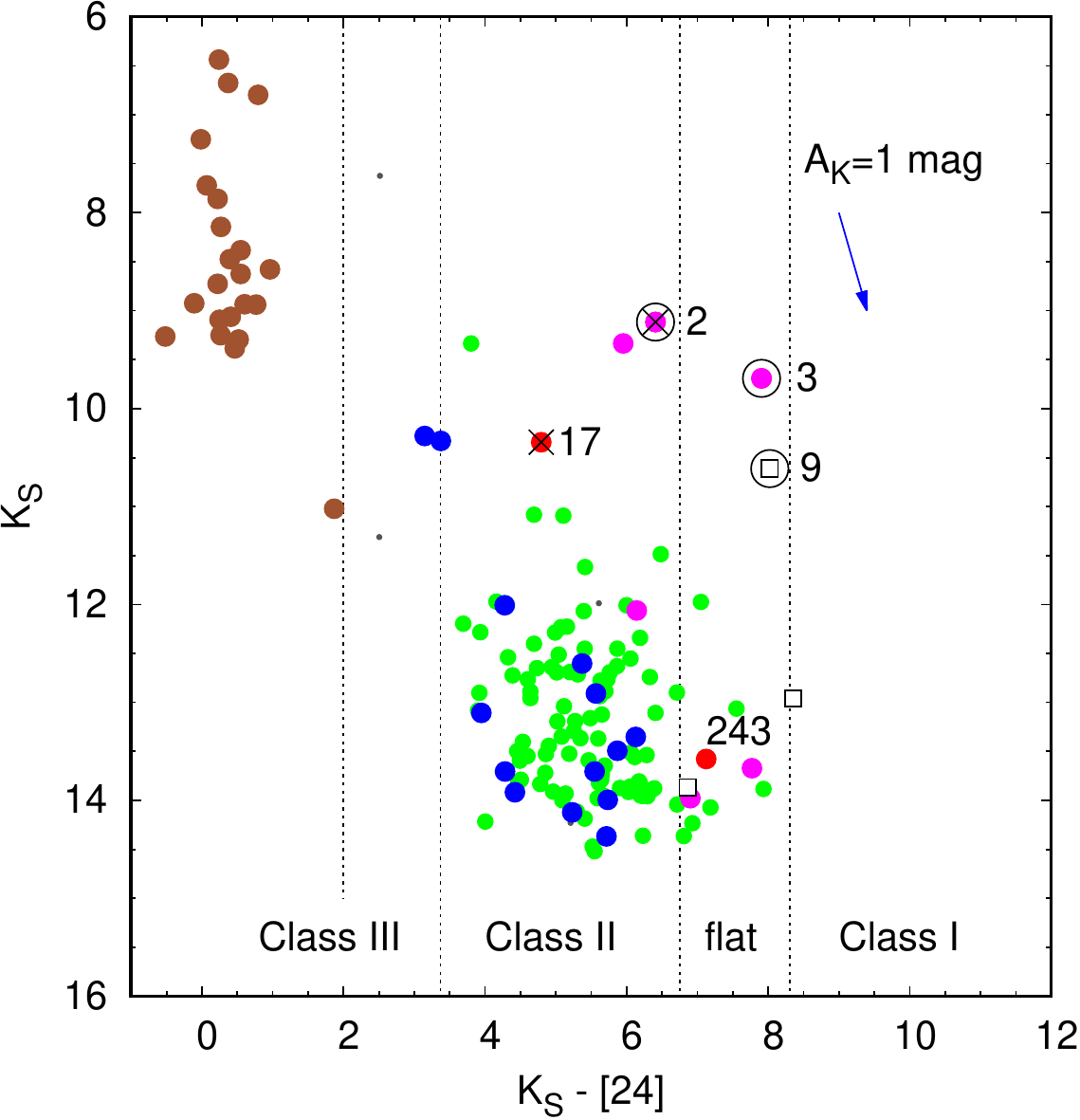}
\caption{$K_{\rm S}$ vs. $K_{\rm S}-$[24] colour-magnitude diagram for observed stars in NGC 2244. Dotted lines denote the divisions between Class I, flat spectrum, Class II, and Class III objects \citep{rebull2007}. The arrow shows the extinction vector for $A_K=1$ mag \citep{flah2007}. Symbols for YSOs are the same as in Fig.~\ref{ccd-58-80-45-58-3}. Additionally, encircled symbols represent HAeBe stars \citep{parksung}, pink dots -- flat spectrum object with $-0.3<\alpha _{\rm IRAC}<0.3$, field stars are marked with brown dots.} 
\label{cmd-K-24-K}
\end{figure}

The $K_{\rm S}$ vs.~$K_{\rm S} - [24]$ colour magnitude diagram for stars in the observed field is shown in Fig.~\ref{cmd-K-24-K}. 
In this diagram YSOs found using the selection criteria of \citet{guter2009} are shown.
As can be seen in the figure, none of Class I star have $K_{\rm S} - [24]>8.31$ mag. Both Class I objects together with 8 Class II objects have $K_{\rm S} - [24]$ between 6.75 and 8.31 mag. According to colour criteria of \citet{rebull2007} they are YSOs with flat spectrum. 
The encircled symbols in Fig.~\ref{cmd-K-24-K} represent HAeBe stars \citep{parksung}. One of them, star 3, according to criteria of \cite{guter2009}, was classified as Class I object and the other HAeBe star, star 2, was classified as Class II object. Third HAeBe star, star 9, are in the area of sources  contaminated by saturated PAH emission objects (Fig.~\ref{ccd-58-80-45-58-3}d)
 
For three stars $K_{\rm S} - [24]$ is between 2 and 3.37 and they are probably diskless YSOs.
We found that 22 stars have $K_{\rm S} - [24]<2$ mag and 21 of them have $K_{\rm S} - [24]<1$ mag.

\begin{figure*}
\centering
\includegraphics[width=12cm]{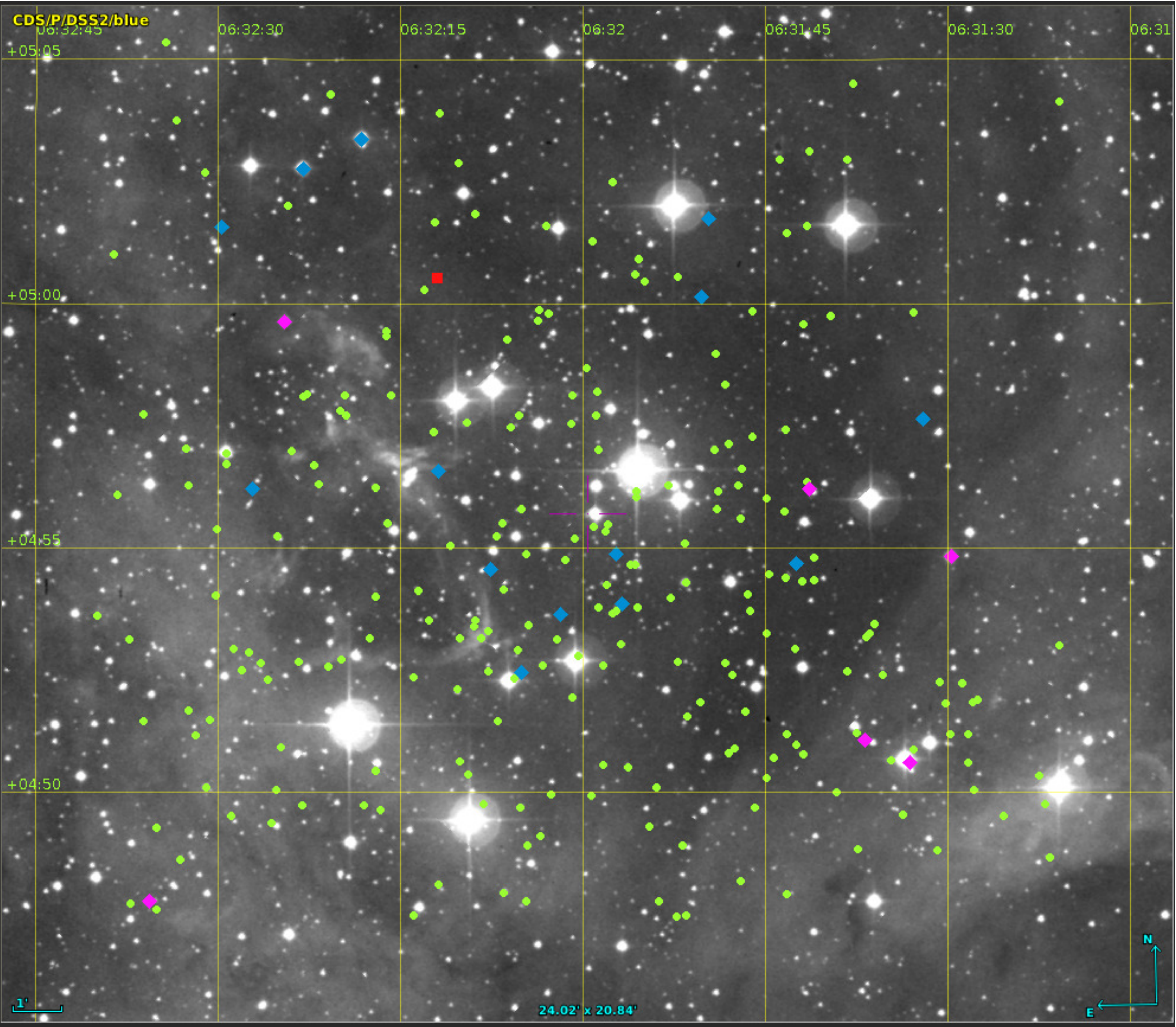}
\caption{Spatial distribution for Class I object (red square), Class II objects (green circles), flat spectrum objects (magenta rhombus) and transition objects (blue rhombus) found from {\it Spitzer} and 2MASS photometry (Sect.~\ref{syso}).} 
\label{map2}
\end{figure*}

The position of YSOs in the map of the sky is shown in Fig.~\ref{map2}. The objects of all classes are distributed homogeneously, without any concentration.

\begin{figure}
\centering
\includegraphics[width=68mm]{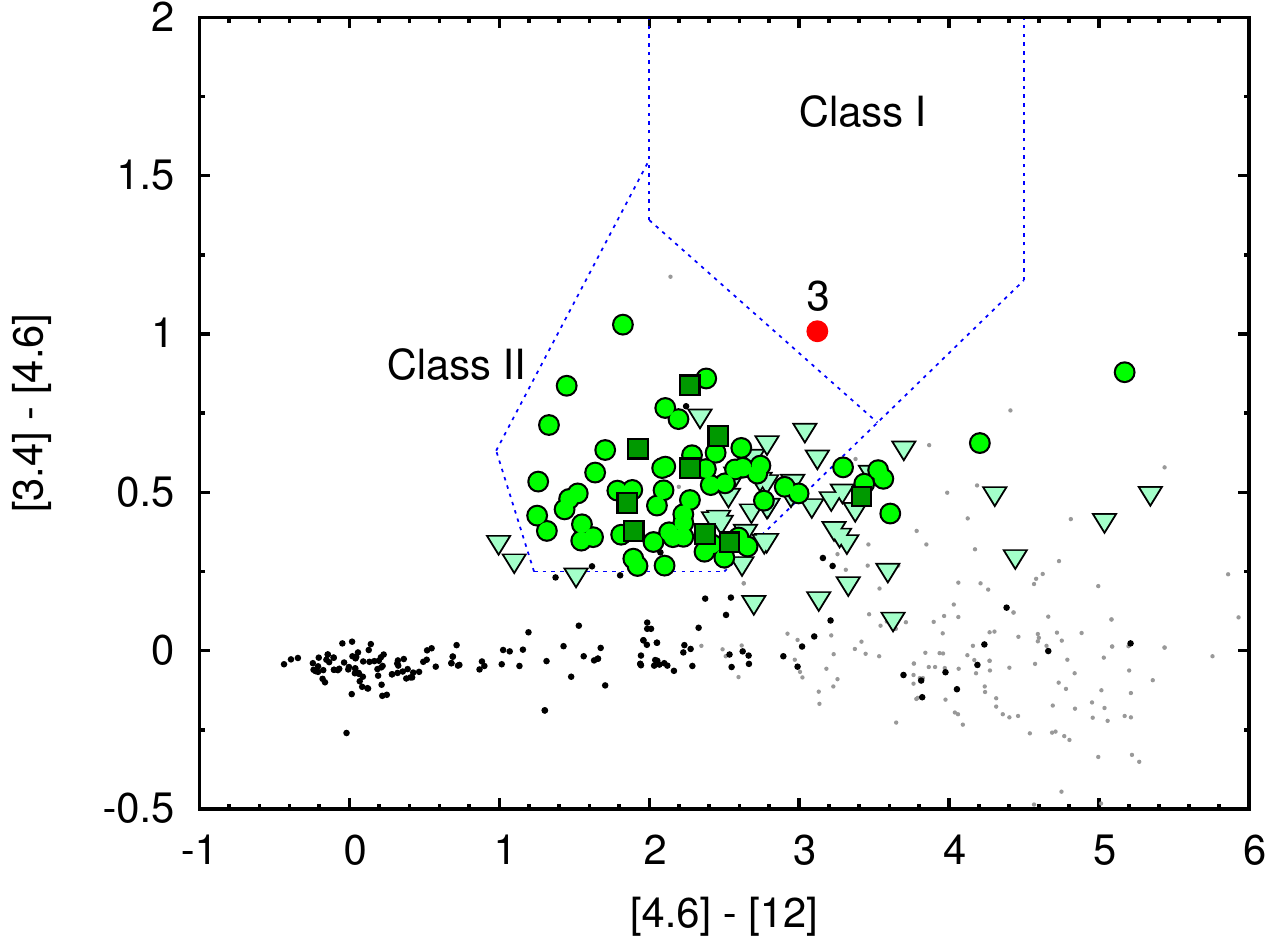}
\caption{WISE colour-colour diagram used for the selection of Class I and Class II objects base on restrictions of \citet{koe2014}. Green symbols represent Class II sources: circles -- common from WISE and IRAC, squares -- only from WISE, triangles -- only from IRAC. Red circle indicate Class I object. Black points are Class III/fiels stars, grey points -- sources that may be extragalactic contaminants.} 
\label{ccd-wise}
\end{figure}

\subsection{\label{salpha}Spectral index \textbfit{$\alpha$}}

The classes of YSOs, described in the introduction of this section, were defined based on $\alpha$ spectral index. We computed $\alpha$ indexes, from the slope of the linear fit to the observed fluxes between the IRAC [3.6] $\mu$m band and the IRAC [8.0] $\mu$m band ($\alpha _{\rm IRAC}$), as it is quite insensitive to extinction \citep{muench2007}. The $\alpha$ indexes of the variable stars are given in Table \ref{tubvi} in Appendix. 

Since $\alpha$ index of HAeBe star 3, classified in Sect.~\ref{ssyso2} as Class I, is less then 0.3, it is rather flat spectrum object. Four other objects which have $\alpha$ between $-0.3$ and $0.3$, we classified as flat spectrum. They are marked with pink dots in Fig.~\ref{cmd-K-24-K}.
One object with $\alpha=-0.31$ and $K_{\rm S} - [24]$ between 6.75 and 8.31 mag we also classified as flat spectrum.
For the other Class I object, star 243,  the $\alpha$ index is equal to 0.98 confirms its previous classification. Another object, HAeBe star 9, also have $\alpha$ index greater then 0.3, such as Class I objects, but according to colours restrictions of \citet{guter2009} is classified as object contaminated by PAH emission. The $\alpha$ indexes of almost all Class II YSOs found in Sect.~\ref{ssyso2}, which are between -1.8 and -0.3, confirmed their classification.

\subsection{WISE photometry}
\label{swise}

Since not for all our variable stars we found {\it Spitzer} counterparts, we identify YSOs based on WISE (Wide-field Infrared Survey Explorer, Wright et al.~2010) photometry as well. This photometry was obtained in in four mid-infrared bands: [3.4], [4.6], [12], and [22] $\mu$m. Using ALLWISE catalog \citep{curti2014} we found 1804 counterparts in the search radius of 2\arcsec.
In order to find YSOs, to the sources with error less then 0.2 mag in [3.4], [4.6] and [12] $\mu$m band, we applied the colour and brightness constraints described in \citet{koe2014}. In this way we found only one Class I object (star 3) and 
9 additional Class II objects (marked with squares in Fig.~\ref{ccd-wise}), 3 of them are variable. 
From the slope of the linear fit to the observed fluxes in WISE [3.4], [4.6] and [12] $\mu$m band, we calculated $\alpha$ indexes ($\alpha _{\rm WISE}$).
The classes of objects and $\alpha _{\rm WISE}$ indexes are written in brackets in Table \ref{tubvi}.

The total number of YSOs found in observed field are summarized in Table \ref{tyso}.

\begin{table}
\centering
\caption{\label{tyso}Number of YSOs found in observed field.}
\begin{tabular}{@{}lcc}\hline
Objects & total & variable \\\hline
Class I & 1 & 1 \\
Class I/\,Flat spectrum & 1 & 1 \\
Flat spectrum$^1$ & 5 & 3 \\
Flat spectrum$^2$ &10 & 3 \\
Class II & 219 & 89 \\
Transition disks & 14 & 7 \\
Class III & 426 & 84 \\\hline
\multicolumn{3}{l}{$^1$ objects classified based on $\alpha$ index;}\\
\multicolumn{3}{l}{$^2$ objects classified based on $K_{\rm S} - [24]$ colour.}\\
\end{tabular}
\end{table}

\section{Characteristics of PMS variable stars}
\label{spms}
The variable PMS stars found in Sect.~\ref{ssin} and \ref{sother} may be WTTSs, CTTSs or HAeBe stars. Their position in colour--magnitude and colour--colour diagrams, near- and mid-IR excess as well as the properties of $H\alpha$ emission line help to understand the possible origin of their variability. 
Therefore, additional archival data of the IPHAS  (Isaac Newton Telescope Photometric H-Alpha Survey) and UKIDSS (United Kingdom Infrared Deep Sky Survey) surveys were used in our analysis.

\subsection{\label{iphas}IPHAS photometry}
Combination of IPHAS $(r^{\prime}-i^{\prime})$ and $(r^{\prime}-H\alpha)$ colour indices allowed to find stars with H$\alpha$ in emission. The method was described by \cite{drew2005} and \cite{baren2011}. The authors provide a grid of colour tracks for stars with different $H\alpha$ emission equivalent widths (EW$_{\rm H\alpha}$), spectral types and reddenings.
We found $r^{\prime}$, $i^{\prime}$ and $H\alpha$-band photometry for about 3700 our objects using IPHAS DR2 catalogue \citep{baren2014} in the search radius of 1\arcsec.

\begin{figure}
\centering
\includegraphics[width=76mm]{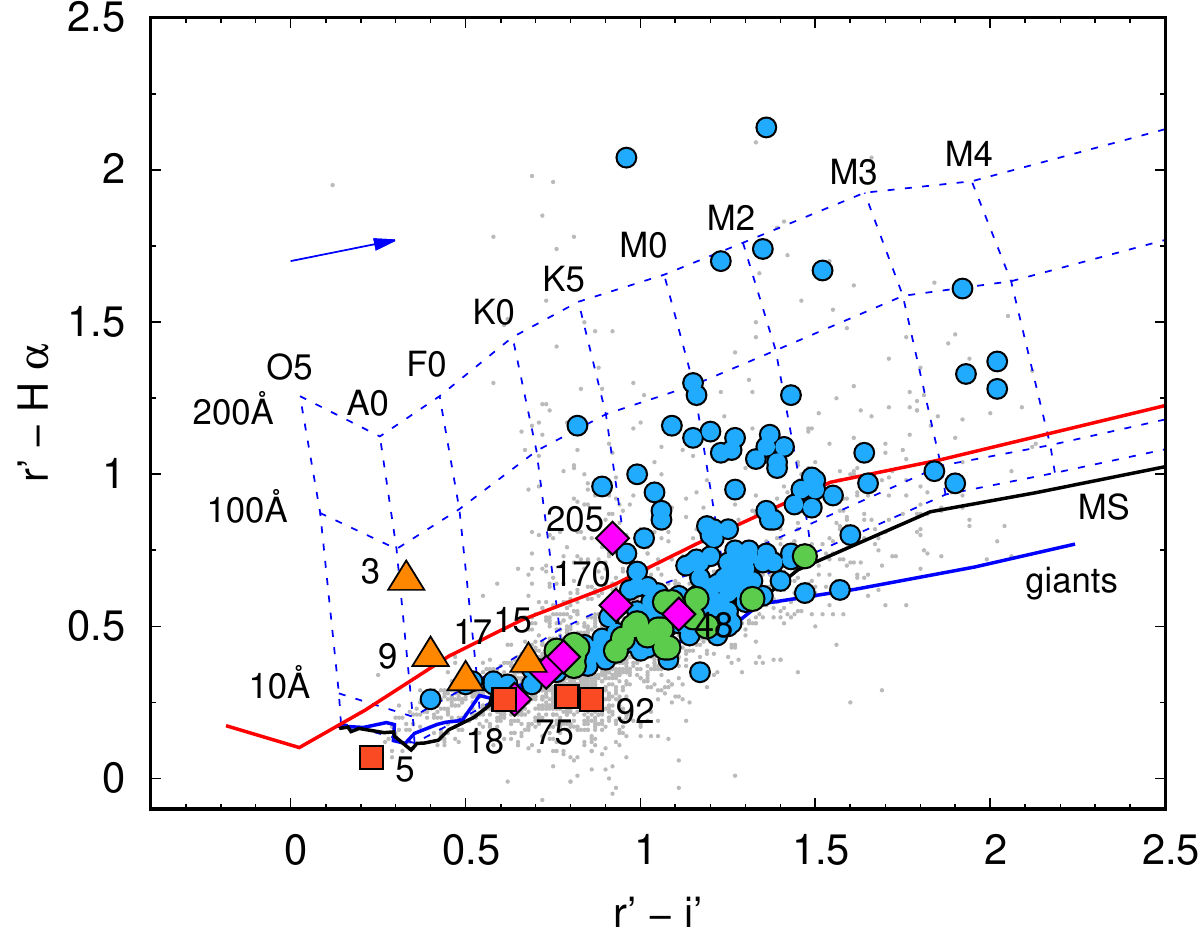}
\caption{IPHAS $(r^{\prime}-H{\alpha})$ vs.~$(r^{\prime}-i^{\prime})$ colour-colour diagram for the observed field. The symbols indicate variable stars found in this paper: pulsating (red squares), eclipsing (pink diamonds), HAeBe (orange triangles), other periodic (green circles) and irregular (blue circles). Synthetic tracks for MS stars and giants (solid black and blue lines), and for stars with photometric EW$_{\rm H\alpha}$=\,10, 100 and 200\,{\AA} for spectral types O5--M6 (dashed blue lines), shifted according to reddening vector for $A_V=1.46$ mag (blue arrow), were taken from \protect\cite{baren2011}. Red line is the selection threshold for CTTSs.}
\label{ccd-bar}
\end{figure}

\begin{figure}
\centering
\includegraphics[width=76mm]{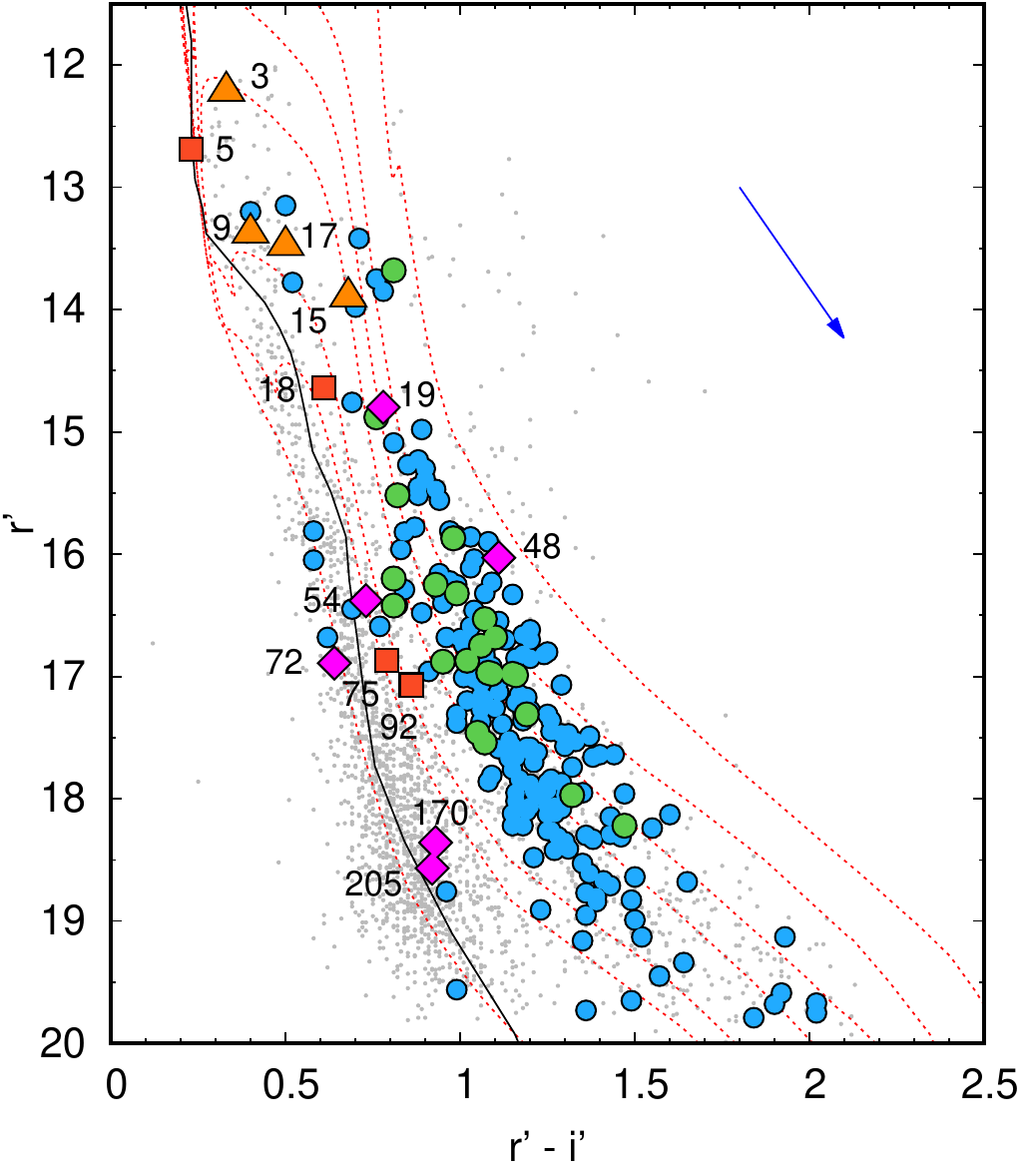}
\caption{IPHAS  $r^{\prime}$ vs.~$(r^{\prime}-i^{\prime})$ colour-magnitude diagram for the observed field. The symbols are the same as in Fig.~\ref{ccd-bar}. The ZAMS relation for Z=0.02 (solid line) was taken from \protect\cite{siess2000}. The isochrones (dotted lines) for 0.1, 0.5, 1, 2, 6, 10, and 100\,Myr for adopted $A_V=1.46$ mag and $(m-M_{\rm V})_0=11.1$ mag were taken from \protect\cite{bress2012}.} 
\label{cmd-bar}
\end{figure}

For 226 variables found in our photometry we identified IPHAS counterparts with photometric errors smaller than 0.1 mag in all three bands. They are shown in the $(r^{\prime}-H{\alpha})$ vs.~$(r^{\prime}-i^{\prime})$ colour-colour and  $r^{\prime}$ vs.~$(r^{\prime}-i^{\prime})$ colour-magnitude diagrams in Fig.~\ref{ccd-bar} and \ref{cmd-bar}, respectively. In order to find CTTS candidates based on  $(r^{\prime}-i^{\prime})$ and $(r^{\prime}-H{\alpha})$ colours, we used the selection threshold described by \cite{baren2011}. For early type stars, this threshold (red line in Fig.~\ref{ccd-bar}) is the same as unreddened 10\,{\AA} track, but for stars later than M0 the empirical threshold from \cite{barrado2003} was taken. We found 143 objects above the threshold. Out of them, 30 variable stars lie above the threshold with 3$\sigma$ confidence level defined by \cite{baren2011}. They are notified with the letters ''IA'' in Table \ref{tubvi} in Appendix. The other stars above the reddened ($E(B-V)=0.47$ mag) 10\,{\AA} track and the objects lying between 0 and 10\,{\AA} track are marked in Table \ref{tubvi} as ''Ia'' and ''Ib'', respectively.
As can be seen in Fig.~\ref{ccd-bar}, most of the variables with strong  $H{\alpha}$ emission have spectral types later than K0. Only two stars above CTTS threshold with spectral type earlier than K0, star 3 and 9, are HAeBe stars reported by \citet{parksung}. The IPHAS magnitudes for variable stars are listed in Table \ref{tir} in Appendix.

\subsection{UKIDSS photometry}
\label{ukidss}
The colour-colour diagram of near-infrared $JHK$ photometry is the other useful tool for understanding YSOs.
Therefore, we decided to use UKIDSS (United Kingdom Infrared Deep Sky Survey) photometry as it is more accurate than 2MASS photometry. The UKIDSS photometry was obtained with the United Kingdom InfraRed Telescope (UKIRT) during Galactic Plane Survey.
Almost for all objects we observed, we found $J$, $H$ and $K$ photometry in the UKIDSS-DR6 catalogue \citep{lucas2008}, assuming a search radius of 1\arcsec. The $JHK$ UKIDSS magnitudes for variable stars are shown in Table \ref{tir} in the Appendix.

\begin{figure}
\centering
\includegraphics[width=76mm]{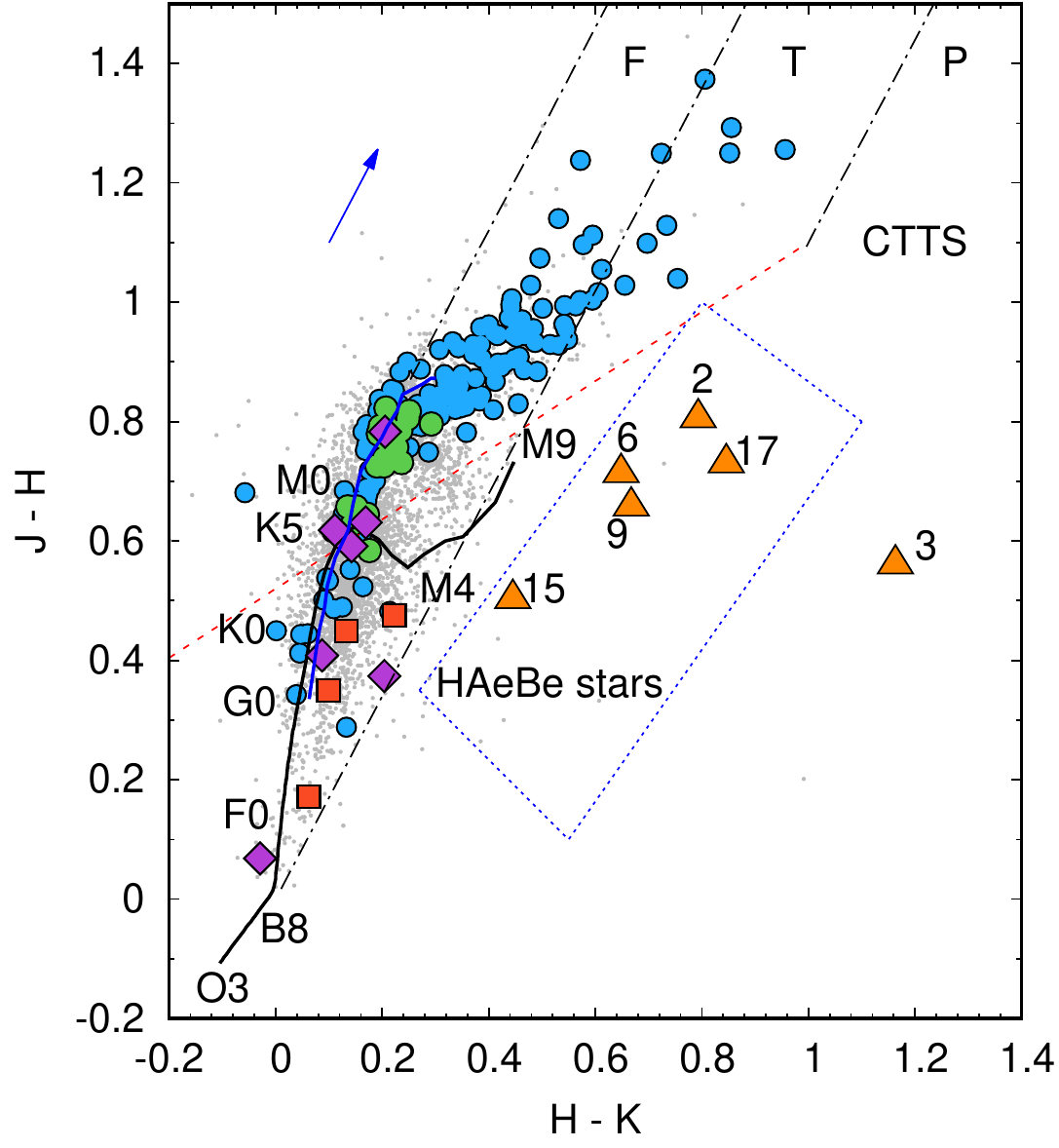}
\caption{The $(J-H)$ {\it vs.} $(H-K)$ UKIDSS colour-colour diagram for the observed field. The symbols are the same as in Fig.~\ref{ccd-bar}. The dashed line is the locus of CTTSs \citep{meyer1997}. Solid black and blue lines denote intrinsic colours for ZAMS and late-type giants, respectively.
The arrow shows the reddening vector for $A_V=1.46$ mag. All the measurements were transformed to the CIT system.} 
\label{ccd-ukidss}
\end{figure}

The $(J-H)$ vs.~$(H-K)$ diagram for variable stars found in Sect.~\ref{svar} is shown in Fig.~\ref{ccd-ukidss}. The dashed line represents the locus of CTTSs determined by \citet{meyer1997}. Since this line was derived in the CIT (California Institute of Technology) system, we transformed UKIDSS photometry to 2MASS photometry according to the transformation relations of  \citet{hewett2006} and then to CIT photometry using relations described by Carpenter on ''2MASS Color Transformations'' web page\footnote{http://www.astro.caltech.edu/\~{}jmc/2mass/v3/transformations/}.
If the brightness of the star in UKIDSS-DR6 catalogue is flagged as "close to saturated", we show 2MASS colours, transformed to CIT system in Fig.~\ref{ccd-ukidss}.

The solid black line in Fig.~\ref{ccd-ukidss}, taken from \citet{pec2013}\footnote{New version were used available at http://www.pas.rochester.edu/ \~{}emamajek/EEM\_dwarf\_UBVIJHK\_colours\_Teff.txt}, represents the ZAMS and the solid blue line, taken from \citet{bebr1988}, denote colours for late-type giants. The arrow shows the reddening vector for $A_V=1.46$ mag. The extinction ratios $A_J/A_V=0.282$, $A_H/A_V=0.175$ and $A_K/A_V=0.112$ we adopted from \citet{rieke1985}. The blue rectangle is the area of dereddened $(J-H)$ and $(H-K)$ colours for HAeBe stars, defined by \citet{hern2005}. As can be seen in this figure, five variable stars we found fall into this box. They will be discussed in Sect.~\ref{sspms}.
 
We identified 208 stars of Class II lying above CTTS locus, 88 of them are variable. Only 20 of Class II objects lie below CTTS line, four of them (2, 9, 15, 17) are variables in HAeBe box in the $(J-H)$ vs.~$(H-K)$ diagram (Fig.~\ref{ccd-ukidss}). 

Following \citet{sug2002} we have drawn three lines parallel to the reddening vector: from the tip (spectral type M4) of the giant branch (left line), from the point $(H-K),(J-H)=(0.0,0.0)$ mag for spectral type A0 of the MS (middle line), and from the tip of the intrinsic CTTS line (right line).  
The objects lying between left and middle line, marked with "F" in Table \ref{tubvi} in Appendix, are either field stars (main-sequence stars or giants) or Class III/Class II sources with small near-infrared (NIR) excesses \citep{Ojha2004}. The sources lying between middle and right line, marked with "T" in  Table \ref{tubvi} in Appendix, are considered to be Class II sources (mostly CTTSs) with large NIR excesses or Herbig Ae/Be stars with small NIR excess. The sources lying redwards of "T" region are most likely Class I objects (protostars) or Herbig Ae/Be stars. 

\subsection{Classification of PMS variable stars} 
\label{sspms}
The age spread of PMS stars in NGC 2244 estimated by \citet{parksung} is between 0.1$-$0.3 and 5.8$-$6.4 Myr, depending on the adopted models. Most of our variables lie between 1 and 6 Myr isochrone and plenty of them are PMS stars (HAeBe stars and TTSs).
The positions of PMS variable stars in the infrared and optical colour-magnitude and colour-colour diagrams helped us to classify them. 

Five variables fall into the HAeBe box in Fig.~\ref{ccd-ukidss}. All of them were classified as Class II objects. Three of these stars, (star 2, 6, and 9) are known HAeBe stars (see Sect.\ref{sother}). 
Fourth known HAeBe, star 3, has strong infrared excess and is placed outside HAeBe box in the $(J-H)$ vs.~$(H-K)$ diagram (Fig.~\ref{ccd-ukidss}). We classified this star as HAeBe Class I/flat spectrum object.
The other two stars falling into the HAeBe box (Fig.~\ref{ccd-ukidss}) are stars 15 and 17. The photometric  EW$_{\rm H\alpha}$ derived from IPHAS photometry of three HAeBe candidates (star 3, 9, and 17) is greater than 10\,{\AA}. For star 15, the photometric EW$_{\rm H\alpha}$ is between 0 and 10 {\AA}.

A common feature of classical T Tauri stars (CTTSs) is the equivalent width of $\rm H\alpha$ (EW$_{\rm H\alpha}$) emission line greater than 10\,{\AA}. 
Using IPHAS photometry we have found 68 variables lying above the 10\,{\AA} track (''IA'' and ''Ia'' in Table \ref{tubvi}). 
Most of them, 45 stars, are Class II objects, 43 of them have spectral types, estimated from $(r^{\prime}-i^{\prime})$ and $(r^{\prime}-H{\alpha})$ colours (Fig.~\ref{ccd-bar}) later than K0 and lie above CTTS locus in $(J-H)$ vs.~$(H-K)$ diagram (Fig.~\ref{ccd-ukidss}). They are the best candidates for CTTS stars and are marked as CTTS in Table \ref{tubvi} in Appendix. Excellent  candidates for CTTSs are also four stars with EW$_{\rm H\alpha}<10$\,{\AA} which large infrared excess placed them in "T" region in Fig.~\ref{ccd-ukidss} and one Class I object, star 243.

Among 48 Class III/field stars, that lie above the 10\,{\AA}  EW$_{\rm H\alpha}$ track in Fig.~\ref{ccd-bar}, 36 stars occupy the region of FGK type main sequence stars, with $(J-H)<0.6$ mag. Seven of them, marked as MS in Table \ref{tubvi}, are variable. The remaining twelve stars can be CTTSs as they lie above CTTSs locus in Fig.~\ref{ccd-ukidss}. Nine of them, marked as CTTS$^\star$ in Table \ref{tubvi}, are variable.  The position of selected CTTS candidates in $r^{\prime}$ vs.~$(r^{\prime}-i^{\prime})$ colour-magnitude diagram (Fig.~\ref{cmd-bar}) is typical for PMSs. Only one star, 222, lie in this diagram close to ZAMS and is placed above 200\,{\AA} track in Fig.~\ref{ccd-bar}. It may be an active field star, but its position in the $V$ vs.~$(V-I_{\rm C})$ diagram a little above 10 Myr isochrone and the membership probability, equal to 74.3\%, does not exclude it from the members of the cluster.

Our classification of other Class II variables lying above CTTS locus in Fig.~\ref{ccd-ukidss} with photometric EW$_{\rm H\alpha}<10$ depends on their spectral index $\alpha$. \citet{lada2006} found that objects with thick disks, typical for CTTSs, are characterised by spectral index $\alpha >-1.8$. They also found  that large fraction (64\%) of WTTSs are diskless stars ($\alpha <-2.56$) and about 22\% of WTTSs have thin disks ($-2.56<\alpha<-1.8$). For this reason, we marked the stars with $\alpha>-1.8$ as CTTS$^\star$ in Table \ref{tubvi} and stars with $\alpha<-1.8$ as WTTS. The Class III variables lying above CTTS locus in Fig.~\ref{ccd-ukidss} and having EW$_{\rm H\alpha}$ between 0 and 10 were also classified as WTTS. The remaining variable stars lying above CTTS line were classified as WTTS$^\star$.
This way, we have found 97 candidates for CTTS stars (49 CTTS and 48 CTTS$^\star$), 68 candidates for WTTS stars (32 WTTS and 36 WTTS$^\star$) and 6 candidates for HAeBe stars (4 of them were known before).

Common feature of YSOs is emission in X-rays. Using catalogue of X-ray sources from {\it Chandra} published by \citet{wang2008}, we identified 482 counterparts in the search radius of 1\arcsec, 182 of them are variable stars. They are marked with letters ''a'', ''b'' or ''c'' in Table\,\ref{tubvi} in Appendix, following the notification used by \citet{wang2008} indicating variability characterization.

\citet{parksung} found ROSAT HRI X-ray counterparts for sixteen stars in NGC 2244. They  classified six of them as PMS stars. Four of them (star 10, 13, 14, and 16) are variable in our photometry. The authors denoted two objects as ''PMS X-ray binary'', one of them is variable star in our data (star 49). \citet{parksung} also selected 14 PMS star and seven PMS candidates based on the strength of ${\rm H\alpha}$ emission line. Since the authors published the positions of stars brighter than 17 mag, we have found only seven counterparts. Six of them, marked with letters ''PS'' in Table\,\ref{tubvi}, are variable stars.

\citet{berg2002} published the results for 138 X-ray sources in NGC 2244 selected from ROSAT PSPC and HRI observations. The majority of them have strong ${\rm H\alpha}$ emission.  Out of them 38 stars are variable in our photometry. They are marked with letters ''BC'' in Table\,\ref{tubvi}. 

\begin{figure}
\centering
\includegraphics[width=76mm]{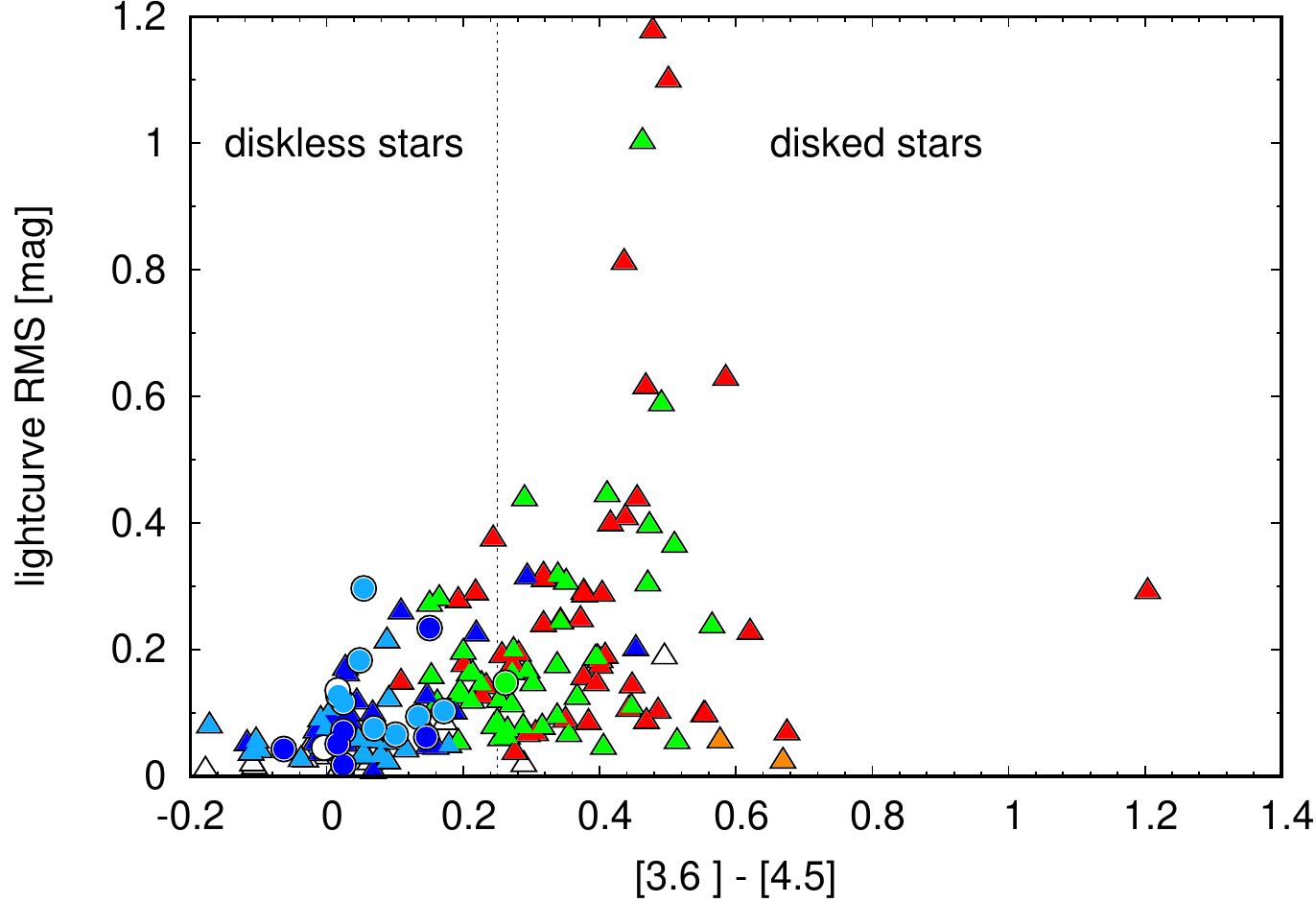}
\caption{The V-filter light curves RMS (variability amplitude) as a function of {\it Spitzer\/} [3.6]$-$[4.5] colour for periodic (dots) and irregular (triangles) variables.  The colours represent variable stars: CTTS (red), CTTS$^\star$ (green), WTTS (dark blue), and WTTS$^\star$ (light blue), described in Sect.~\ref{sspms}. Dashed line is the boundary between diskless and disked stars ([3.6]$-$[4.5]$=$0.25 mag), adopted from \citet{dutta2018}.} 
\label{rms}
\end{figure}

On average, the amplitudes of light curve variations of CTTSs are larger than those of WTTSs \citep{lata2012}. The variability of CTTSs is usually irregular, caused by accretion processes and the changes in their disks. The changes in light curves of WTTSs are often periodic, caused by cool spots.
\citet{dutta2018} show that variability amplitude of  stars with infrared excess is larger for those that have disks. Following the authors, we calculated the observed root mean square (RMS=$\sigma_{\rm obs}$) in the light curves, expressed as: 
\begin{equation}
\label{eq:sigma_obs}
   \sigma_{\rm obs}^2 = {n\;\sum_{i=1}^{n} w_i(m_i - \bar{m})^2 \over
                         (n-1)\;\sum_{i=1}^{n}w_i},
\end{equation}
where $m_i$ and $\bar{m}$ are respectively individual and average $V$-filter 
magnitudes, and  $w_i = 1.0/\sigma_i^2$ is the weight assigned to each observation with photometric uncertainties $\sigma_i$. The RMS as a function of [3.6]$-$[4.5] colour is shown in Fig.~\ref{rms}.
As can be seen in this figure, most stars classified as CTTS or CTTS$^\star$ have larger amplitudes then WTTSs. They also have the [3.6]$-$[4.5] colours greater than 0.25 mag which means, according to \citet{dutta2018}, they are disked stars.

\section{Summary} 
\label{ssum}

We presented multi-wavelength photometric study of young open cluster NGC 2244. Using $UBVI_{\rm C}$ photometry we have identified 245 variable stars. The time series photometry of only one of them was published before.
Based on near- and mid-infrared excess and photometric equivalent width of $H\alpha$ emission line we classify PMS variables. 
Among 211 stars irregular stars we have found 96 CTTS candidates, 54 WTTS candidates and 6 HAeBe stars. 
We have also identified  23 periodic variables with periods greater than one day. Fourteen of them we have classified as WTTS candidates and one star as CTTS$^\star$ candidate. 
The relation between mass of the star and its rotational period found in many young open clusters, e.g.~NGC 2264, NGC 7380, NGC 1893 and Be 59 is not visible in our sample of periodic variables. We have shown that stars with infrared excess have, on average, higher amplitudes. Similar tendency was found in NGC 2282 \citep{dutta2018}.
  
We have found four new $\delta$ Scuti stars. All of them are likely non-members. 
In the observed field there is one star reported in the literature as $\beta$ Cephei-type star. The small amplitude of its variability is however below our detection threshold. Unfortunately, the saturation of bright stars makes impossible to verify if in such a young cluster there are any $\beta$ Cephei or SPB stars. 

In addition to the well-known eclipsing binary V578 Mon, we have found six other eclipsing systems. Two of them are probably the cluster members.

Based on {\it Spitzer}, UKIDSS, 2MASS and WISE photometry we have identified YSOs in our photometry. As a result, one variable was classified as Class I object, one  as Class I/flat spectrum object, 91 as Class II objects, 4 as flat spectrum sources and 7 as transition disk objects. In comparison to \citet{balog2007} we have identified small number of Class I objects. 
This is due to the fact that we classified only the sources detected from our photometry, while most of these objects are not visible in the optical bands.

Due to the short period of observations, we were not able to find variable PMS stars such as UXors, FUors or EXors, although some candidates were found to have the range of variations greater than 1 mag and the shape of light curves similar to those type of variables. It is not out of the question that some of our HAeBe stars could be classified as UXors, and some of our identified CTTSs as FUors or EXors, in the future.

A large number of identified PMS variable stars (165 TTS and 6 HAeBe stars) can be used to verify models of PMS stars.  Since the variability of WTTSs is caused by cool spots, the amplitudes, periods and shape of light curves can change over time \citep{cohen2004}. It would be interesting to check their variability after several years. Some additional observation can also be useful to check variability among the remaining Class II objects: to find the periods of long period variables and to verify the existence of early-type pulsating stars.

\section*{Acknowledgements}
This paper uses observations made at the Cerro Tololo Inter-American Observatory (CTIO). 
We thank anonymous referees for comments that have helped us to improve the paper. 
This work was supported by NCN grant No. 2016/21/B/ST9/01126. 

This research has made use of the VizieR catalogue access tool, CDS, Strasbourg, France. The original description of the VizieR service was published in A\&AS 143, 23.

This research has made use of the WEBDA database, operated at the Department of Theoretical Physics and Astrophysics of the Masaryk University.

This paper makes use of data obtained as part of the INT Photometric H-alpha Survey of the Northern Galactic Plane (IPHAS, www.iphas.org) carried out at the Isaac Newton Telescope (INT). The INT is operated on the island of La Palma by the Isaac Newton Group in the Spanish Observatorio del Roque de los Muchachos of the Instituto de Astrofisica  de Canarias. All IPHAS data are processed by the Cambridge Astronomical Survey Unit, at the Institute of Astronomy in Cambridge. The bandmerged DR2 catalogue was assembled at the Centre for Astrophysics Research, University of Hertfordshire, supported by STFC grant ST/J001333/1.

This work is based in part on data obtained as part of the UKIRT Infrared Deep Sky Survey.
    
This publication makes use of data products from the Wide-field Infrared Survey Explorer, which is a joint project of the University of California, Los Angeles, and the Jet Propulsion Laboratory/California Institute of Technology, funded by the National Aeronautics and Space Administration. 
 
This publication makes use of data products from the Two Micron All Sky Survey, which is a joint project of the University of Massachusetts and the Infrared Processing and Analysis Center/California Institute of Technology, funded by the National Aeronautics and Space Administration and the National Science Foundation.

This publication makes use of data products from the Wide-field Infrared Survey Explorer, which is a joint project of the University of California, Los Angeles, and the Jet Propulsion Laboratory/California Institute of Technology, funded by the National Aeronautics and Space Administration.

This research has also made use of IRAF. IRAF is distributed by the National Optical Astronomy Observatory, which is operated by the Association of Universities for Research in Astronomy, Inc., under cooperative agreement with the National Science Foundation. 

This work has made use of data from the European Space Agency (ESA) mission {\it Gaia} (https://www.cosmos.esa.int/gaia), processed by the {\it Gaia} Data Processing and Analysis Consortium (DPAC, https://www.cosmos.esa.int/web/gaia/dpac/consortium). Funding for the DPAC has been provided by national institutions, in particular the institutions participating in the {\it Gaia} Multilateral Agreement.

We are indebted to Prof.~Andrzej Pigulski and Przemysław Mikołajczyk for his comments made upon reading the manuscript. 
We thank Ewa Niemczura, Stefan Meingast and Ivanka Stateva who helped to carry out some additional observations which were, however, not included to the analysis.
\bibliographystyle{mn2e}
\bibliography{ngc2244} 
\bsp

\appendix
\section{}
In Table \ref{tubvi} we present $UBVI_{\rm C}$ photometry and coordinates of variable stars found in NGC 2244 together with some important information about these stars. In Table  \ref{tir} we present 2MASS, UKIDSS, Spitzer and IPHAS infrared  photometry of variable stars.

Phase diagrams of other periodic variables described in  Sect.3.3 are shown in Appendix \ref{sinph1} and the light curves of irregular variables variables described in Sect.~3.4. are shown in Appendix \ref{other}.
\clearpage
 
\begin{table*}
\scriptsize 
 \centering
 \begin{minipage}{160mm}
 \caption{\label{tubvi}$UBVI_{\rm C}$ photometry and coordinates of variable stars found in NGC 2244 together with some important information about these stars.}
\begin{tabular}{@{}rccccccccccccccc}
 \hline
 Star & R.Asc. & Dec. & $V$  & $V$--$I_{\rm C}$ & $B$--$V$ & $U$--$B$ & X$^{(1)}$ & H$\alpha ^{(2)}$ & Cl.$^{(3)}$ & $\alpha ^{(4)}$ & Var.$^{(5)}$ & NIR$^{(6)}$ &  Prob.$^{(7)}$ & Other$^{(8)}$ & Notes$^{(9)}$ \\
  & (2000.0) & (2000.0) & mag  & mag & mag & mag &  & &  & & type & reg. & [\% ] & name &\\
 \noalign{\vskip1pt}\hline\noalign{\vskip1pt}
1 &  6:32:00.61 & 4:52:40.9 &   8.530 & 0.254 & 0.156 & -0.746 & a & -- & III & -2.61 & ECL & - & 48.6 & V578 Mon & B0V+B1V \\
2 &  6:31:33.11 & 4:50:35.5 &  12.304 & 0.587 & 0.344 & -0.177 & - & BC PS & FS & -0.25 & HAeBe & - & 87.5 & HD 259012B & B4/A1Ve \\
3 &  6:31:29.76 & 4:54:49.2 &  12.421 & 0.453 & 0.328 & -0.256 & a & BC PS IA & I/FS & 0.24 & HAeBe & - & 84.7 & W74 & B7Ve \\
4 &  6:31:25.39 & 5:02:08.8 &  12.836 & 1.051 & 0.747 & -0.189 & - & -- & -- & --- & --- & - & 0.0 & W62 & B3 \\
5 &  6:31:30.88 & 4:58:12.4 &  12.892 & 0.439 & 0.415 & 0.296 & - & -- & III & -2.93 & DSCT & - & 27.1 & W69 & A2 \\
6 &  6:31:40.02 & 5:05:56.8 &  13.084 & 1.302 & 0.987 & 0.115 & - & BC & (II) & (-0.48) & HAeBe & - & 0.0 & W106 & F3Ve \\
\end{tabular}
 \end{minipage}
 \end{table*} 
\begin{table*}
\scriptsize 
 \centering
 \begin{minipage}{160mm}
 \contcaption{}
 \begin{tabular}{@{}rccccccccccccccc}
 \hline
 Star & R.Asc. & Dec. & $V$  & $V$--$I_{\rm C}$ & $B$--$V$ & $U$--$B$ & X$^{(1)}$ & H$\alpha ^{(2)}$ & Cl.$^{(3)}$ & $\alpha ^{(4)}$ & Var.$^{(5)}$ & NIR$^{(6)}$ &  Prob.$^{(7)}$ & Other$^{(8)}$ & Notes$^{(9)}$ \\
  & (2000.0) & (2000.0) & mag  & mag & mag & mag &  & &  & & type & reg. & [\% ] & name  &\\
 \noalign{\vskip1pt}\hline\noalign{\vskip1pt}
71 &  6:32:14.74 & 4:50:23.7 &  17.314 & 2.036 & 1.677 & --- & a & -- & III & -2.73 & WTTS* & F & 88.5 & -- &  \\
72 &  6:31:34.31 & 5:04:16.0 &  17.349 & 1.131 & 0.854 & 0.267 & - & -- & -- & --- & ECL & - & 0.0 & -- &  nm\\
73 &  6:32:05.93 & 4:57:27.7 &  17.350 & 2.077 & 1.493 & 0.686 & a & Ia & II & -0.73 & CTTS & F & 29.1 & -- &  \\
74 &  6:31:57.34 & 4:53:42.3 &  17.383 & 1.874 & 1.465 & --- & a & -- & II & -1.85 & --- & F & 89.7 & -- &  \\
75 &  6:31:43.92 & 5:02:38.9 &  17.404 & 1.410 & 1.056 & 0.454 & - & -- & -- & --- & DSCT & - & 1.1 & -- &  nm\\
76 &  6:32:07.82 & 4:52:28.5 &  17.407 & 1.843 & 1.439 & 0.565 & c & IA & II & -1.16 & CTTS & F & 88.9 & -- &  \\
\end{tabular}
 \end{minipage}
 \end{table*} 
\begin{table*}
\scriptsize 
 \centering
 \begin{minipage}{160mm}
 \contcaption{}
 \begin{tabular}{@{}rccccccccccccccc}
 \hline
 Star & R.Asc. & Dec. & $V$  & $V$--$I_{\rm C}$ & $B$--$V$ & $U$--$B$ & X$^{(1)}$ & H$\alpha ^{(2)}$ & Cl.$^{(3)}$ & $\alpha ^{(4)}$ & Var.$^{(5)}$ & NIR$^{(6)}$ &  Prob.$^{(7)}$ & Other$^{(8)}$ & Notes$^{(9)}$ \\
  & (2000.0) & (2000.0) & mag  & mag & mag & mag &  & &  & & type & reg. & [\% ] & name  &\\
 \noalign{\vskip1pt}\hline\noalign{\vskip1pt}
141 &  6:31:59.63 & 4:52:16.8 &  18.264 & 2.245 & --- & --- & a & Ib & III & -2.51 & WTTS & F & 88.3 & -- &  \\
142 &  6:31:39.62 & 4:59:45.1 &  18.270 & 2.454 & 1.640 & --- & a & Ia & II & -1.53 & CTTS & F & 74.9 & -- &  \\
143 &  6:31:49.81 & 4:49:06.8 &  18.277 & 2.268 & 1.611 & --- & - & Ib & III & -2.71 & WTTS & F & --- & -- &  \\
144 &  6:31:59.70 & 4:55:36.0 &  18.280 & 2.225 & --- & --- & - & -- & III & -2.66 & WTTS* & F & 89.1 & -- &  \\
145 &  6:31:48.05 & 4:50:48.3 &  18.287 & 2.343 & 1.389 & --- & - & IA & II & -0.54 & CTTS & F & 0.0 & -- &  \\
146 &  6:31:43.57 & 4:59:49.1 &  18.310 & 2.238 & 1.632 & --- & b & -- & -- & --- & --- & F & 57.7 & -- &  \\
\end{tabular}
 \end{minipage}
 \end{table*} 
\begin{table*}
\scriptsize 
 \centering
 \begin{minipage}{160mm}
 \contcaption{}
 \begin{tabular}{@{}rccccccccccccccc}
 \hline
 Star & R.Asc. & Dec. & $V$  & $V$--$I_{\rm C}$ & $B$--$V$ & $U$--$B$ & X$^{(1)}$ & H$\alpha ^{(2)}$ & Cl.$^{(3)}$ & $\alpha ^{(4)}$ & Var.$^{(5)}$ & NIR$^{(6)}$ &  Prob.$^{(7)}$ & Other$^{(8)}$ & Notes$^{(9)}$ \\
  & (2000.0) & (2000.0) & mag  & mag & mag & mag &  & &  & & type & reg. & [\% ] & name  &\\
 \noalign{\vskip1pt}\hline\noalign{\vskip1pt}
211 &  6:31:32.87 & 4:59:49.5 &  19.590 & 2.092 & --- & --- & a & -- & II & -1.33 & CTTS & T & 86.9 & -- &  nm\\
212 &  6:32:24.01 & 4:56:59.3 &  19.593 & 2.594 & --- & --- & c & IA & II & -1.31 & CTTS & F & 56.8 & -- &  \\
213 &  6:32:31.92 & 4:51:09.0 &  19.615 & 2.336 & --- & --- & - & Ia & II & -1.60 & CTTS & F & 12.7 & -- &  \\
214 &  6:32:22.70 & 4:58:08.4 &  19.617 & 2.586 & --- & --- & a & IA & II & -1.28 & CTTS & F & --- & -- &  \\
215 &  6:31:29.84 & 4:51:10.7 &  19.645 & 2.604 & --- & --- & a & IA & II & -1.04 & CTTS & T & --- & -- &  \\
216 &  6:31:25.46 & 4:49:31.0 &  19.648 & 2.426 & --- & --- & c & Ia & II & -1.19 & CTTS & F & 67.8 & -- &  \\
\end{tabular}
 \end{minipage}
 \end{table*}

\begin{table*}
\scriptsize 
 \centering
 \begin{minipage}{160mm}
 \caption{\label{tir}2MASS, UKIDSS, Spitzer and IPHAS infrared  photometry of variable stars.}
 \begin{tabular}{@{}rccccccccccccccc}
 \hline
 Star & Name & $J$ & $H$ & $K_S$  & $J$ & $H$ & $K$ & $[3.6]$ & $[4.5]$ & $[5.8]$ & $[8.0]$ & $[24]$ & $r$ & $i$ & $H_{\alpha}$\\
  & & mag & mag & mag & mag & mag & mag & mag & mag & mag & mag & mag & mag & mag & mag\\
 \noalign{\vskip1pt}\hline\noalign{\vskip1pt}
1 & J063200.60+045240.9 & 8.152 & 8.125 & 8.120 & 10.825 & 11.668 & 9.523 & 8.376 & 8.311 & 8.197 & 8.184 & --- & --- & --- & ---\\
2 & J063133.11+045035.4 & 10.770 & 9.942 & 9.115 & 10.715 & 10.823 & 9.641 & 7.713 & 7.136 & 6.435 & 5.484 & 2.707 & --- & --- & ---\\
3 & J063129.76+045449.1 & 11.449 & 10.887 & 9.690 & 11.371 & 11.148 & 9.843 & 7.422 & 6.753 & 6.061 & 4.740 & 1.783 & 12.22 & 11.89 & 11.57\\
4 & J063125.42+050209.3 & 10.977 & 10.711 & 10.544 & 10.913 & 11.005 & 10.560 & --- & --- & --- & --- & --- & --- & --- & ---\\
5 & J063130.87+045812.3 & 12.047 & 11.908 & 11.812 & 12.020 & 11.926 & 11.879 & 11.862 & 11.859 & 11.963 & 11.912 & --- & 12.69 & 12.46 & 12.62\\
6 & J063140.00+050556.4 & 9.684 & 8.956 & 8.274 & 10.209 & 11.272 & 9.477 & --- & --- & --- & --- & --- & --- & --- & ---\\
\end{tabular}
 \end{minipage}
 \end{table*} 
\begin{table*}
\scriptsize 
 \centering
 \begin{minipage}{160mm}
 \begin{tabular}{@{}rccccccccccccccc}
 \hline
 Star & Name & $J$ & $H$ & $K_S$  & $J$ & $H$ & $K$ & $[3.6]$ & $[4.5]$ & $[5.8]$ & $[8.0]$ & $[24]$ & $r$ & $i$ & $H_{\alpha}$\\
  & & mag & mag & mag & mag & mag & mag & mag & mag & mag & mag & mag & mag & mag & mag\\
 \noalign{\vskip1pt}\hline\noalign{\vskip1pt}
71 & J063214.74+045023.6 & 13.794 & 12.906 & 12.676 & 13.780 & 12.999 & 12.723 & 12.728 & 12.581 & 12.604 & 12.609 & --- & 16.55 & 15.44 & 16.01\\
72 & J063134.30+050416.0 & --- & --- & --- & 15.138 & 14.812 & 14.535 & --- & --- & --- & --- & --- & 16.89 & 16.25 & 16.63\\
73 & J063205.92+045727.6 & 13.449 & 12.502 & 12.068 & 13.622 & 12.735 & 12.176 & 11.887 & 11.212 & 10.869 & 9.991 & 6.671 & 16.70 & 15.50 & 15.90\\
74 & J063157.34+045342.3 & 13.958 & 13.091 & 12.824 & 13.934 & 13.208 & 12.910 & 12.233 & 12.060 & 11.874 & 11.360 & --- & 16.40 & 15.45 & 15.96\\
75 & J063143.91+050238.9 & 14.923 & 14.384 & 14.376 & 14.858 & 14.449 & 14.256 & --- & --- & --- & --- & --- & 16.87 & 16.08 & 16.60\\
76 & J063207.82+045228.4 & 14.030 & 13.208 & 12.764 & 14.067 & 13.225 & 12.683 & 11.990 & 11.504 & 11.249 & 10.485 & 8.159 & 16.74 & 15.73 & 15.95\\
\end{tabular}
 \end{minipage}
 \end{table*} 
\begin{table*}
\scriptsize 
 \centering
 \begin{minipage}{160mm}
 \begin{tabular}{@{}rccccccccccccccc}
 \hline
 Star & Name & $J$ & $H$ & $K_S$  & $J$ & $H$ & $K$ & $[3.6]$ & $[4.5]$ & $[5.8]$ & $[8.0]$ & $[24]$ & $r$ & $i$ & $H_{\alpha}$\\
  & & mag & mag & mag & mag & mag & mag & mag & mag & mag & mag & mag & mag & mag & mag\\
 \noalign{\vskip1pt}\hline\noalign{\vskip1pt}
141 & J063159.62+045216.8 & 14.161 & 13.353 & 13.160 & 14.168 & 13.439 & 13.127 & 12.868 & 12.845 & 12.745 & 12.589 & --- & 16.95 & 15.86 & 16.39\\
142 & J063139.61+045945.0 & 14.117 & 13.132 & 12.650 & 14.099 & 13.192 & 12.627 & 12.005 & 11.655 & 11.484 & 10.823 & 7.916 & 17.64 & 16.20 & 16.74\\
143 & J063149.80+044906.8 & 14.372 & 13.501 & 13.216 & 14.389 & 13.630 & 13.303 & 13.024 & 12.993 & 13.035 & 12.890 & --- & 17.47 & 16.18 & 16.81\\
144 & J063159.69+045535.9 & 14.320 & 13.471 & 13.292 & 14.352 & 13.574 & 13.277 & 13.151 & 13.063 & 13.135 & 12.960 & --- & 17.57 & 16.27 & 16.99\\
145 & J063148.05+045048.3 & 14.152 & 13.253 & 12.763 & 14.311 & 13.427 & 12.843 & 12.194 & 11.725 & 11.129 & 10.210 & 7.033 & 17.53 & 16.20 & 16.48\\
146 & J063143.56+045949.0 & 14.450 & 13.652 & 13.360 & 14.478 & 13.720 & 13.427 & --- & --- & --- & --- & --- & 17.58 & 16.38 & 16.99\\
\end{tabular}
 \end{minipage}
 \end{table*} 
\begin{table*}
\scriptsize 
 \centering
 \begin{minipage}{160mm}
 \begin{tabular}{@{}rccccccccccccccc}
 \hline
 Star & Name & $J$ & $H$ & $K_S$  & $J$ & $H$ & $K$ & $[3.6]$ & $[4.5]$ & $[5.8]$ & $[8.0]$ & $[24]$ & $r$ & $i$ & $H_{\alpha}$\\
  & & mag & mag & mag & mag & mag & mag & mag & mag & mag & mag & mag & mag & mag & mag\\
 \noalign{\vskip1pt}\hline\noalign{\vskip1pt}
211 & J063132.86+045949.4 & 14.692 & 13.605 & 12.884 & 14.831 & 13.623 & 12.586 & 11.782 & 11.281 & 11.074 & 10.411 & 7.183 & 17.53 & 16.46 & 17.04\\
212 & J063224.01+045659.1 & 14.971 & 14.072 & 13.710 & 14.895 & 14.096 & 13.721 & 13.190 & 12.786 & 12.508 & 11.833 & --- & 18.30 & 16.94 & 17.21\\
213 & J063231.91+045109.0 & 15.632 & 14.511 & 13.928 & 15.446 & 14.433 & 13.815 & 13.146 & 12.828 & 12.641 & 12.037 & 8.787 & 18.80 & 17.41 & 17.78\\
214 & J063222.70+045808.2 & 15.048 & 14.198 & 13.645 & 15.040 & 14.213 & 13.630 & 12.673 & 12.296 & 12.020 & 11.294 & 7.954 & 18.67 & 17.26 & 17.58\\
215 & J063129.84+045110.6 & 15.015 & 13.671 & 12.738 & 14.962 & 13.799 & 12.645 & 11.434 & 10.849 & 10.390 & 9.858 & 6.408 & 19.13 & 17.61 & 17.46\\
216 & J063125.45+044930.9 & 15.345 & 14.550 & 14.154 & 15.206 & 14.393 & 13.981 & 13.424 & 13.167 & 12.756 & 12.007 & --- & 18.84 & 17.45 & 17.80\\
\end{tabular}
 \end{minipage}
 \end{table*}

\begin{figure*}
\centering
\includegraphics[width=42pc]{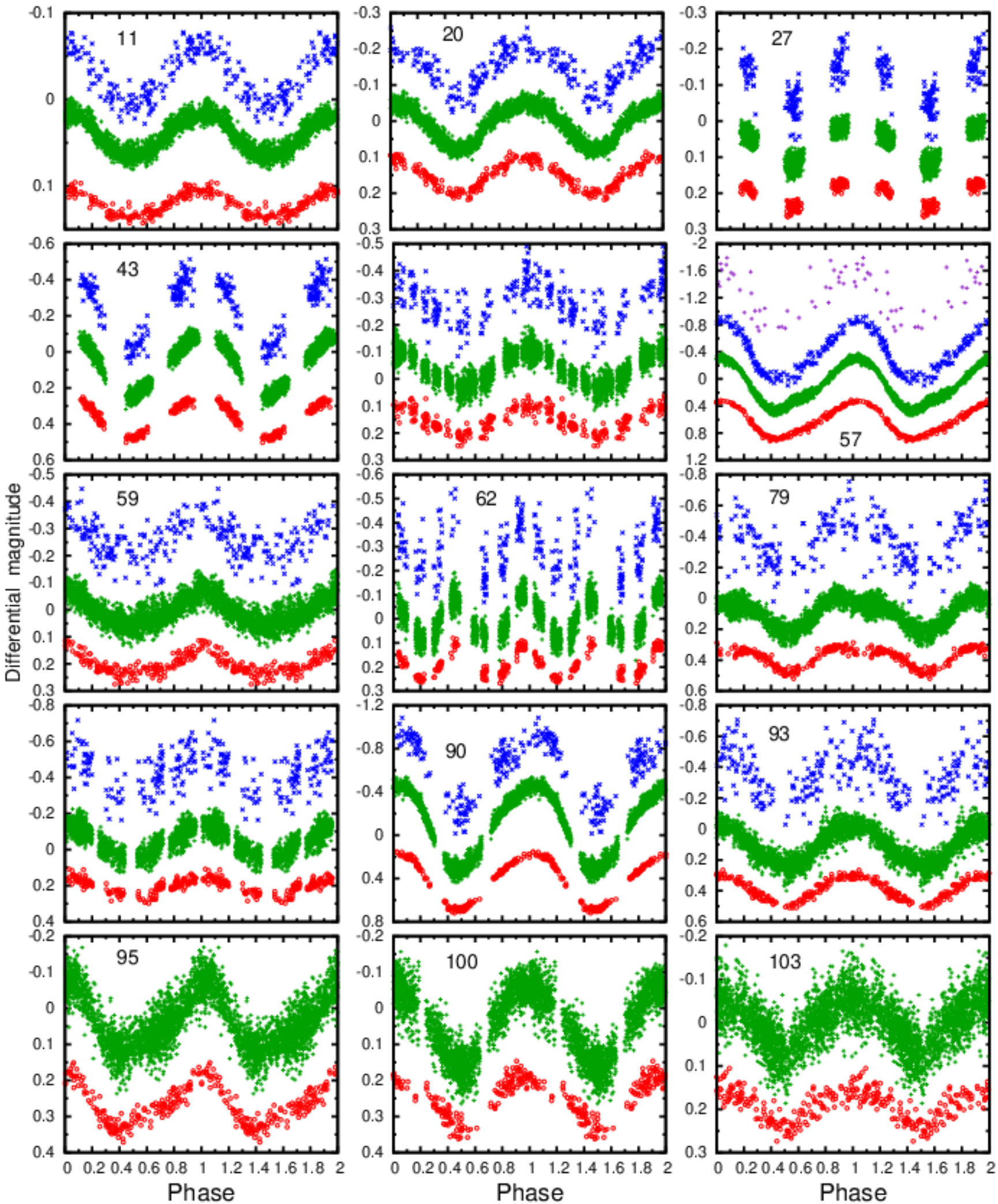}
\caption{Phase diagrams of $B$ (blue), $V$ (green) and $I$ (red) observations of other periodic variables found in observed field of NGC 2244 (see Sect.~3.3).} 
\label{sinph1}
\end{figure*}
\begin{figure*}
\centering
\includegraphics[width=42pc]{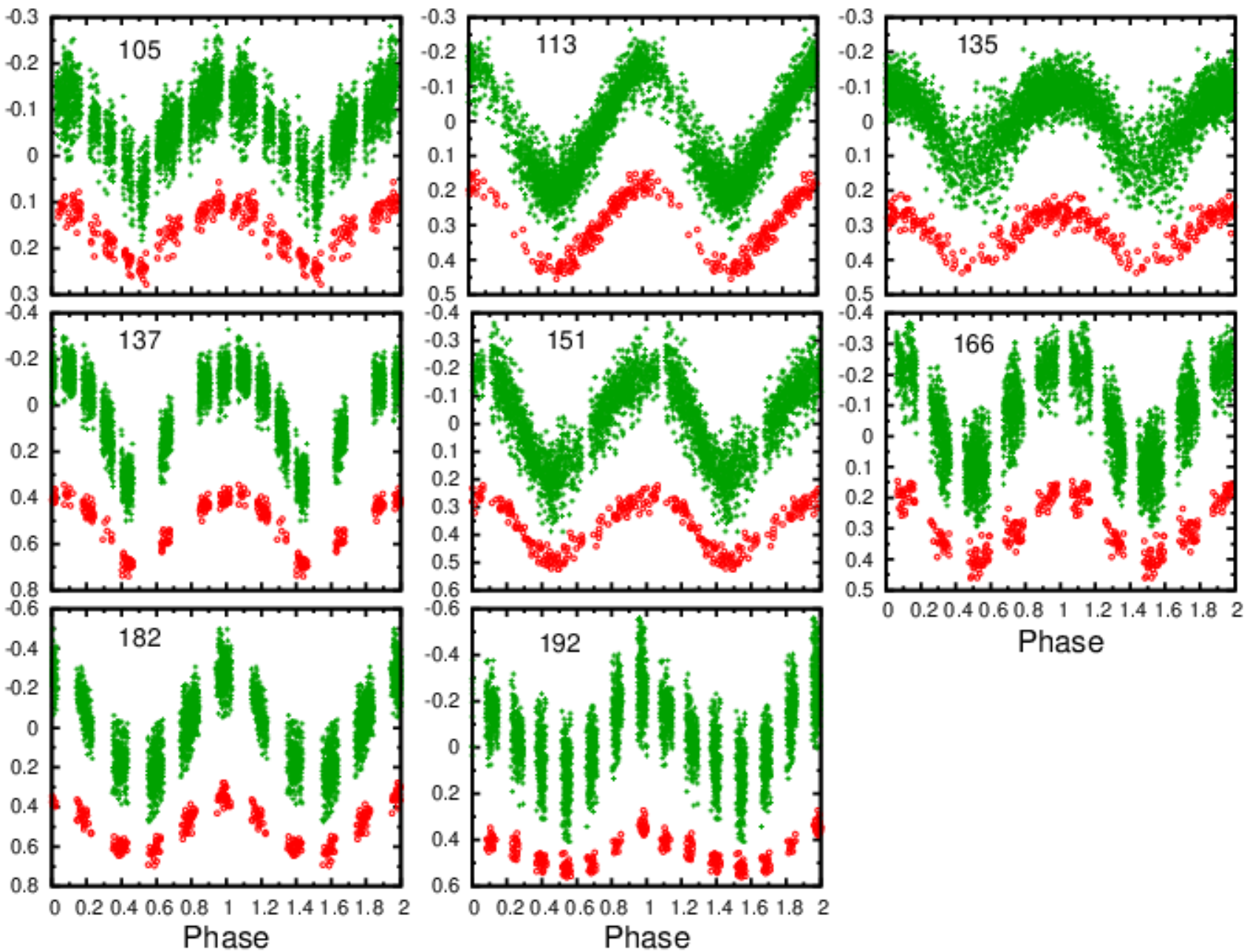}
  \contcaption{}
\end{figure*}

\begin{figure*}
\centering
\includegraphics[width=42pc]{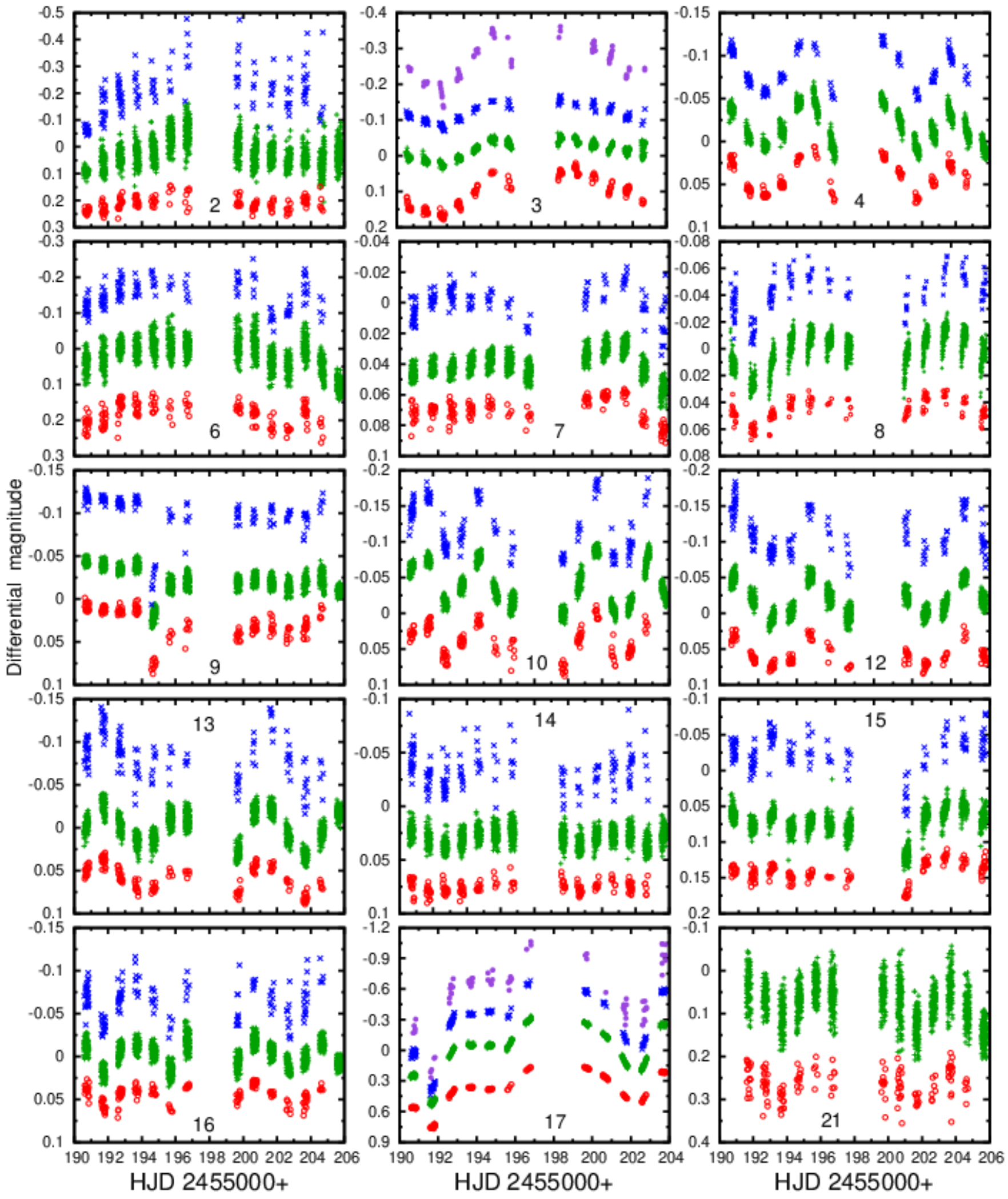}
  \caption{Light curves of $B$ (blue), $V$ (green) and $I$ (red) observations of irregular variables found in observed field of NGC 2244(see Sect.3.4).}
\label{other}
\end{figure*}
\begin{figure*}
\centering
\includegraphics[width=42pc]{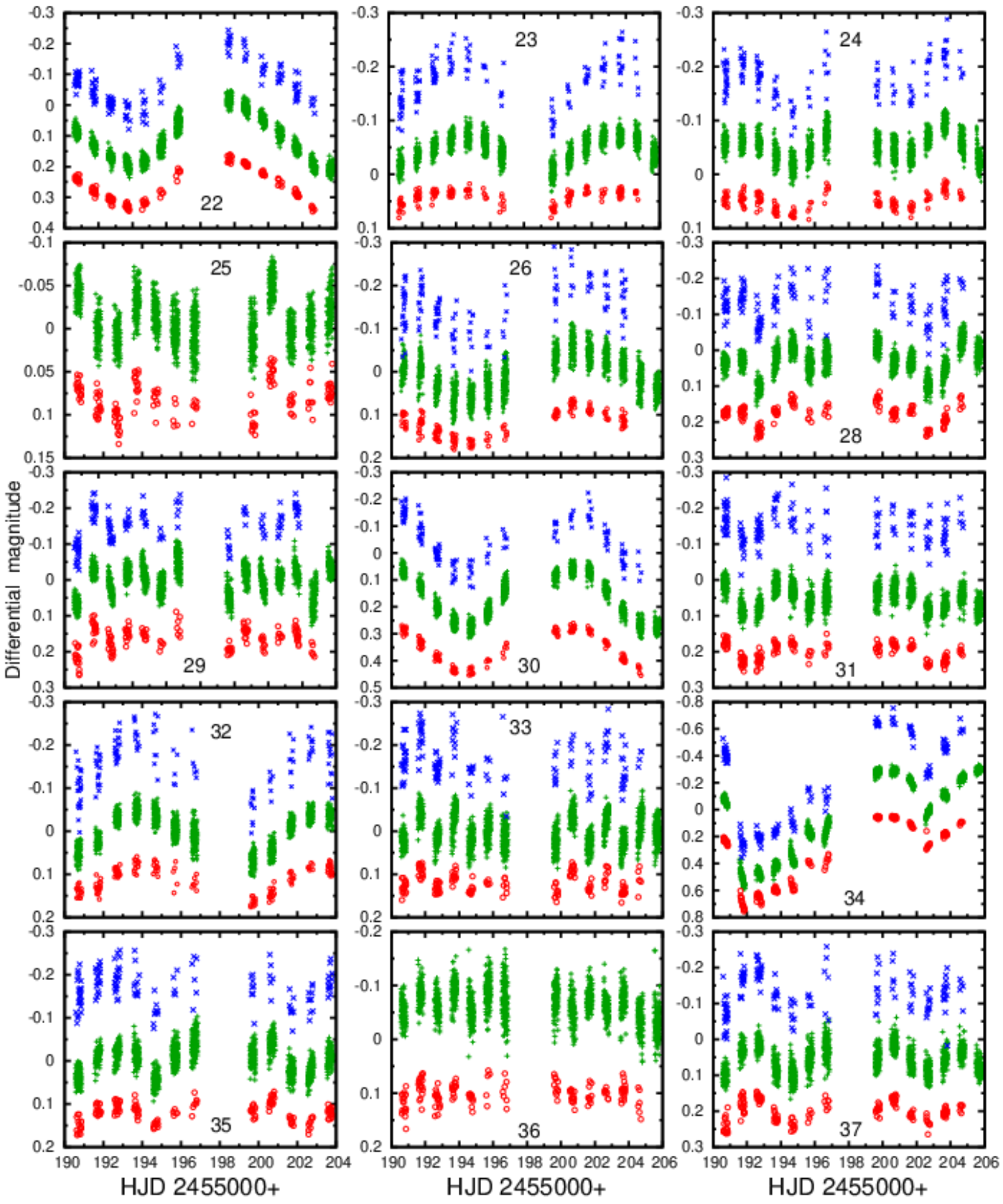}
  \contcaption{}
\end{figure*}
\begin{figure*}
\centering
\includegraphics[width=42pc]{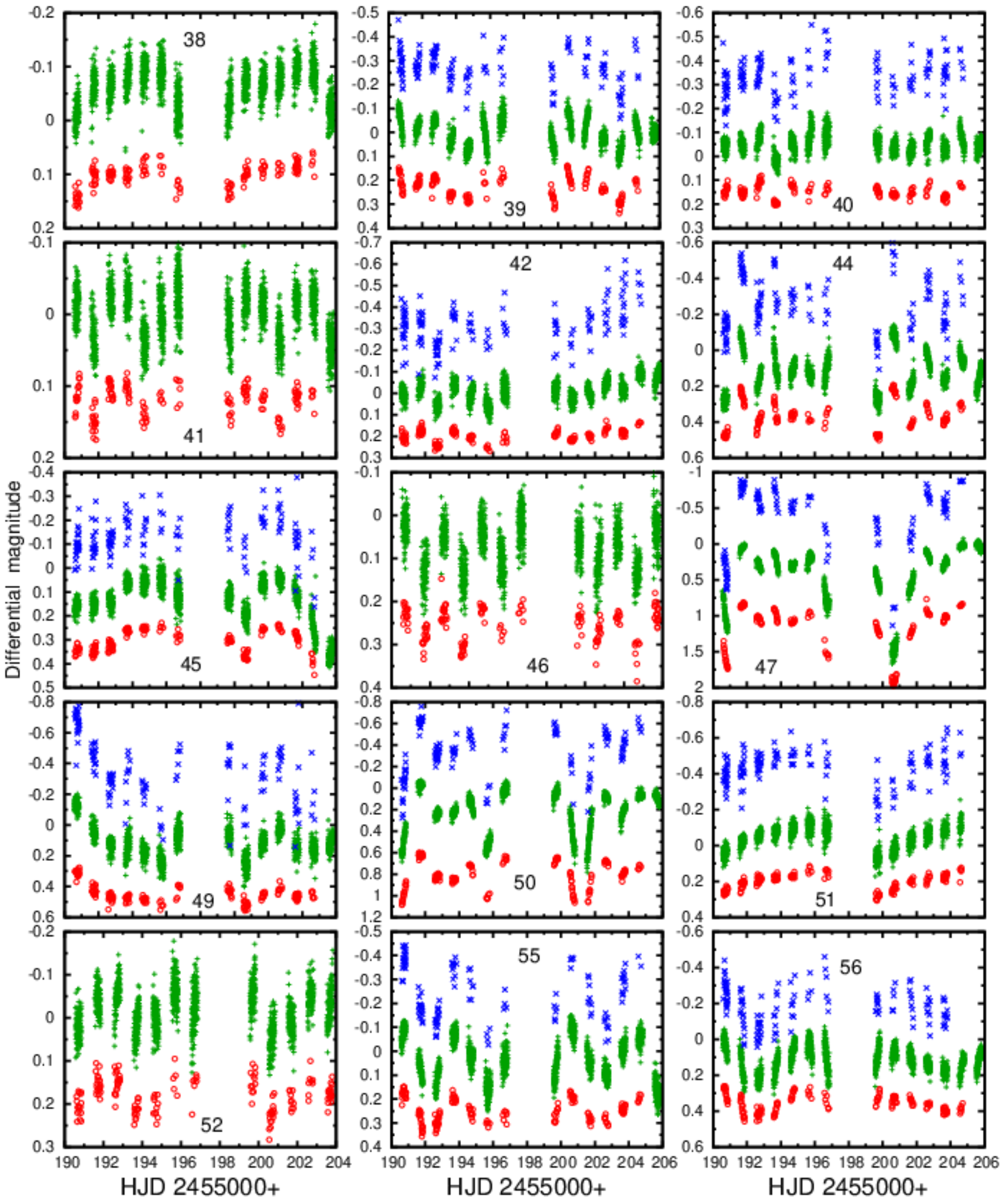}
  \contcaption{}
\end{figure*}
\begin{figure*}
\centering
\includegraphics[width=42pc]{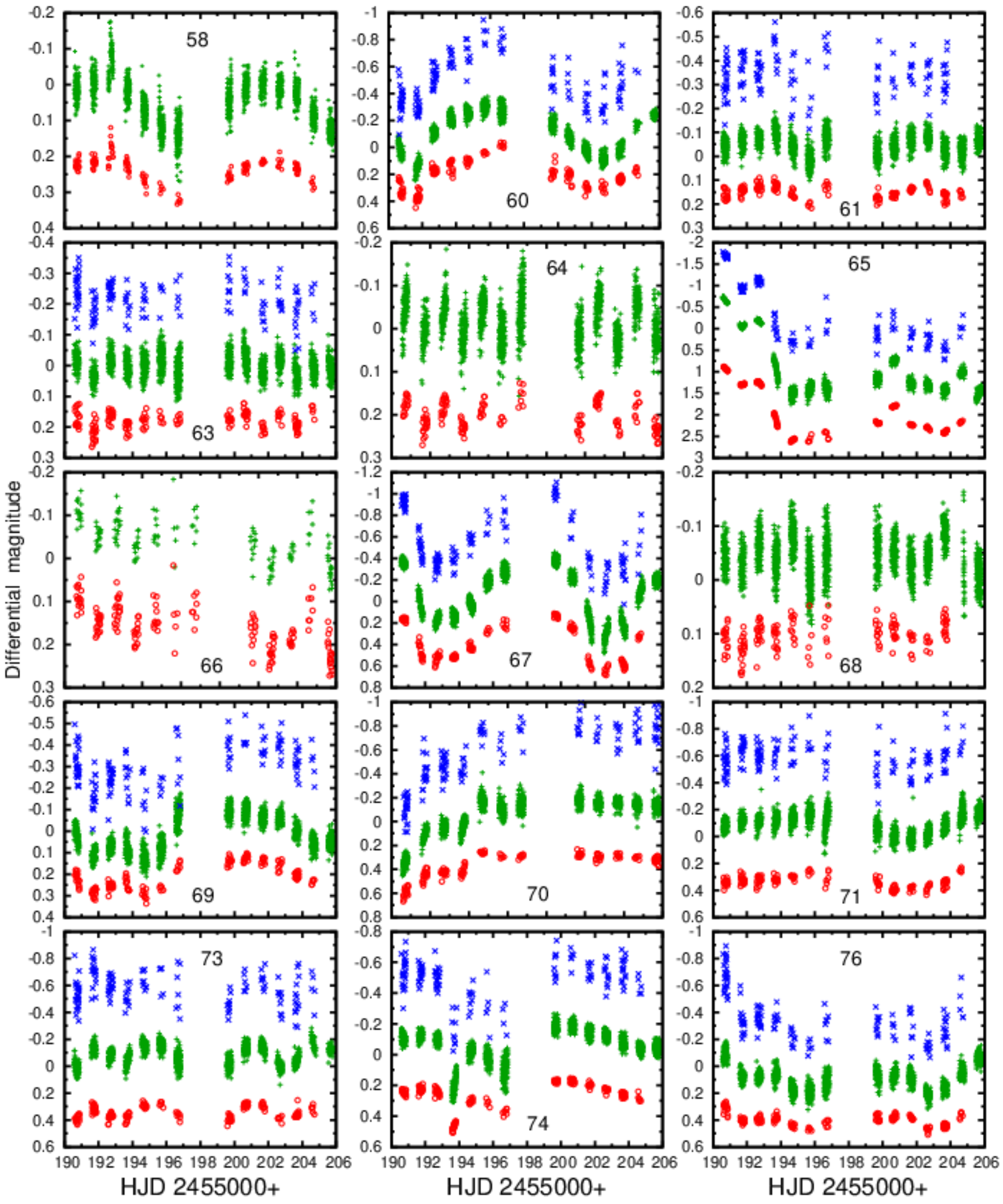}
  \contcaption{}
\end{figure*}
\begin{figure*}
\centering
\includegraphics[width=42pc]{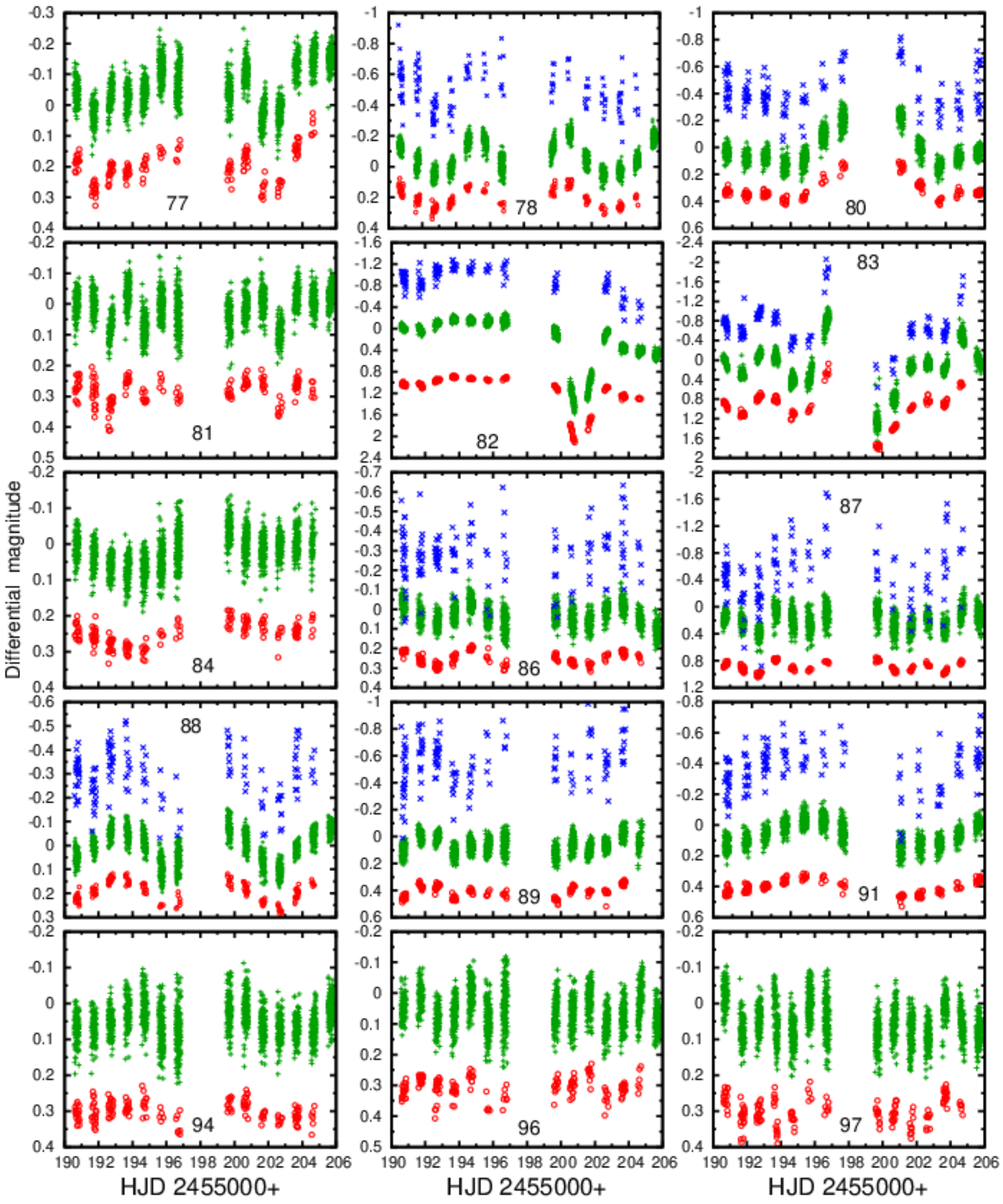}
  \contcaption{}
\end{figure*}
\begin{figure*}
\centering
\includegraphics[width=42pc]{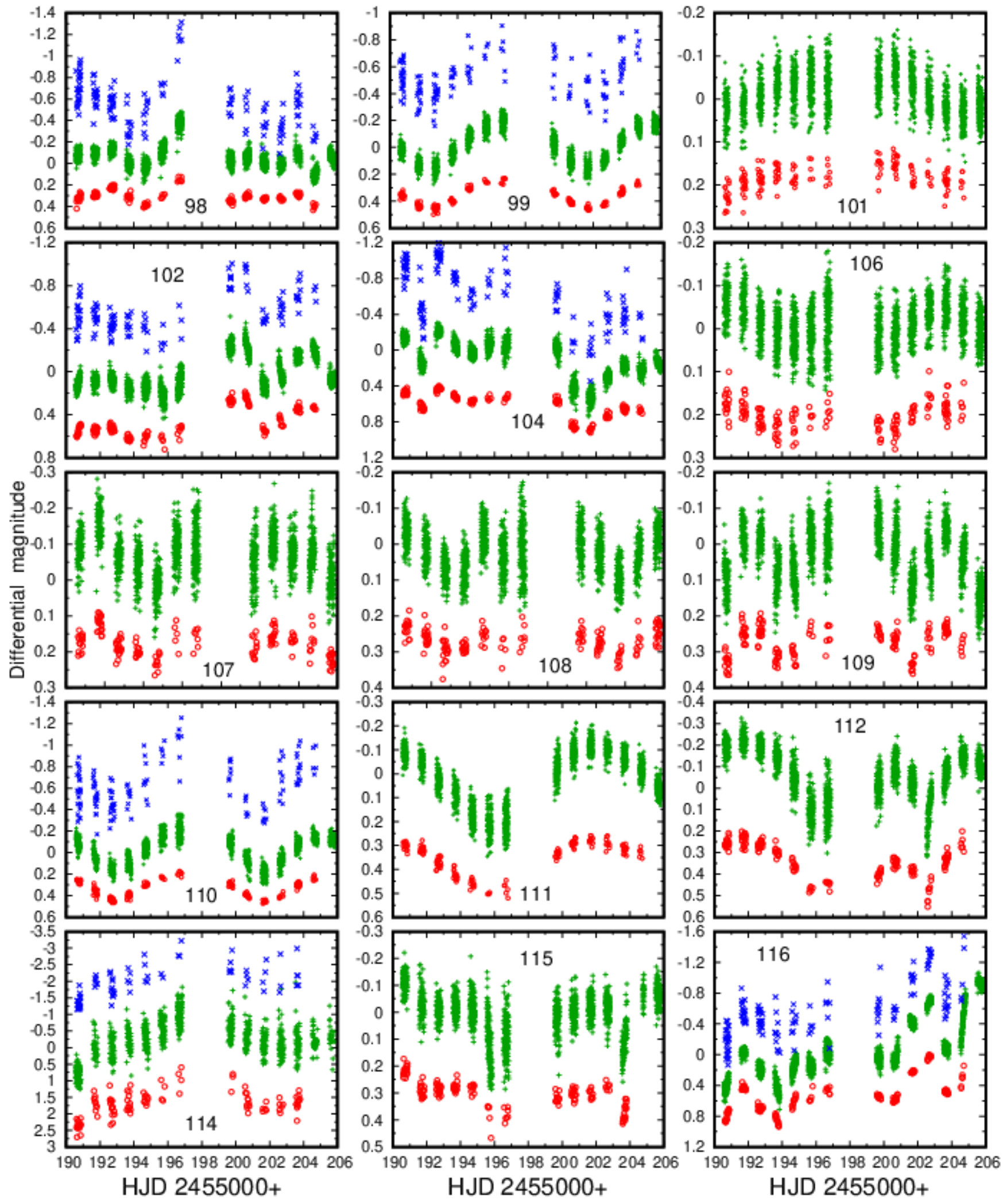}
  \contcaption{}
\end{figure*}
\begin{figure*}
\centering
\includegraphics[width=42pc]{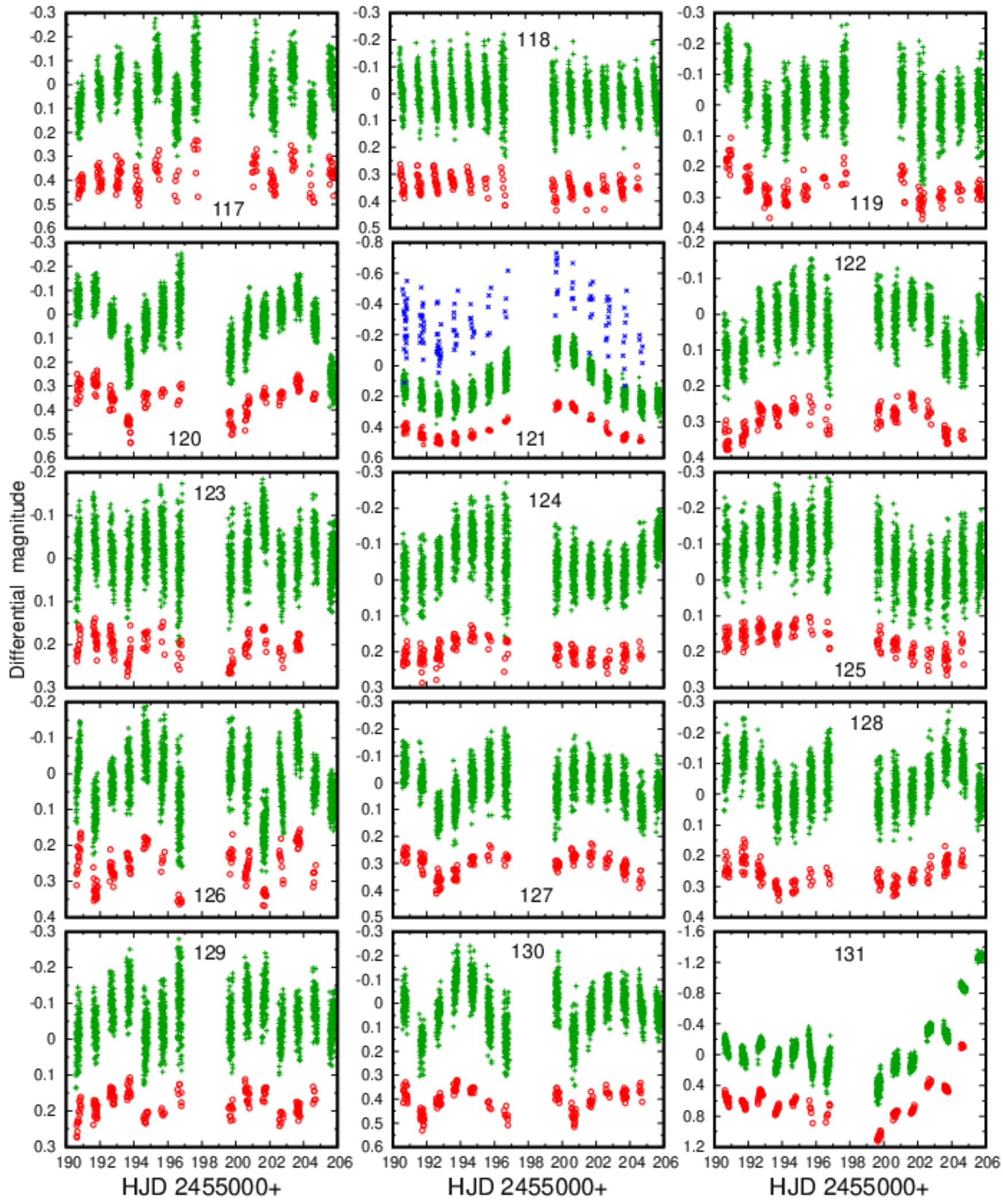}
  \contcaption{}
\end{figure*}
\begin{figure*}
\centering
\includegraphics[width=42pc]{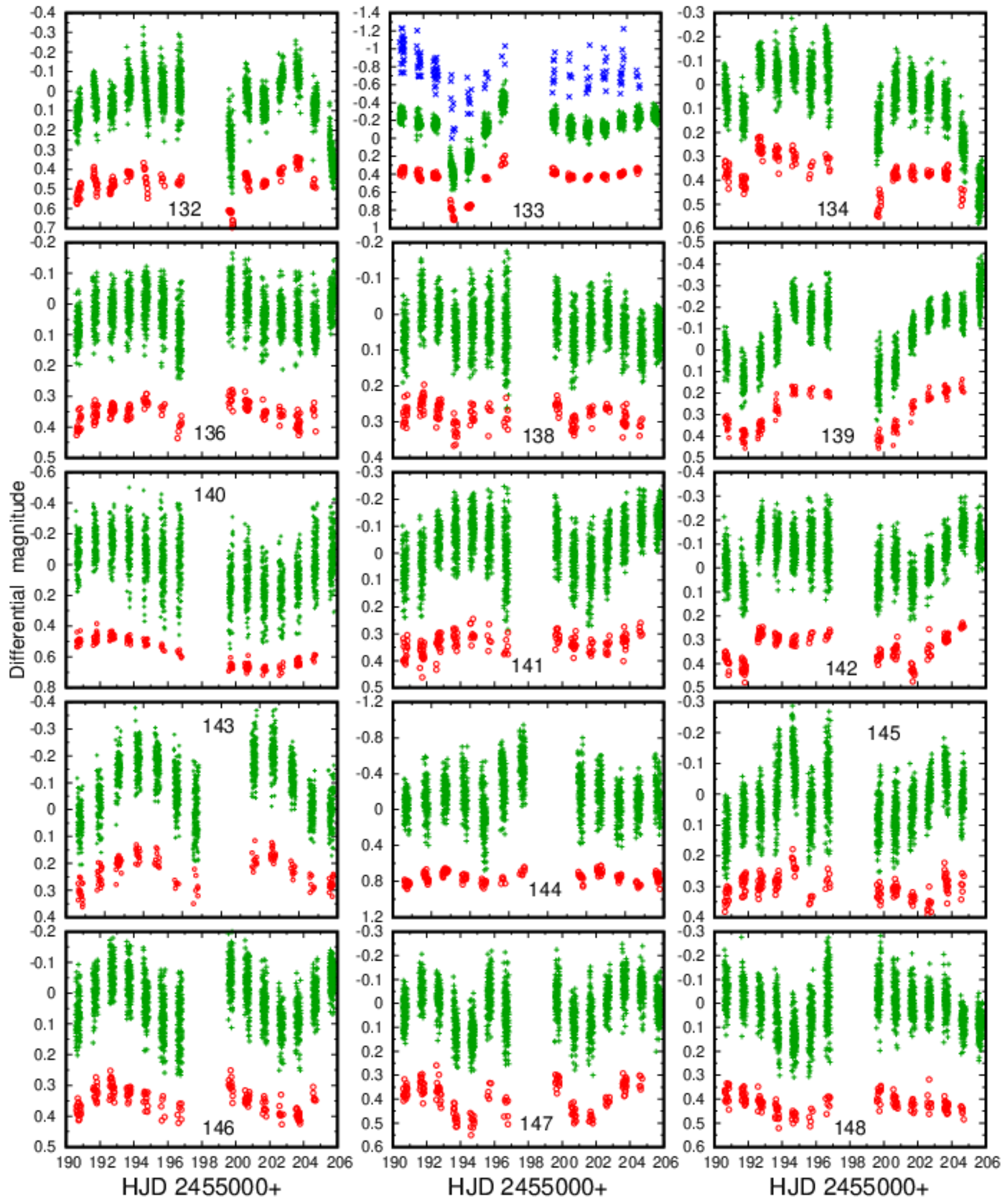}
  \contcaption{}
\end{figure*}
\begin{figure*}
\centering
\includegraphics[width=42pc]{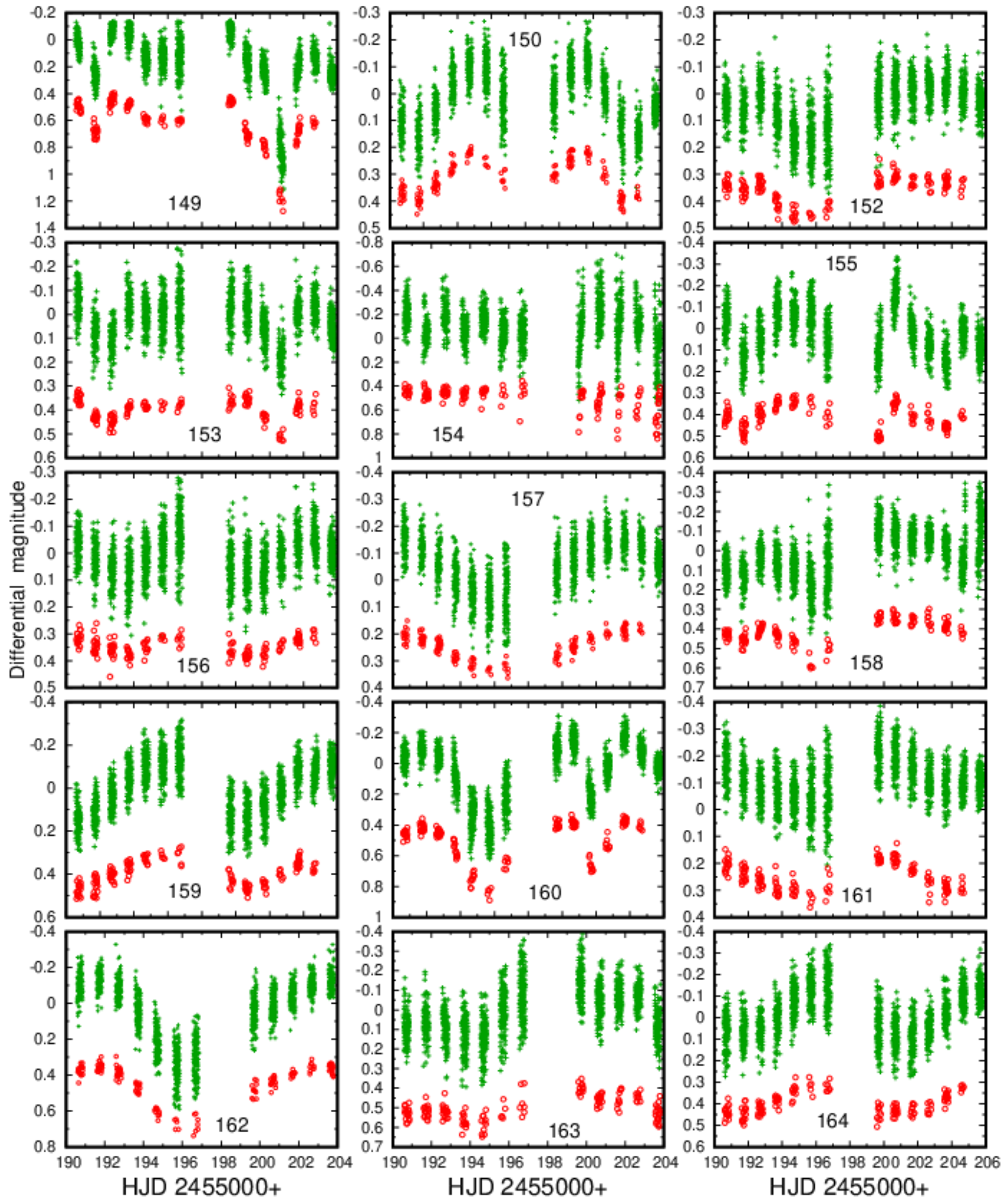}
  \contcaption{}
\end{figure*}
\begin{figure*}
\centering
\includegraphics[width=42pc]{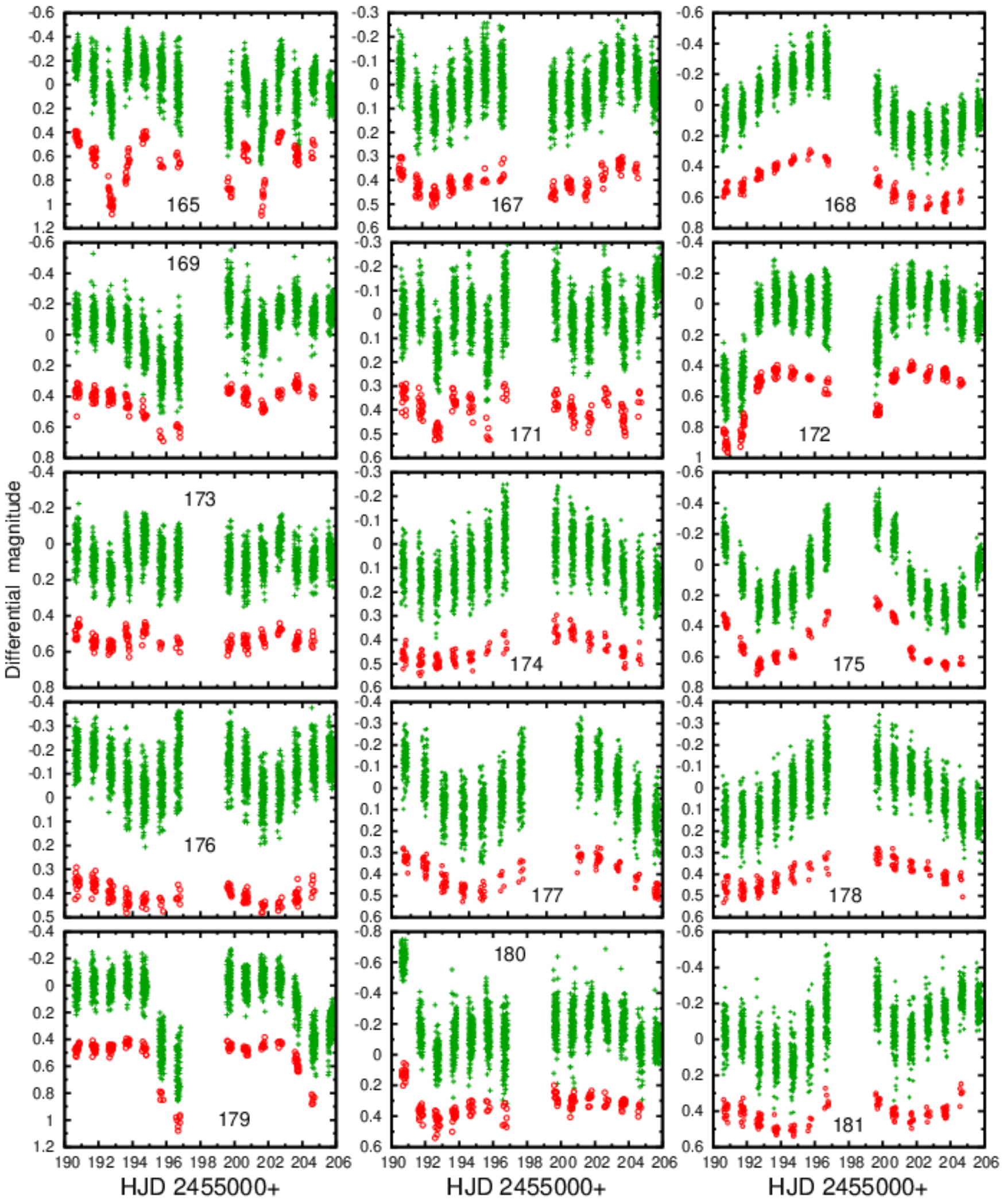}
  \contcaption{}
\end{figure*}
\begin{figure*}
\centering
\includegraphics[width=42pc]{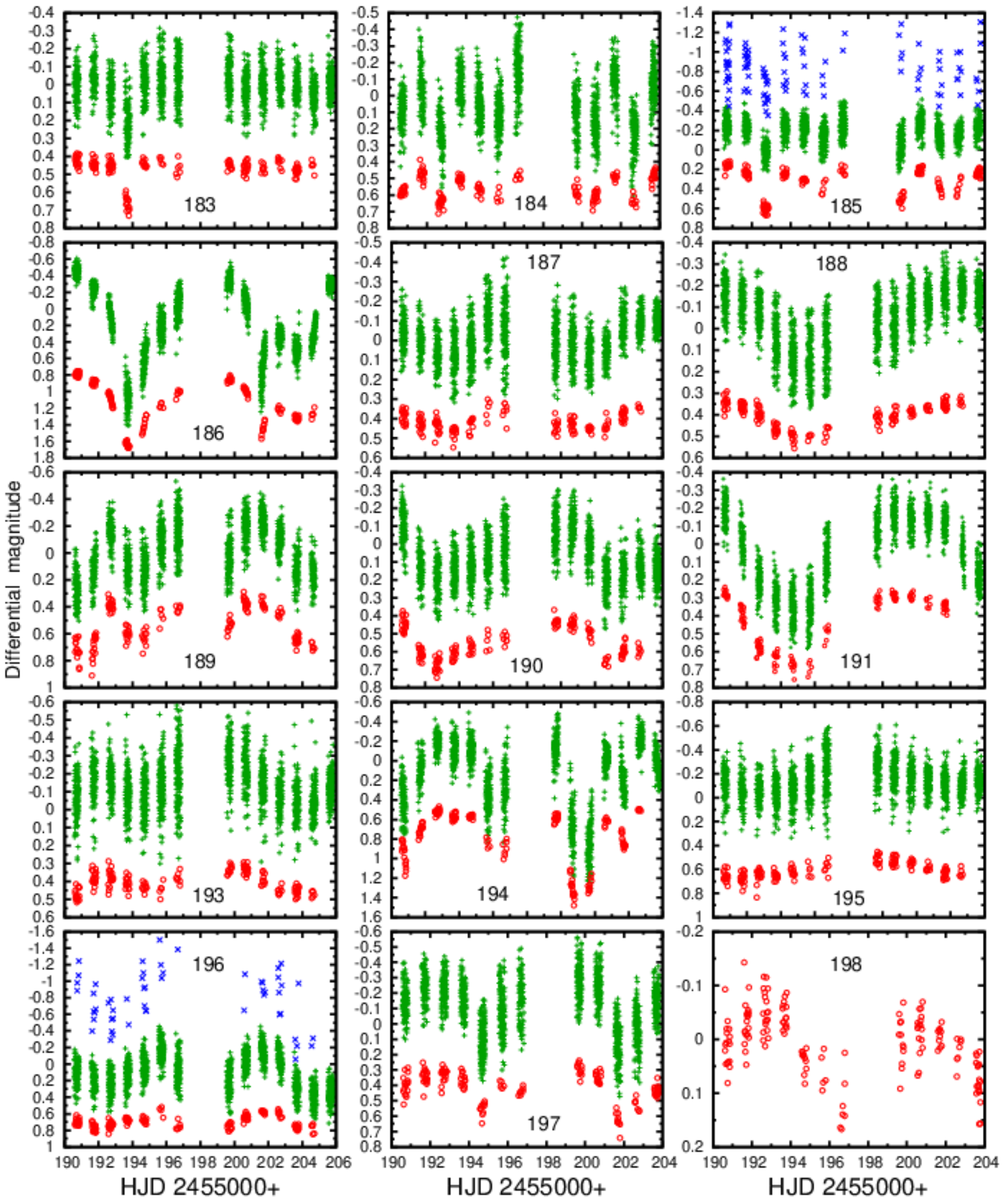}
  \contcaption{}
\end{figure*}
\begin{figure*}
\centering
\includegraphics[width=42pc]{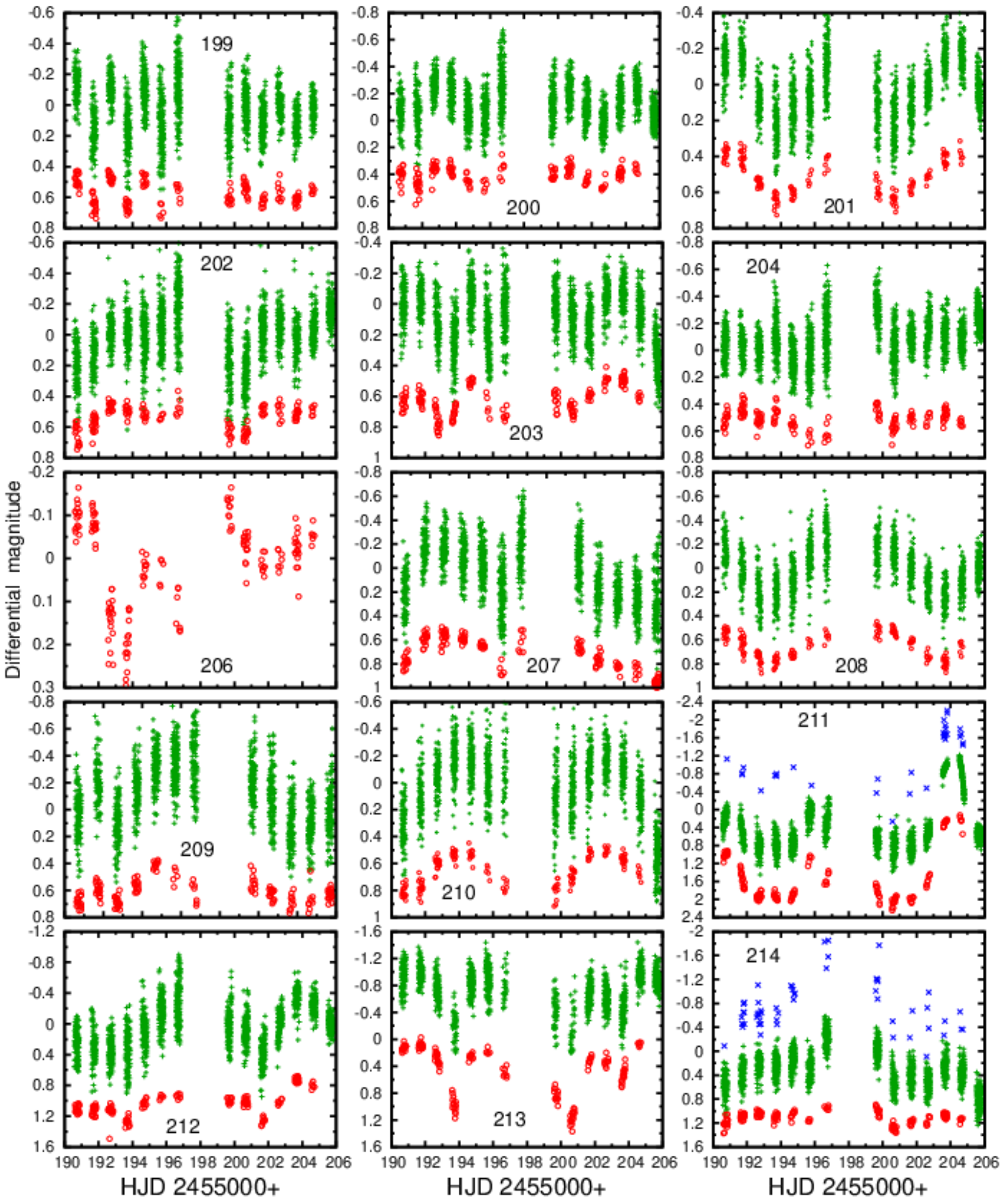}
  \contcaption{}
\end{figure*}
\begin{figure*}
\centering
\includegraphics[width=42pc]{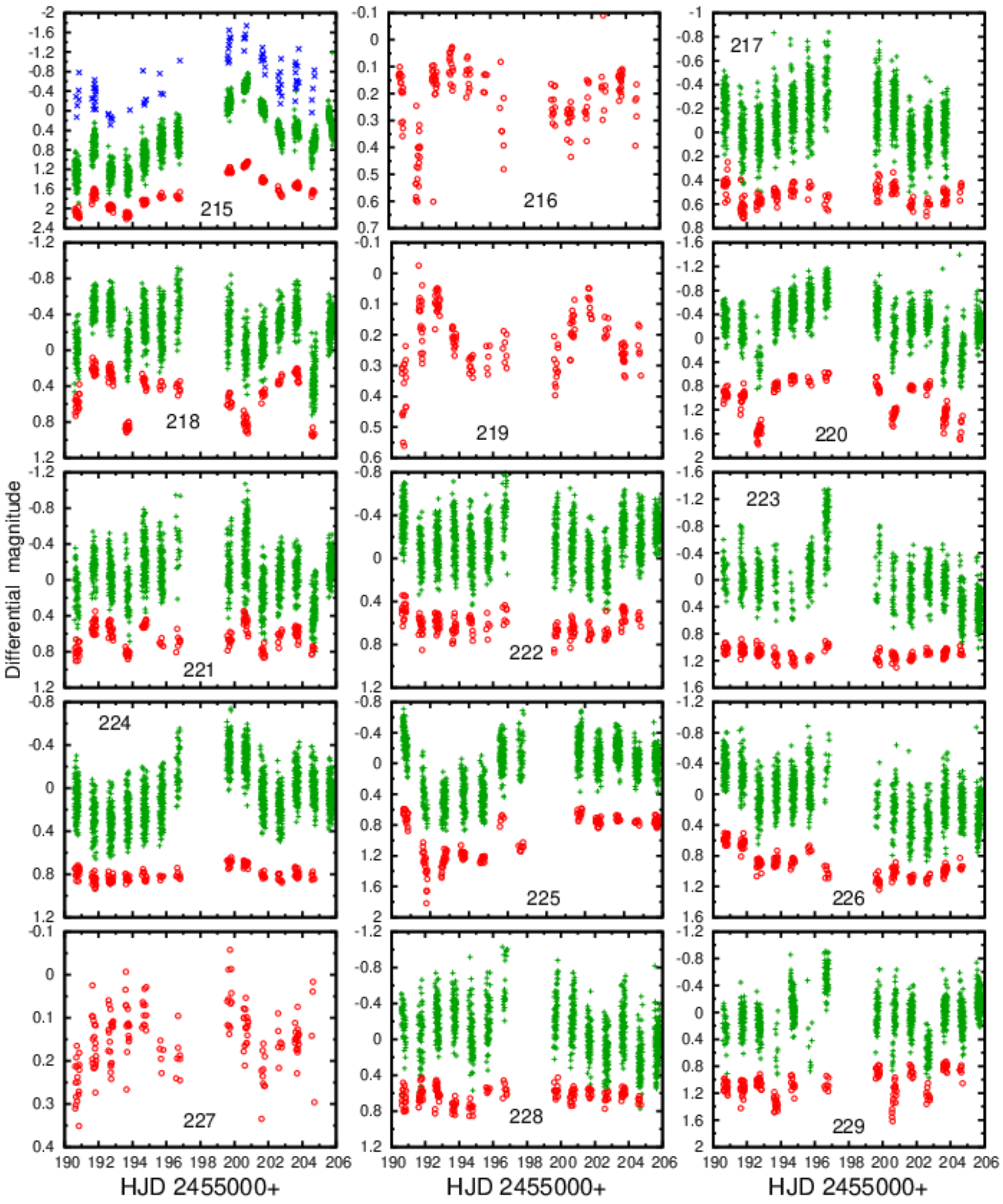}
  \contcaption{}
\end{figure*}
\begin{figure*}
\centering
\includegraphics[width=42pc]{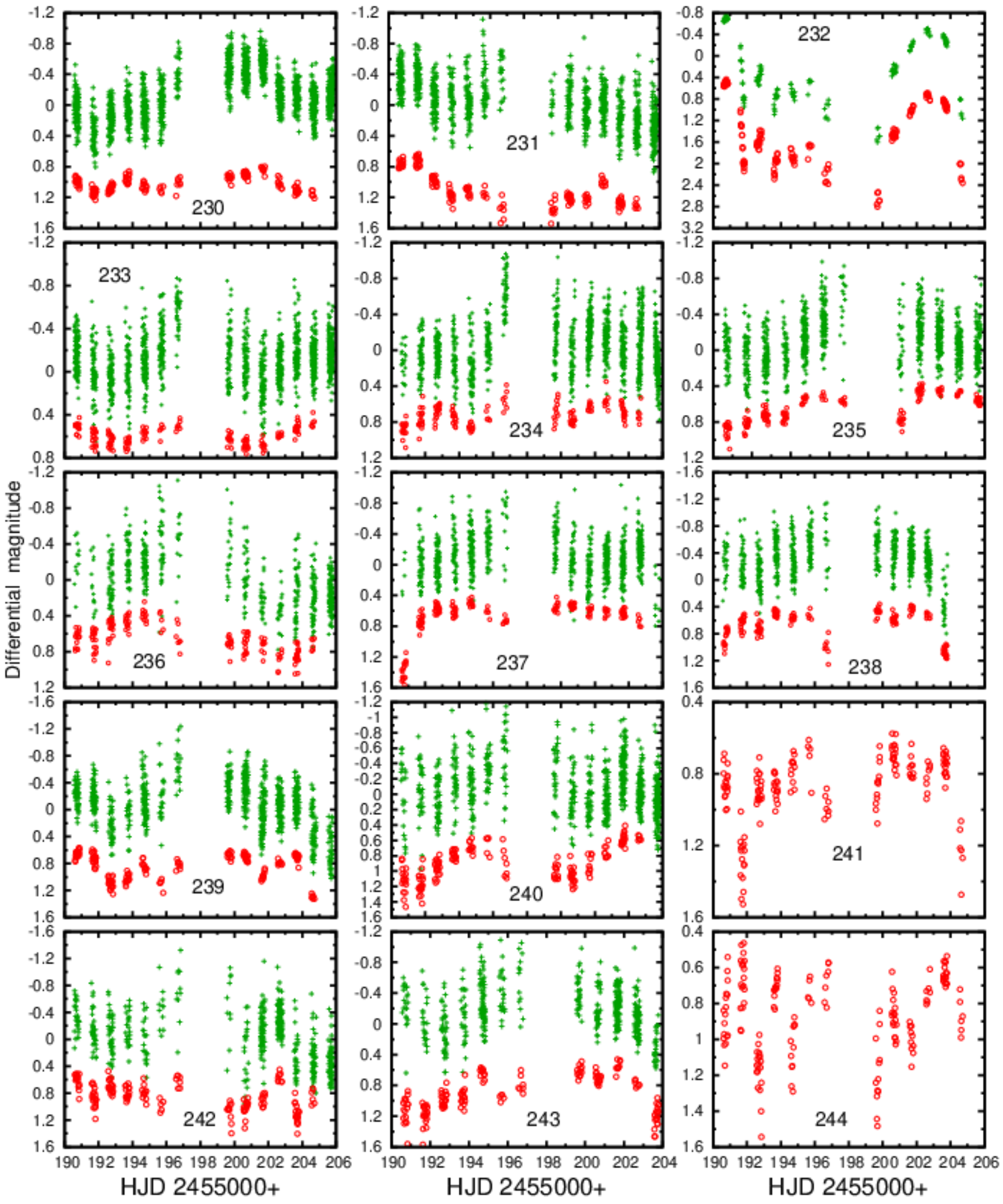}
  \contcaption{}
\end{figure*}
\begin{figure*}
\centering
\includegraphics[width=14pc]{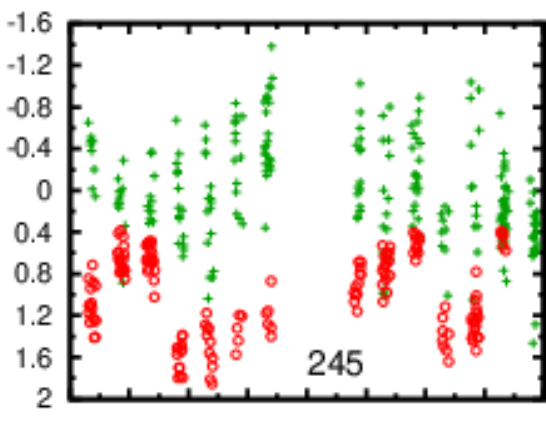}
  \contcaption{}
\end{figure*}

\label{lastpage}
\end{document}